\documentclass[english,5p]{elsarticle}
\usepackage[applemac]{inputenc}
\usepackage{babel}
\usepackage{amsmath}
\usepackage{graphicx}
\usepackage{subfigure}
\usepackage[draft]{hyperref}
\usepackage{color}

\makeatletter

\biboptions{authoryear}

\makeatother

\begin{document}
\begin{frontmatter}

\title{Cluster Failure Revisited: Impact of First Level Design \\and Data Quality on Cluster False Positive Rates}

\author[mymainaddress,mysecondaryaddress,mythirdaddress]{Anders Eklund\corref{mycorrespondingauthor}}
\cortext[Anders Eklund]{Corresponding author}
\ead{anders.eklund@liu.se}

\author[mymainaddress,mythirdaddress]{Hans Knutsson}

\author[myfourthaddress,myfifthaddress,mysixthaddress]{Thomas E. Nichols}

\address[mymainaddress]{Division of Medical Informatics, Department of Biomedical Engineering,
Link\"{o}ping University, Link\"{o}ping, Sweden}

\address[mysecondaryaddress]{Division of Statistics \& Machine Learning, Department of Computer
and Information Science, Link\"{o}ping University, Link\"{o}ping, Sweden}

\address[mythirdaddress]{Center for Medical Image Science and Visualization (CMIV), Link\"{o}ping University, Link\"{o}ping, Sweden}

\address[myfourthaddress]{Big Data Institute, University of Oxford, Oxford, United Kingdom}

\address[myfifthaddress]{Wellcome Trust Centre for Integrative Neuroimaging (WIN-FMRIB), University of Oxford, Oxford, United Kingdom}

\address[mysixthaddress]{Department of Statistics, University of Warwick, Coventry, United Kingdom}

\begin{abstract}

Methodological research rarely generates a broad interest, yet our work on the validity of cluster inference methods for functional magnetic resonance imaging (fMRI) created intense discussion on both the minutia of our approach and its implications for the discipline. In the present work, we take on various critiques of our work and further explore the limitations of our original work. We address issues about the particular event-related designs we used, considering multiple event types and randomisation of events between subjects. We consider the lack of validity found with one-sample permutation (sign flipping) tests, investigating a number of approaches to improve the false positive control of this widely used procedure. {\color{black} We found that the combination of a two-sided test and cleaning the data using ICA FIX resulted in nominal false positive rates for all datasets, meaning that data cleaning is not only important for resting state fMRI, but also for task fMRI.} Finally, we discuss the implications of our work on the fMRI literature as a whole, estimating that at least 10\% of the fMRI studies have used the most problematic cluster inference method (P = 0.01 cluster defining threshold), and how individual studies can be interpreted in light of our findings. These additional results underscore our original conclusions, on the importance of data sharing and thorough evaluation of statistical methods on realistic null data.

\end{abstract}
\end{frontmatter}

\section{Introduction}

In our previous work~\citep{eklundPNAS} we used freely available resting state functional magnetic resonance imaging (fMRI) data to evaluate the validity of standard fMRI inference methods. {\color{black} Group analyses involving only healthy controls were used to empirically estimate the degree of false positives, after correcting for multiple comparisons, based on the idea that a two-sample t-test using only healthy controls should lead to nominal false positive rates (e.g. 5\%). By considering resting state fMRI as null task fMRI data, the same approach was used to evaluate the statistical methods for one-sample t-tests. Briefly, we found that parametric statistical methods (e.g. Gaussian random field theory (GRFT)) perform well for voxel inference, where each voxel is separately tested for significance, but the combination of voxel inference and familywise error (FWE) correction is seldom used due to its low statistical power. For this reason, the false discovery rate (FDR) is in neuroimaging~\citep{fdr} often used to increase statistical power. For cluster inference, where groups of voxels are tested together by looking at the size of each cluster, we found that parametric methods perform well for a high cluster defining threshold (CDT) (p = 0.001) but result in inflated false positive rates for low cluster defining thresholds (e.g. p = 0.01). GRFT is for cluster inference based on two additional assumptions, compared to GRFT for voxel inference, and we found that these assumptions are violated in the analyzed data. First, the spatial autocorrelation function (SACF) is assumed to be Gaussian, but real fMRI data have a SACF with a much longer tail. Second, the spatial smoothness is assumed to be constant over the brain, which is not the case for fMRI data. The non-parametric permutation test is not based on these assumptions~\citep{winkler} and therefore produced nominal results for all two-sample t-tests, but in some cases failed to control FWE for one-sample t-tests.}

\subsection{Related work}

Our paper has generated intense discussions regarding cluster inference in fMRI~\citep{eklundPNAS2,coxPNAS,coxbiorxiv,kessler,friston,Kaundinya1,etac}, 
  on the validity of using resting state fMRI data as null data~\citep{slotnick,slotnick2,nichols}, how the spatial resolution can affect parametric cluster inference~\citep{mueller}, how to obtain residuals with a Gaussian spatial autocorrelation function~\citep{Kaundinya2}, {\color{black} how to model the long-tail spatial autocorrelation function~\citep{coxbiorxiv}}, as well as how different MR sequences can change the spatial autocorrelation function and thereby cluster inference~\citep{wald}. Furthermore, some of our results have been reproduced and extended~\citep{coxbiorxiv,kessler,friston,mueller}, using the same freely available fMRI data~\citep{biswal2,openfmri} and our processing scripts available on github\footnote{\label{note1}https://github.com/wanderine/ParametricMultisubjectfMRI}. Cluster based methods have now also been evaluated for surface-based group analyses of cortical thickness, surface area and volume (using FreeSurfer)~\citep{greve}, with a similar conclusion that the non-parametric permutation test showed good control of the FWE for all settings, while traditional Monte Carlo methods fail to control FWE for some settings. 
  
\subsection{Realistic first level designs}  

The event related paradigms (E1, E2) used in our study were criticized by some for not being realistic designs, as only a single regressor was used~\citep{slotnick} and the rest between the events was too short. The concern here is that this design may have a large transient at the start (due to the delay of the hemodynamic response function) and then only small variation (due to the short interstimulus interval), which may be overly-sensitive to transients at the start of the acquisition (Figure 1 (a), however, shows this is not really the case). Another criticism was that exactly the same task was used for all subjects~\citep{friston}, meaning that our "false positives" actually reflect consistent pretend-stimulus-linked behavior over subjects. Yet another concern was if the first few volumes (often called dummy scans) in each fMRI dataset were included in the analysis or not\footnote{http://www.ohbmbrainmappingblog.com/blog/keep-calm-and-scan-on, comment by John Ashburner}, which can affect the statistical analyses. This last point we can definitively address, as according to a NITRC document\footnote{http://www.nitrc.org/docman/view.php/296/716/fcon\_1000\_Preprocessing.pdf}, the first 5 time points of each time series were discarded for all data included in the 1000 functional connectomes project release. {\color{black}In the Methods section we therefore describe new analyses based on two new first level designs.}

\subsection{Non-parametric inference}

Non-parametric group inference is now available in the AFNI function 3dttest++~\citep{coxPNAS,coxbiorxiv}, meaning that the three most common fMRI softwares now all support non-parametric group inference (SPM users can use the SnPM toolbox (http://warwick.ac.uk/snpm), and FSL users can use the randomise function~\citep{winkler}). Permutation tests cannot only be applied to simple designs such as one-sample and two-sample t-tests, but to virtually any regression model with independent errors~\citep{winkler}. To increase statistical power, permutation tests enable more advanced thresholding approaches~\citep{tfce} as well as the use of multivariate approaches with complicated null distributions~\citep{friman,stelzer}. 

The non-parametric permutation test produced nominal results for all two sample t-tests, but not for the one sample t-tests~\citep{eklundPNAS}, and the Oulu data were more problematic compared to Beijing and Cambridge. {\color{black}As described in the Methods section, we investigated numerous ways to achieve nominal results, and finally concluded that (physiological) artifacts are a problem for one-sample t-tests, especially for the Oulu data. This is a good example of the challenge of validating statistical methods. One can argue that real fMRI data are essential since they contain all types of noise~\citep{retroicor,lund,respiratory,chang,grevenoise} which are difficult to simulate. From this perspective, the Oulu data are helpful since they highlight the problem of (physiological) noise. On the other hand one can argue that a pure fMRI simulation~\citep{simulation} is better, since the researcher then can control all parameters of the data and independently test different settings. From this perspective the Oulu data should be avoided, since the assumption of no consistent activation over subjects is violated by the (physiological) noise, however data quality varies dramatically over sites and studies and we expect there is plenty of data collected that has quality comparable to Oulu.}

\subsection{Implications}

The original publication~\citep{eklundPNAS} inadvertently implied that a large, unspecified proportion of the fMRI literature was affected by our findings, principally the severe inflation of false positive risk for a CDT of p = 0.01; this was clarified in a subsequent correction~\citep{eklundPNAScorrection}. In the Discussion we consider the interpretation of our findings and their impact on the literature as a whole. We estimate that at least 10\% of 23,000 published fMRI studies have used the problematic CDT p = 0.01.

\section{Methods}

\subsection{New paradigms}
 
\begin{figure*}
\subfigure[]{
\includegraphics[scale=0.45]{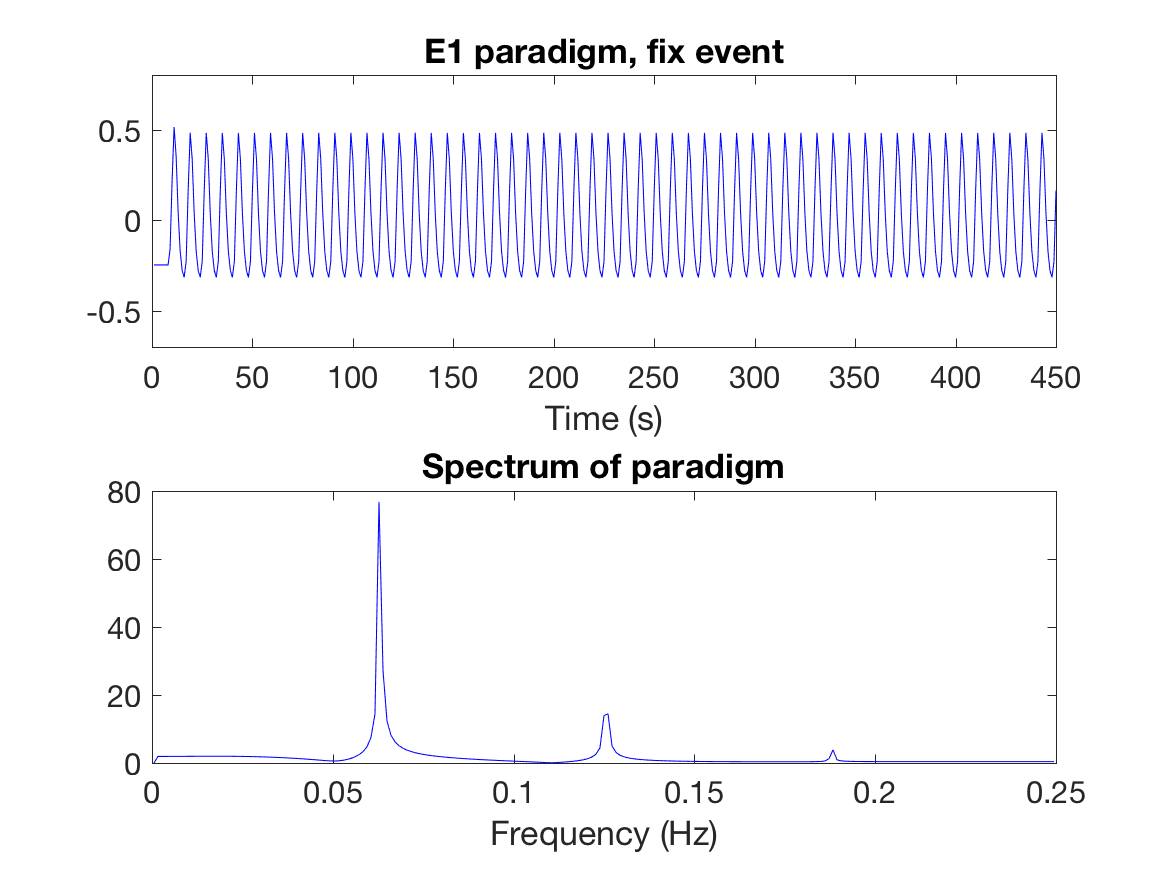}
}
\hspace{-0.4cm}
\subfigure[]{
\includegraphics[scale=0.45]{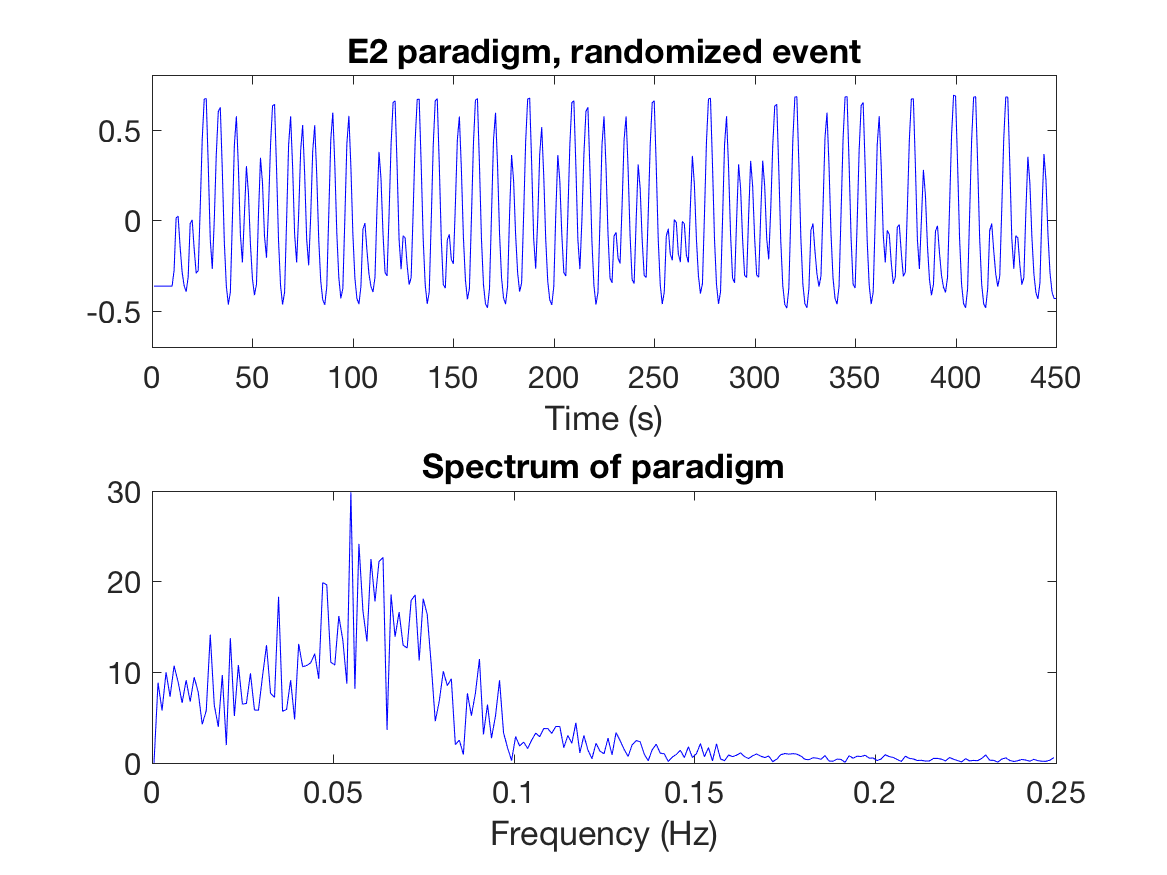}
}
\subfigure[]{
\includegraphics[scale=0.45]{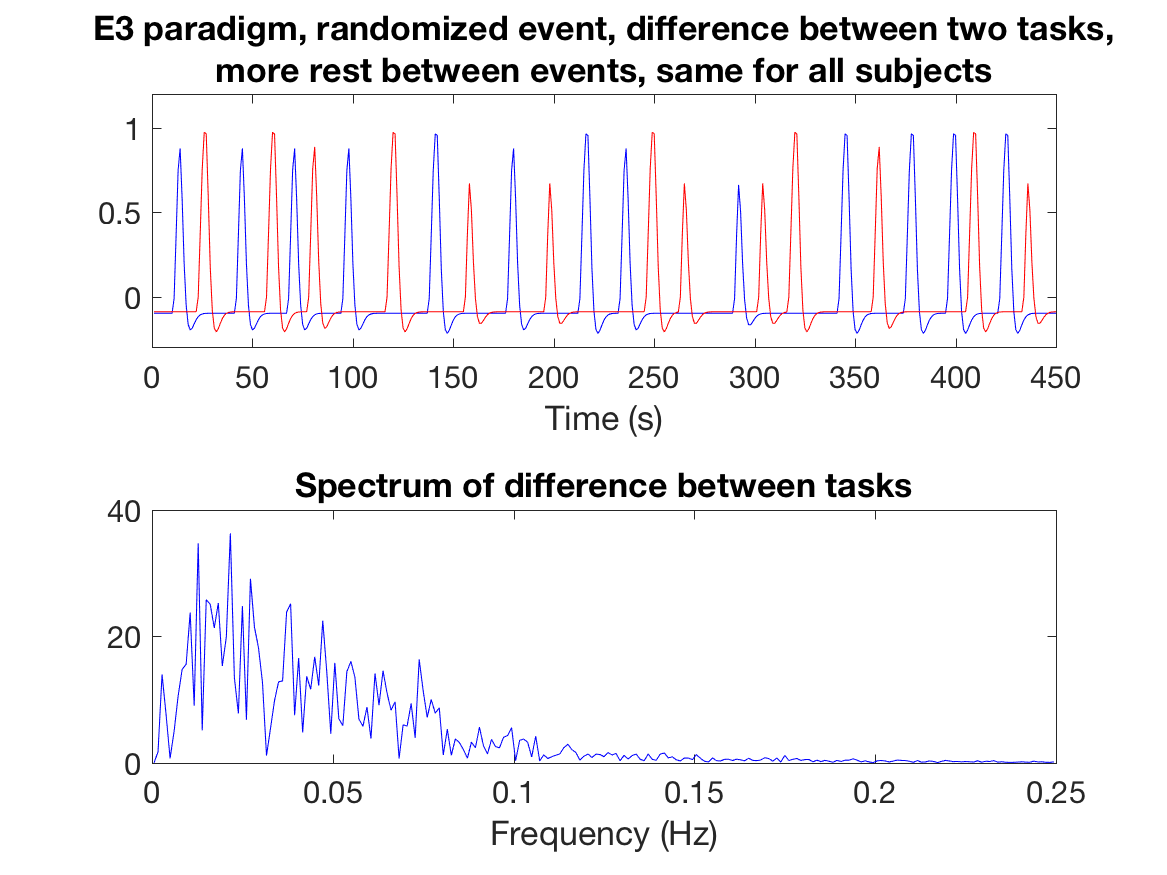}
}
\subfigure[]{
\includegraphics[scale=0.45]{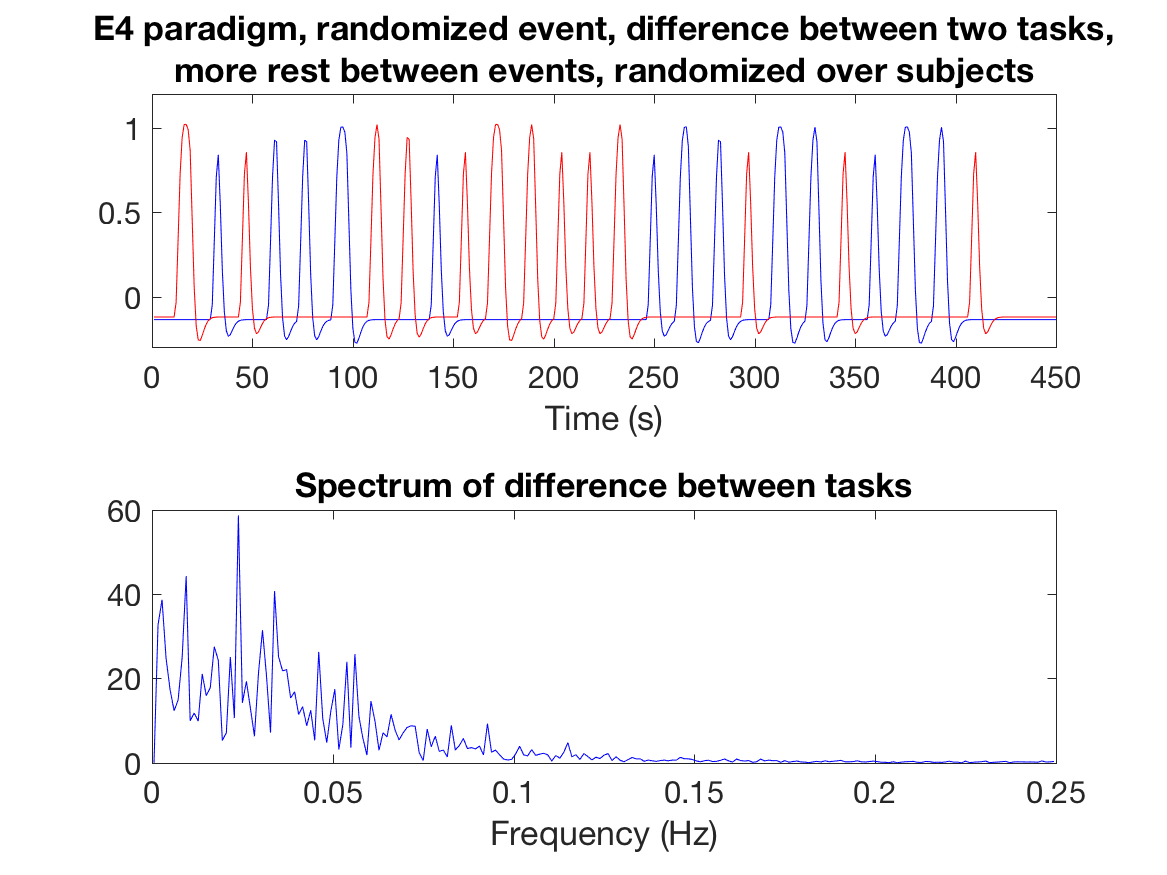}
}
\caption{\emph{A comparison of the paradigms used in the original paper (a) E1, b) E2) and the new paradigms used in this paper (c) E3,  d) E4), for the Beijing datasets. A single task was used for both E1 and E2, while two pretended tasks where used for E3 and E4 (and all first level analyses tested for a difference in activation between these two tasks). Paradigms E1, E2 and E3 are the same for all subjects, while E4 is randomized over subjects.}}
\label{fig:paradigms}
\end{figure*}

To address the concerns regarding realistic first level designs, we have made new analyses using two new event related paradigms, called E3 and E4. For both E3 and E4, two pretended tasks are used instead of a single task, and each first level analysis tests for a difference in activation between the two tasks. Additionally, the rest between the events is longer. {\color{black} For Beijing data, 13 events were used for each of the two tasks. Each task is 3 - 7 seconds long, and the rest between each event is 11 - 13 seconds. For Cambridge data, 11 events of 3 - 6 seconds were used for each task. For Oulu data, 13 events of 3 - 6 seconds were used for each task.} See Figure~\ref{fig:paradigms} for a comparison between E1, E2, E3 and E4. For E4, the regressors are randomized over subjects, such that each subject has {\color{black} the same number of events for each task, but the order and the timing of the events is different for every subject.} For E3 the same regressors are used for all subjects. 

First level analyses as well as group level analyses were performed as in the original study~\citep{eklundPNAS}, using the same data (Beijing, Cambridge, Oulu) from the 1000 functional connectomes project~\citep{biswal2}. {\color{black} Analyses were performed with SPM 8~\citep{spm}, FSL 5.09~\citep{fsl} and AFNI 16.3.14~\citep{afni}.} Familywise error rates were estimated for different levels of smoothing (4 mm - 10 mm), one-sample as well as two-sample t-tests, and two cluster defining thresholds (p = 0.01 and p = 0.001). Group analyses using 3dMEMA in AFNI were not performed, as the results for 3dttest++ and 3dMEMA were very similar in the original study~\citep{eklundPNAS}. Another difference is that cluster thresholding for AFNI was performed using the new ACF (autocorrelation function) option in 3dClustSim~\citep{coxbiorxiv}, which uses a long-tail spatial ACF instead of a Gaussian one. To be able to compare the AFNI results for the new paradigms (E3, E4) and the old paradigms (B1, B2, E1, E2), the group analyses for the old paradigms were re-evaluated using the ACF option (note, that this ACF AFNI option still assumes a stationary spatial autocorrelation structure). {\color{black}Interested readers are referred to our github account for further details.}

\subsection{Using ICA-FIX for denoising}

We investigated numerous ways to achieve nominal familywise error rates for the one-sample (sign flipping) permutation test; 

\begin{enumerate}

\item Applying the~\citet{yeo} transform (signed Box-Cox) to reduce skew (as the sign flipping test is based on an assumption of  symmetric errors)

\item Using robust regression (in every permutation) to suppress the influence of outliers~\citep{wager,woolrich,mumford}

\item Using two-sided tests instead of one-sided

\item Increasing the number of head motion regressors from 6 to 24

\item Using bootstrap instead of sign flipping, and

\item Including the global mean as a covariate in each first level analysis~\citep{gsr1,gsr2} (which is normally not done for task fMRI).\end{enumerate} While some of these approaches resulted in nominal familywise error rates for a subset of the parameter combinations, no approach worked well for all settings and datasets.  {\color{black} In our original study we only used one-sided tests, but this is based on an implicit assumption that a random regressor is equally likely to be positively or negatively correlated with resting state fMRI data. Additionally, most fMRI studies that use a one-sample t-test take advantage of a one-sided test to increase statistical power~\citep{twosided}.}

To understand the spatial distribution of clusters we created images of prevalence of false positive clusters, computed by summing the binary maps of FWE-significant clusters over the random analyses. In our original study, we found a rather structured spatial distribution for the two-sample t-test (supplementary Figure 18 in~\citet{eklundPNAS}), with large clusters more prevalent in the posterior cingulate. We have now created the same sort of maps for one-sample t-tests, with a small modification: to increase the number of clusters observed, we created clusters at a CDT of p = 0.01 for both increases and decreases on a given statistic map. As discussed in the Results, there appears to be physiological artefacts which ideally would be remediated by respiration or cardiac time series modelling~\citep{retroicor,lund,respiratory,chang,bollmann}, but unfortunately the 1000 functional connectomes datasets~\citep{biswal2} do not have these physiological recordings.

To suppress the influence of artifacts, we therefore instead applied ICA FIX {\color{black}(version 1.065)} in FSL~\citep{icafix1,icafix2,griffanti} to all 499 subjects, to remove ICA components that correspond to noise or artifacts. We applied 4 mm of spatial smoothing for MELODIC \citep{melodic}, and used the classifier weights for standard fMRI data avalable in ICA FIX (trained for 5 mm smoothing). To use ICA FIX for 8 or 10 mm of smoothing would require re-training the classifier. The cleanup was performed using the aggressive (full variance) option instead of the default less-aggressive option, and motion  confounds were also included in the cleanup.  {\color{black} To study the effect of re-training the ICA FIX classifier specifically for each dataset (Beijing, Cambridge, Oulu), instead of using the pre-trained weights available in ICA FIX, we manually labeled the ICA components of 10 subjects for each dataset (giving a total of 350 - 450 ICA components per dataset). Indeed, a large portion of the ICA components are artefacts that are similar across subjects. Interested readers are referred to the github account for ICA FIX processing scripts and the re-trained classifier weights for Beijing, Cambridge and Oulu}.

First level analyses for B1, B2, E1, E2, E3 and E4 were performed {\color{black} using FSL} for all 499 subjects after ICA FIX, with {\color{black} motion correction and} smoothing turned off. Group level analyses were finally performed using the non-parametric one-sample t-test available in BROCCOLI~\citep{broccoli}.

\section{Results}

\subsection{New paradigms}

\begin{figure*}
\subfigure[]{
\includegraphics[scale=0.425]{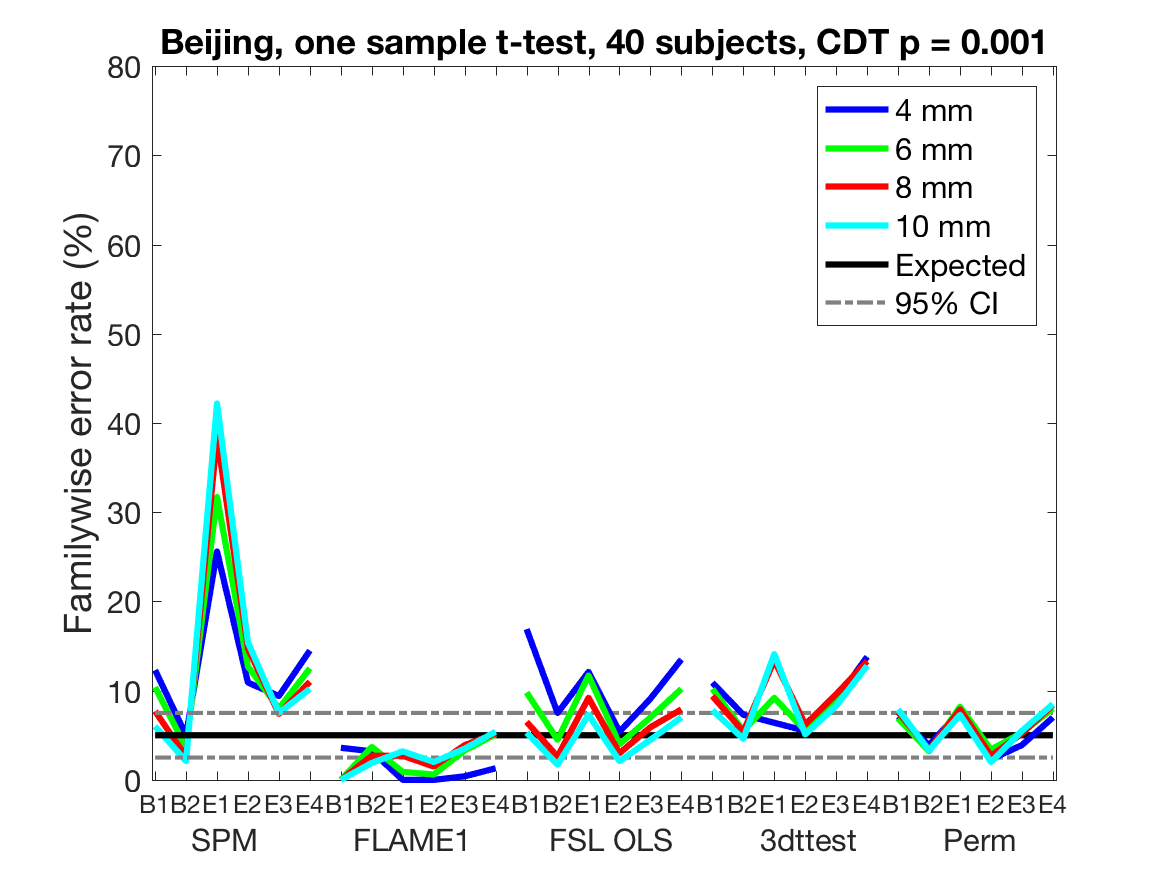}
}
\subfigure[]{
\includegraphics[scale=0.425]{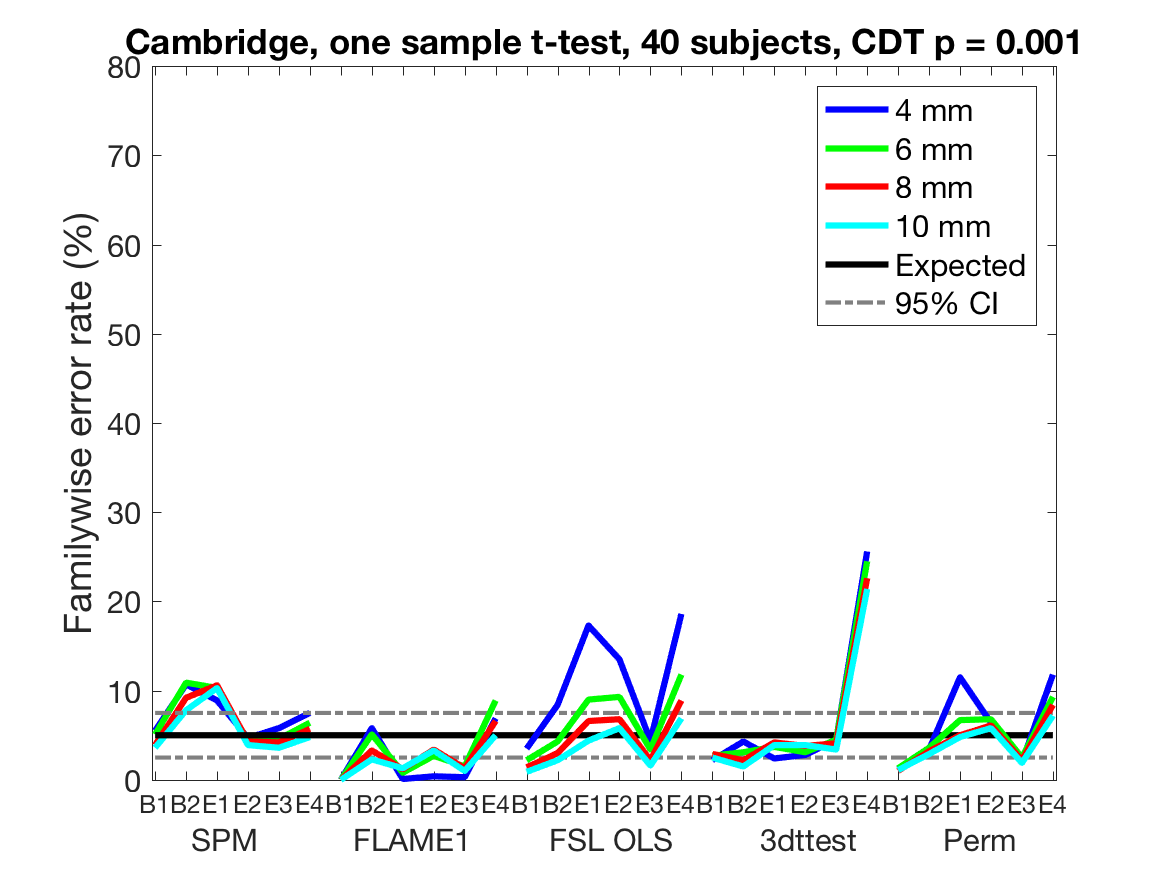}
}
\subfigure[]{
\includegraphics[scale=0.425]{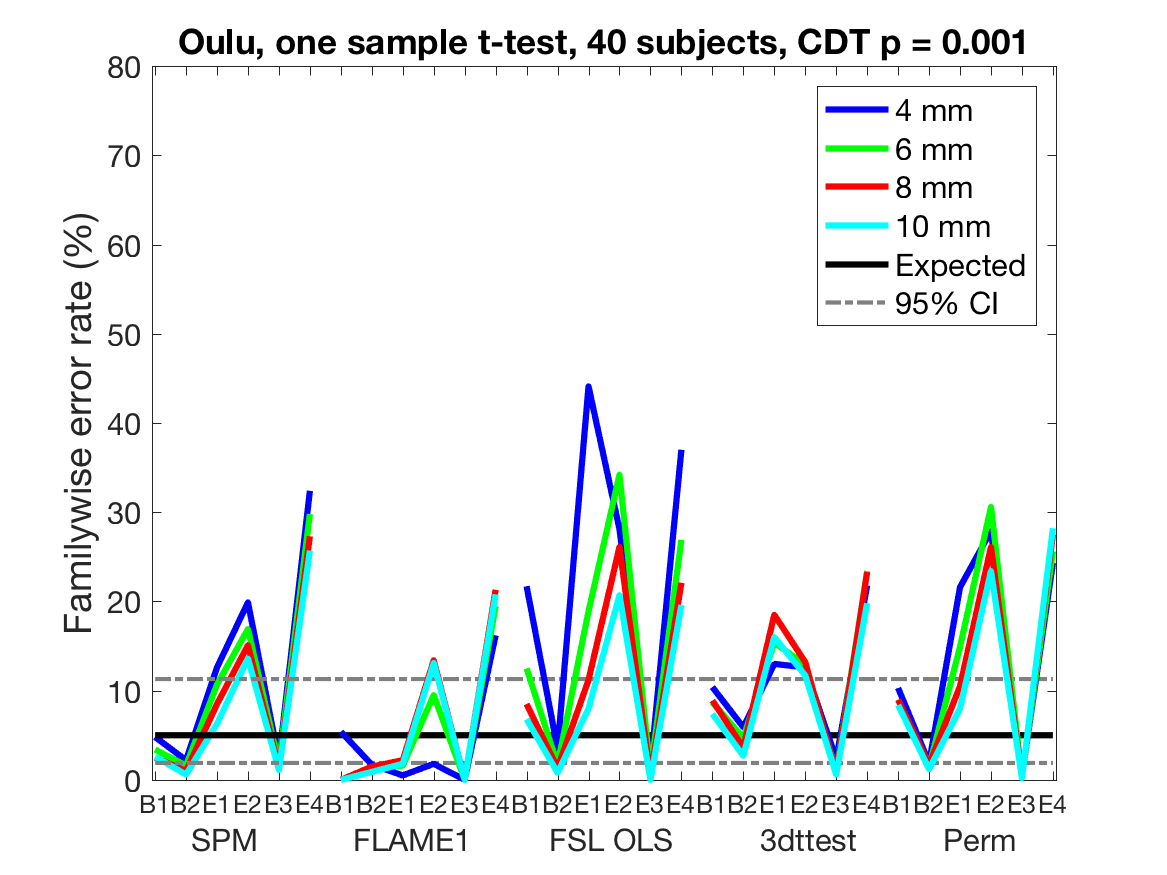}
}
\caption{\emph{Results for one sample t-test and cluster-wise inference using a cluster defining threshold (CDT) of p = 0.001, showing estimated familywise error rates for 4 - 10 mm of smoothing and six different activity paradigms (old paradigms B1, B2, E1, E2 and new paradigms E3, E4), for SPM, FSL, AFNI and a permutation test. These results are for a group size of 40. Each statistic map was first thresholded using a CDT of p = 0.001, uncorrected for multiple comparisons, and the surviving clusters were then compared to a FWE-corrected cluster extent threshold, $p_{FWE} = 0.05$. The estimated familywise error rates are simply the number of analyses with any significant group activations divided by the number of analyses (1,000). \textbf{(a)} results for Beijing data \textbf{(b)} results for Cambridge data \textbf{(c)} results for Oulu data.}}
\label{fig:fwe_cluster_onesample_40_subjects_cdt2}
\end{figure*}

\begin{figure*}
\subfigure[]{
\includegraphics[scale=0.425]{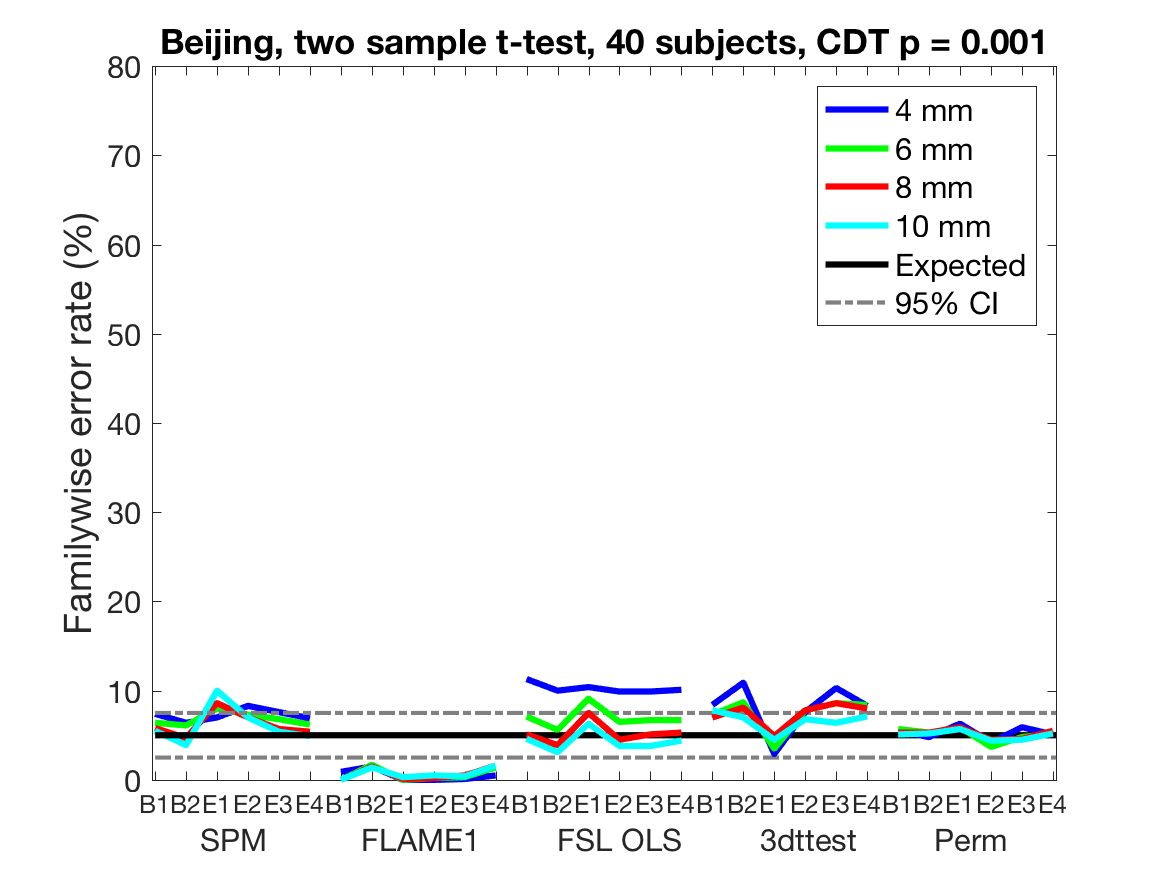}
}
\subfigure[]{
\includegraphics[scale=0.425]{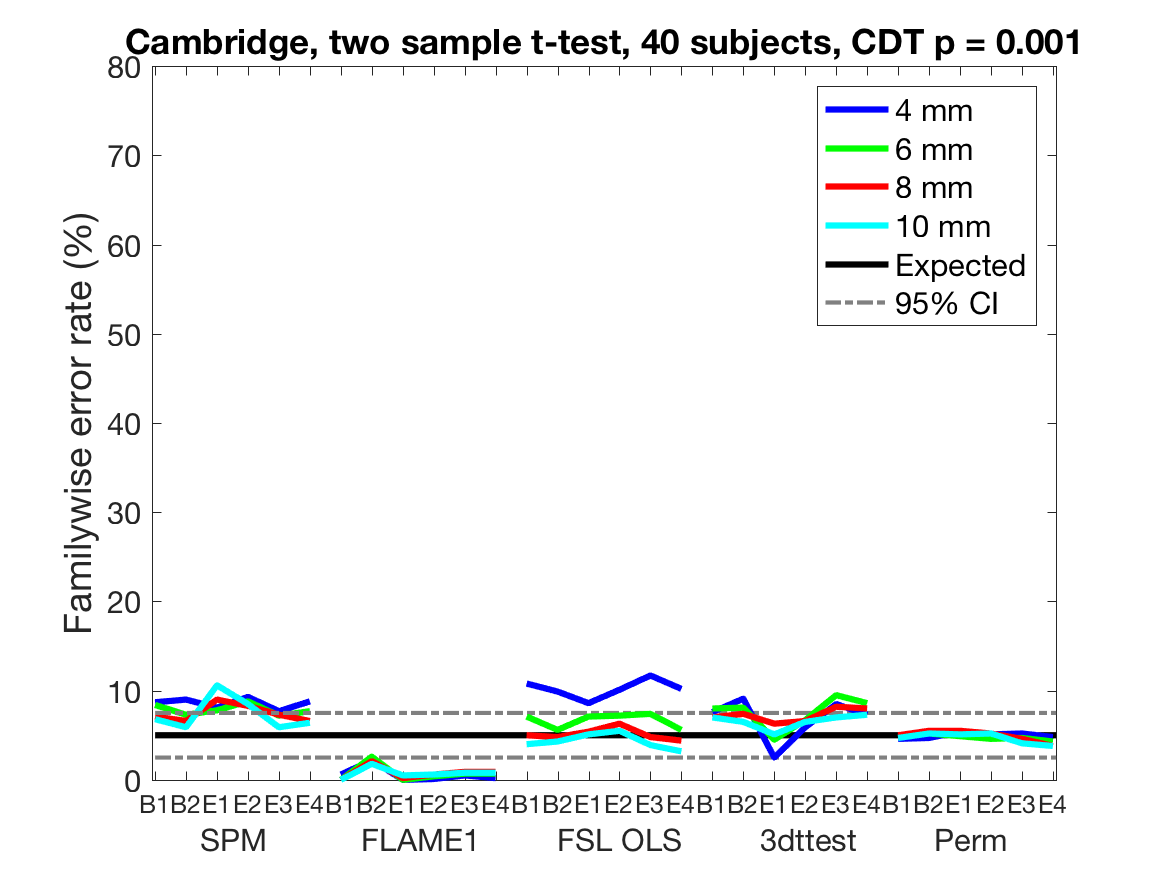}
}
\subfigure[]{
\includegraphics[scale=0.425]{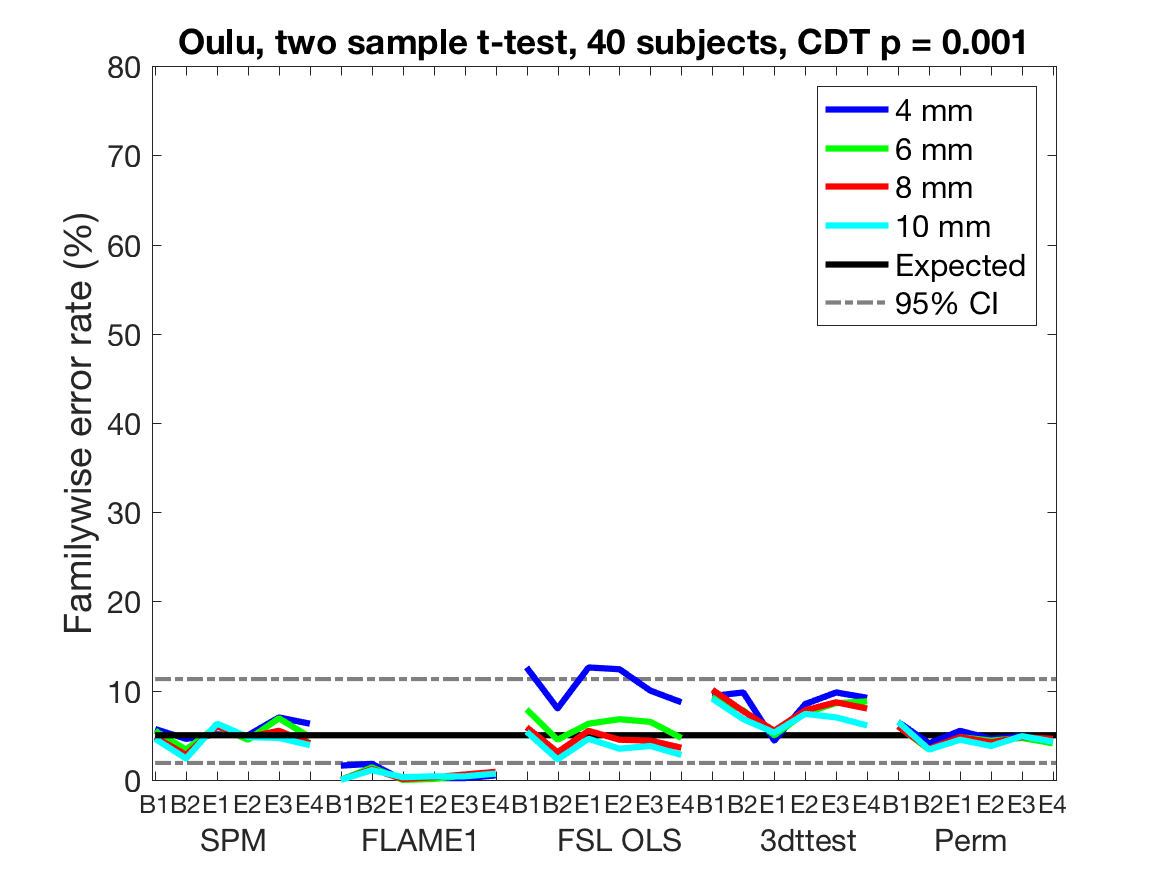}
}
\caption{\emph{Results for two sample t-test and cluster-wise inference using a cluster defining threshold (CDT) of p = 0.001, showing estimated familywise error rates for 4 - 10 mm of smoothing and six different activity paradigms (old paradigms B1, B2, E1, E2 and new paradigms E3, E4), for SPM, FSL, AFNI and a permutation test. These results are for a group size of 20, giving a total of 40 subjects. Each statistic map was first thresholded using a CDT of p = 0.001, uncorrected for multiple comparisons, and the surviving clusters were then compared to a FWE-corrected cluster extent threshold, $p_{FWE} = 0.05$. The estimated familywise error rates are simply the number of analyses with any significant group activations divided by the number of analyses (1,000). \textbf{(a)} results for Beijing data \textbf{(b)} results for Cambridge data \textbf{(c)} results for Oulu data.}}
\label{fig:fwe_cluster_twosample_40_subjects_cdt2}
\end{figure*}

Figures~\ref{fig:fwe_cluster_onesample_40_subjects_cdt2} -~\ref{fig:fwe_cluster_twosample_40_subjects_cdt2} show estimated familywise error (FWE) rates for the two new paradigms (E3, E4), for 40 subjects in each group analysis and a CDT of p = 0.001. Figures~\ref{fig:fwe_cluster_onesample_40_subjects_cdt1} -~\ref{fig:fwe_cluster_twosample_40_subjects_cdt1} show the FWE rates for a CDT of p = 0.01. The four old paradigms (B1, B2, E1, E2) are included as well for the sake of comparison. In brief, the new paradigm with two pretend tasks (E3) does not lead to lower familywise error rates, compared to the old paradigms. Likewise, randomising task events over subjects (E4) has if anything worse familywise error rates compared to not randomising the task over subjects. As noted in our original paper, the very low FWE of FSL's FLAME1 is anticipated behavior when there is zero random effects variance.  When fitting anything other than a one-sample group model this conservativeness may not hold; in particular, we previously reported on two-sample null analyses on task data, where each sample has non-zero but equal effects, and found that FLAME1's FWE was equivalent to that of FSL OLS~\citep{eklundPNAS}.

{\color{black} By looking at Figure~\ref{fig:fwe_cluster_onesample_40_subjects_cdt2} it is possible to compare the parametric methods (who are simultaneously affected by non-Gaussian SACF, non-stationary smoothness and physiological noise) and the non-parametric permutation test (only affected by physiological noise, as no assumptions are made regarding the SACF and stationary smoothness). For the Beijing data the permutation test performs rather well, while all parametric approaches struggle despite a strict cluster defining threshold. It is also clear that the Oulu data is more problematic compared to Beijing and Cambridge.}

\subsection{ICA denoising}

\begin{figure*}
\subfigure[]{
\includegraphics[scale=0.425]{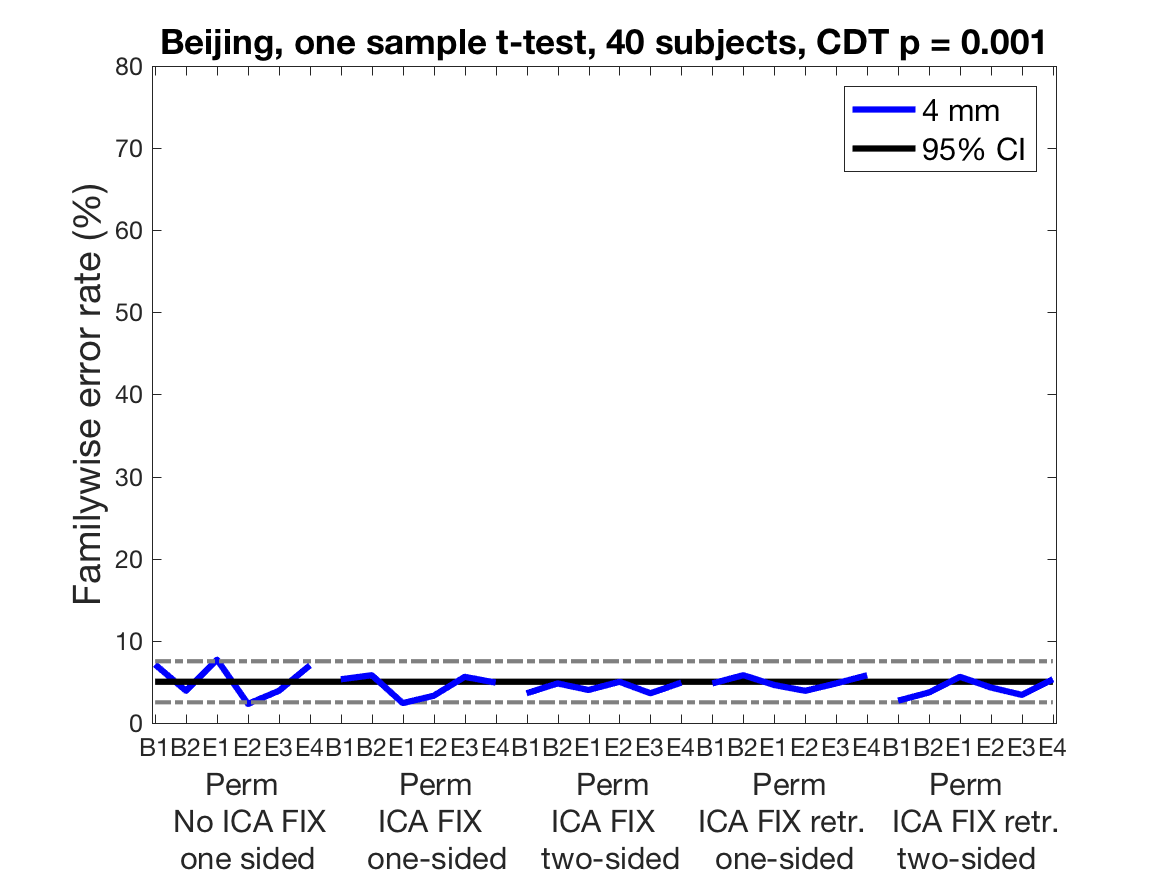}
}
\subfigure[]{
\includegraphics[scale=0.425]{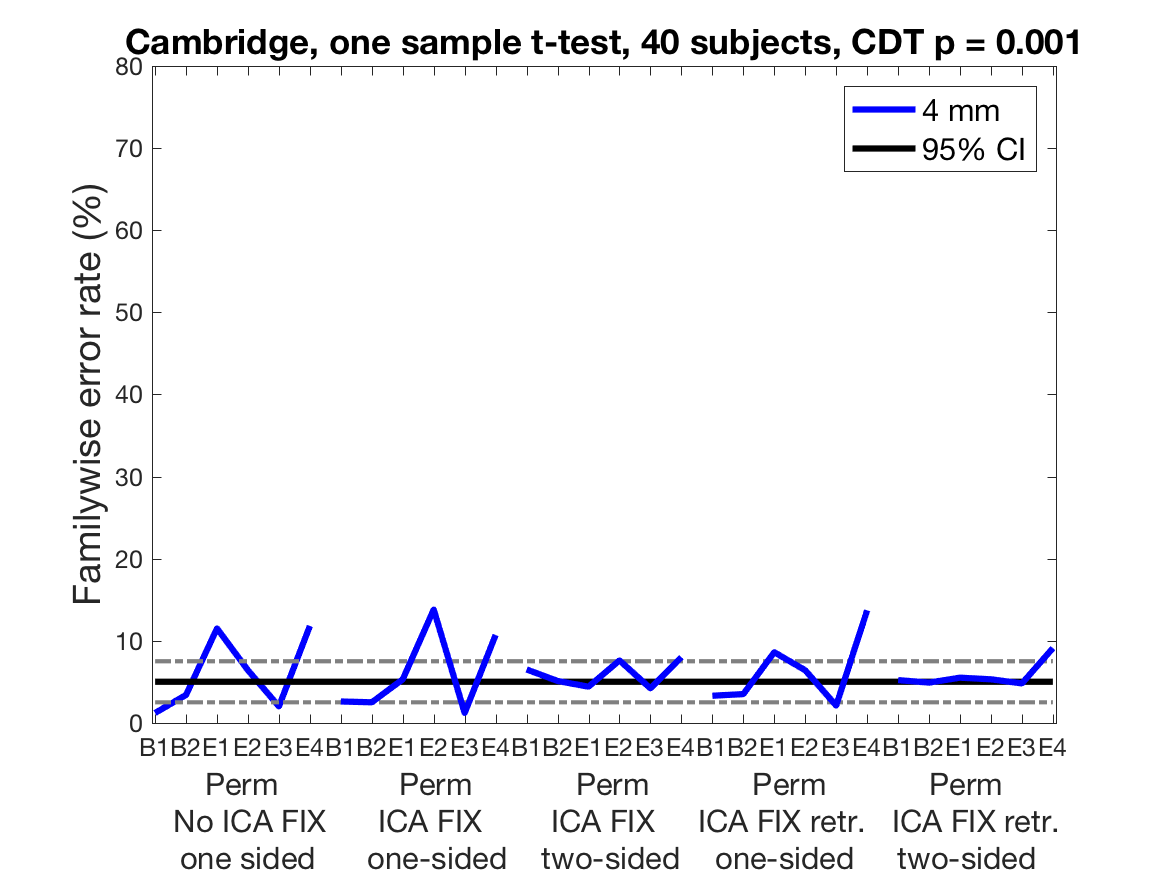}
}
\subfigure[]{
\includegraphics[scale=0.425]{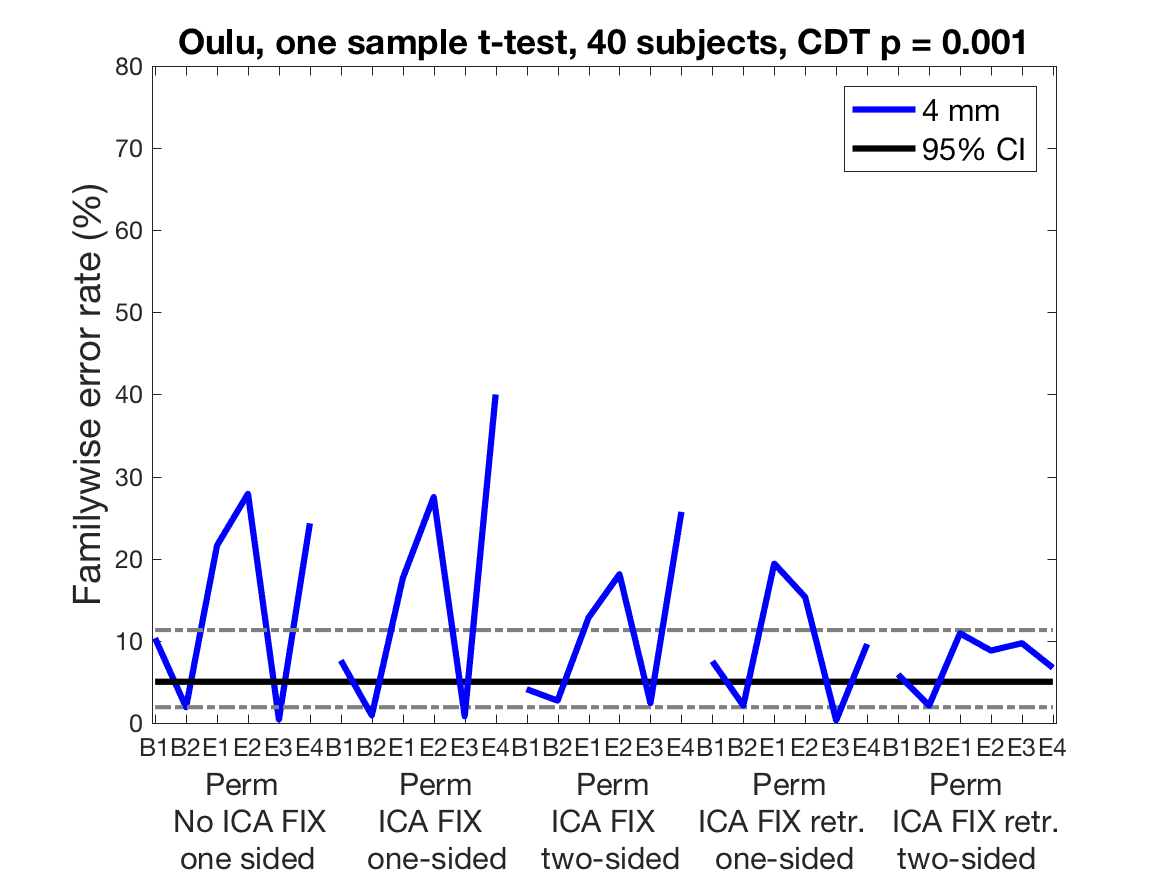}
}
\caption{\emph{Results for non-parametric (sign flipping) one-sample t-tests for cluster-wise inference using a cluster defining threshold (CDT) of p = 0.001, for no ICA FIX, pre-trained ICA FIX and re-trained ICA FIX. \textbf{(a)} results for Beijing data \textbf{(b)} results for Cambridge data \textbf{(c)} results for Oulu data. Results are only shown for 4 mm smoothing, as other smoothing levels would require re-training the ICA FIX classifier. For both Beijing and Cambridge, the pre-trained classifier weights for ICA FIX are sufficient to achieve nominal false positive rates, while it is necessary to re-train the ICA FIX classifier specifically for the Oulu data (a possible explanation is that the Oulu data have a spatial resolution of 4 x 4 x 4.4 mm$^3$, while ICA FIX for standard fMRI data is pre-trained on data with a spatial resolution of 3.5 x 3.5 x 3.5 mm$^3$).}}
\label{fig:fwe_perm_ICAFIX}
\end{figure*}

{\color{black}
Figure~\ref{fig:fwe_perm_ICAFIX} shows familywise error rates for the non-parametric one-sample t-test, for no ICA FIX, pre-trained ICA FIX and re-trained ICA FIX, for one-sided as well as two-sided tests.} For the Beijing data, the familywise error rates are almost within the 95\% confidence interval even without ICA FIX, and come even closer to the expected 5\% after ICA FIX. For the Cambridge data, it is necessary to combine ICA FIX with a two-sided test to achieve nominal results (only using a two-sided test is not sufficient). For the Oulu data, neither ICA FIX in isolation nor in combination with two-sided inference was sufficient to bring false positives to a nominal rate. {\color{black} However, re-training the ICA FIX classifier specifically for the Oulu dataset finally resulted in nominal false positive rates.}

\begin{figure*}
\subfigure[]{
\includegraphics[scale=0.425]{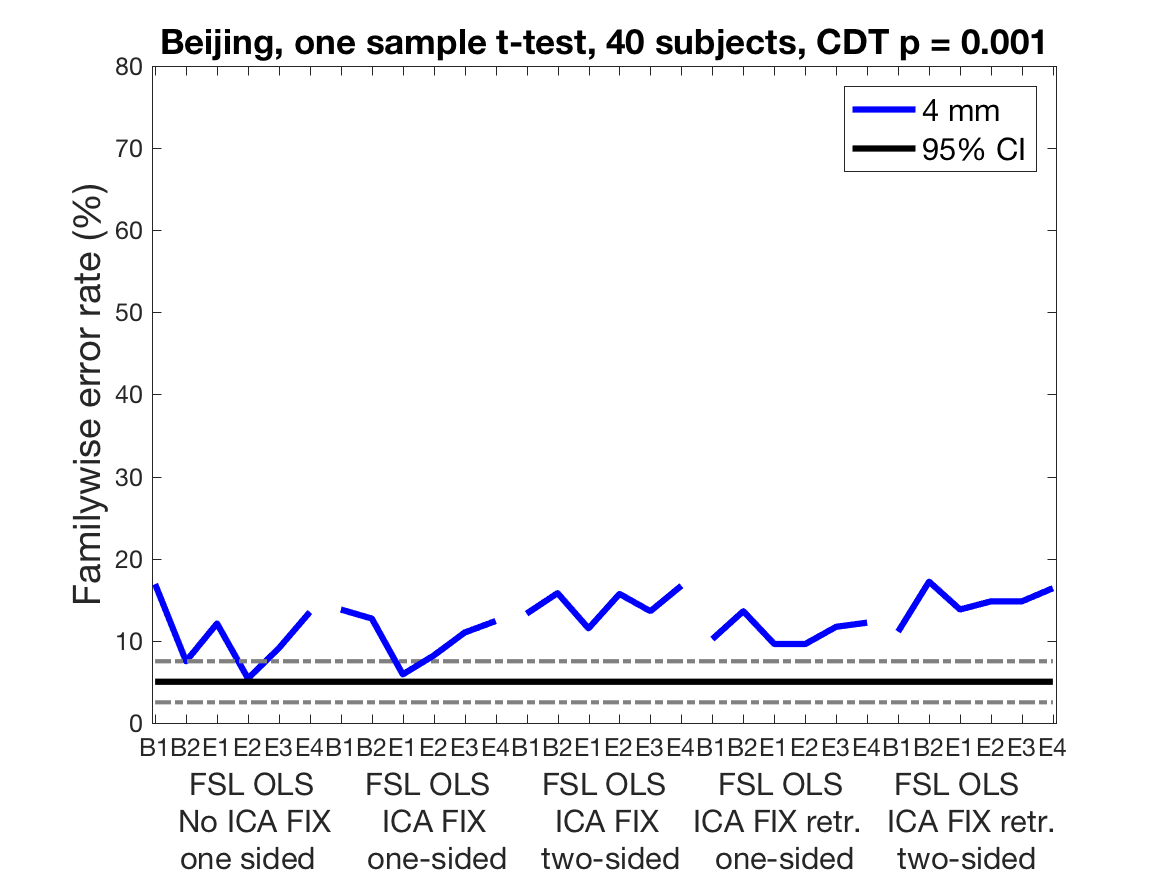}
}
\subfigure[]{
\includegraphics[scale=0.425]{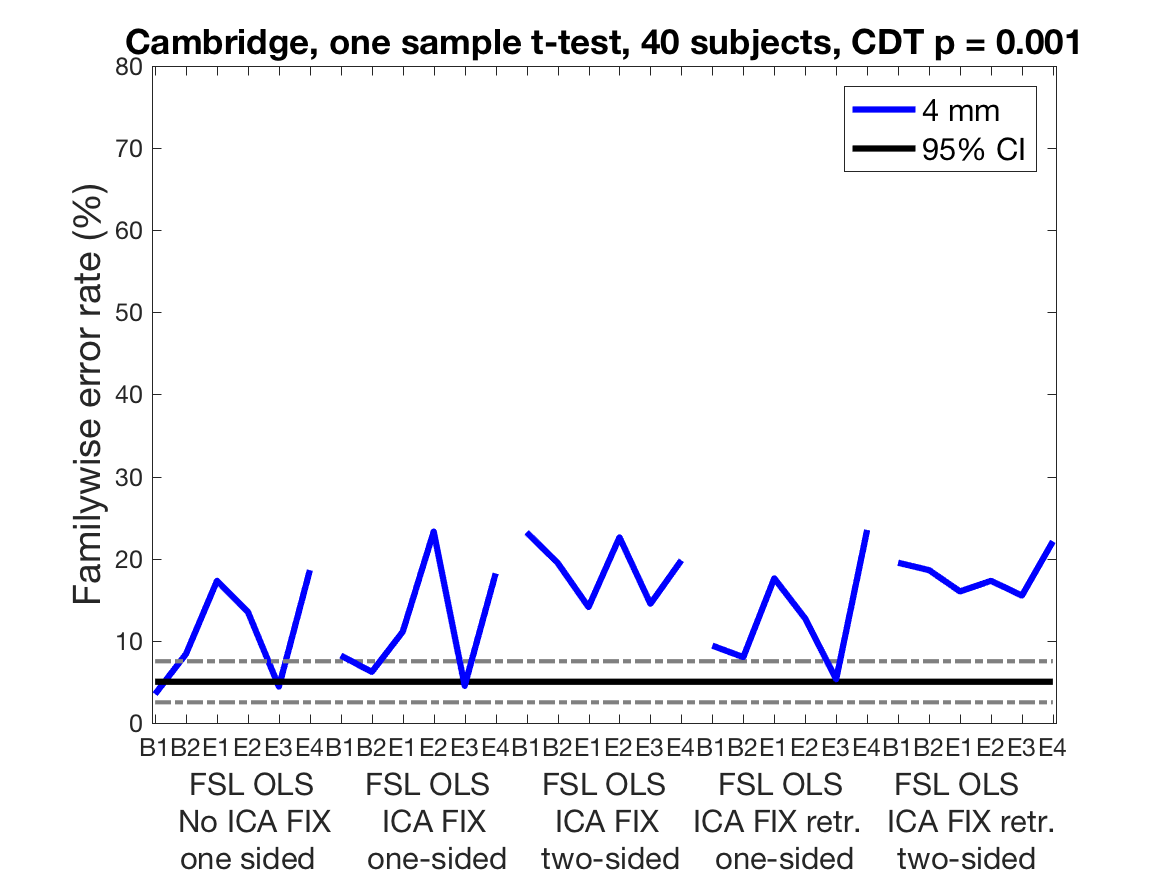}
}
\subfigure[]{
\includegraphics[scale=0.425]{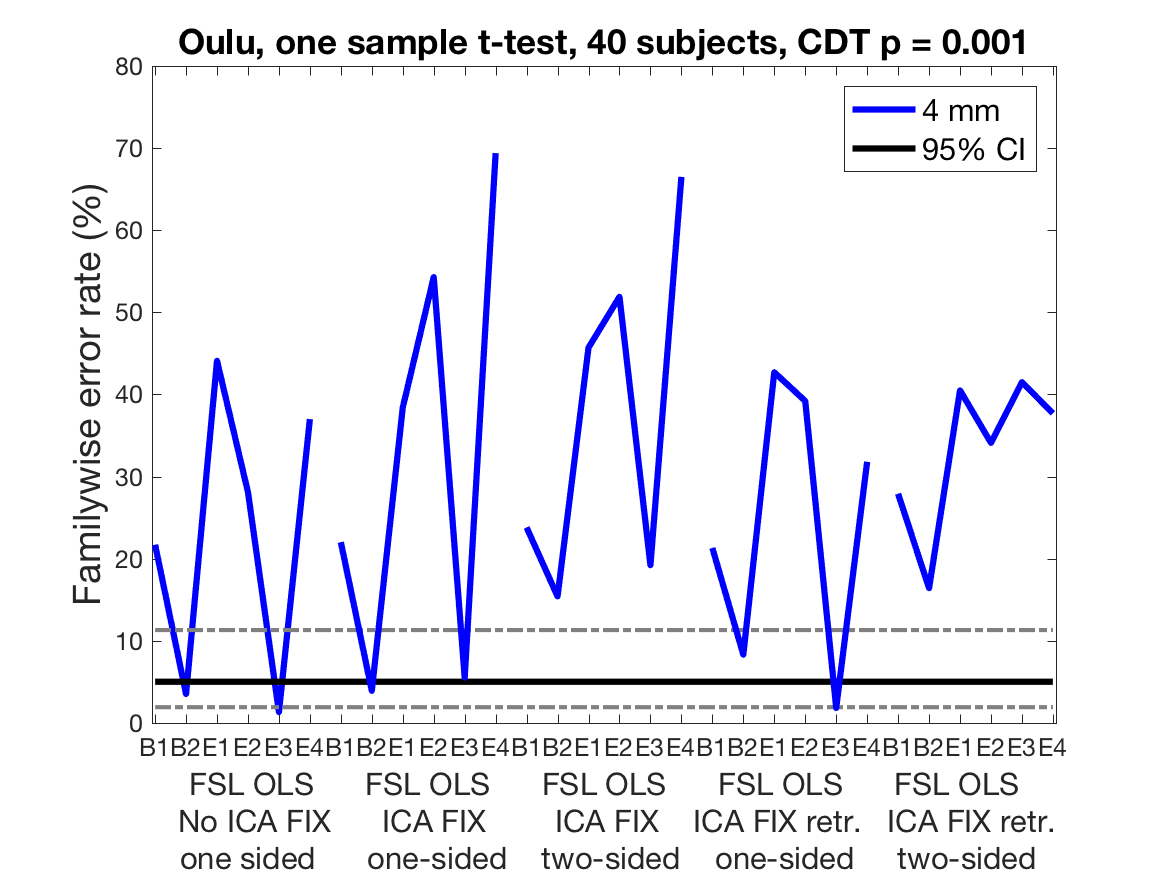}
}
\caption{\emph{Results for FSL OLS one-sample t-tests for cluster-wise inference using a cluster defining threshold (CDT) of p = 0.001, for no ICA FIX, pre-trained ICA FIX and re-trained ICA FIX. \textbf{(a)} results for Beijing data \textbf{(b)} results for Cambridge data \textbf{(c)} results for Oulu data. Results are only shown for 4 mm smoothing, as other smoothing levels would require re-training the ICA FIX classifier.}}
\label{fig:fwe_FSLOLS_ICAFIX}
\end{figure*}

{\color{black}
To test if using ICA FIX also results in nominal familywise error rates for FSL OLS, we performed group analyses for no ICA FIX, pre-trained ICA FIX and re-trained ICA FIX, for one-sided as well as two-sided tests, see Figure~\ref{fig:fwe_FSLOLS_ICAFIX}. Since ICA FIX cleaning and all first level analyses were performed using FSL, we only performed the group analyses using FSL. Clearly, using ICA FIX does not lead to nominal familywise error rates for FSL OLS, and using a two-sided test leads to even higher familywise error rates compared to a one-sided test. A possible explanation is that (two) parametric tests for p = 0.025 are even more inflated compared to parametric tests for p = 0.05. To test this hypothesis, we performed 18,000 one-sided one-sample group analyses (three datasets and six activity paradigms, 1,000 analyses each, for first level analyses with no ICA FIX, CDT p = 0.001) with a FWE significance threshold of 1\%; false positives at FWE 1\% should occur 1/5=0.2$\times$ as often with FWE 5\% results.  We found nominal FWE 1\% false positives occurred at a rate 0.694$\times$ the 5\% FWE results. That is, the relative inflation of false positives for parametric methods is much higher for more stringent significance thresholds.}

\begin{figure*}

\subfigure[]{
\includegraphics[scale=0.45]{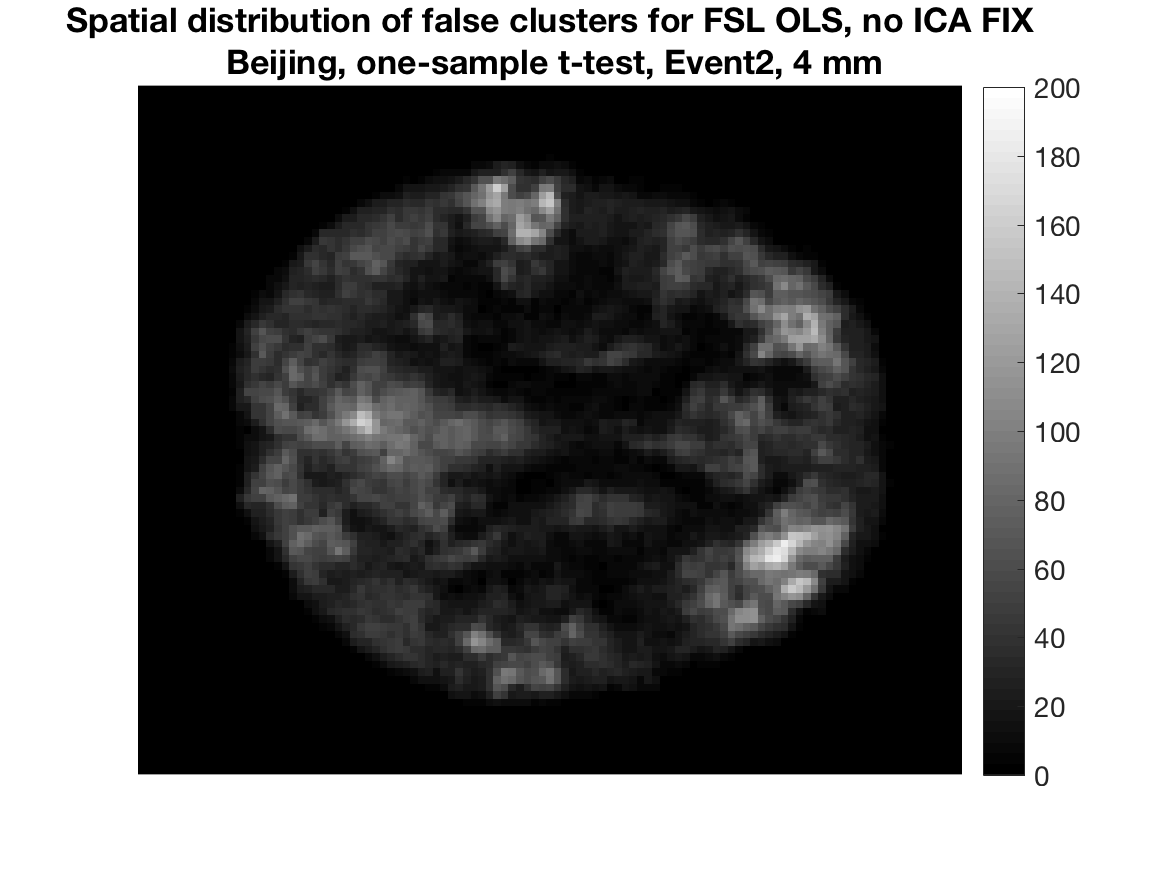}
}
\subfigure[]{
\includegraphics[scale=0.45]{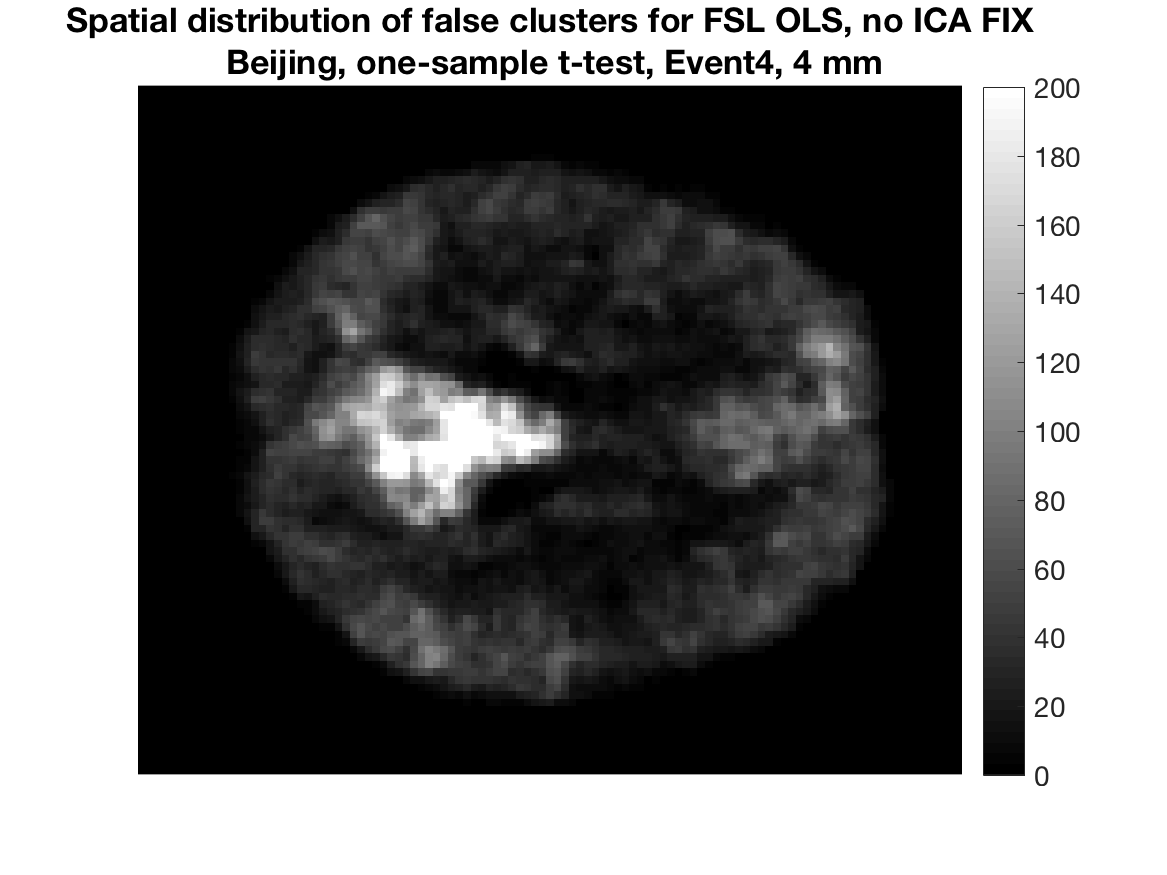}
}
\subfigure[]{
\includegraphics[scale=0.45]{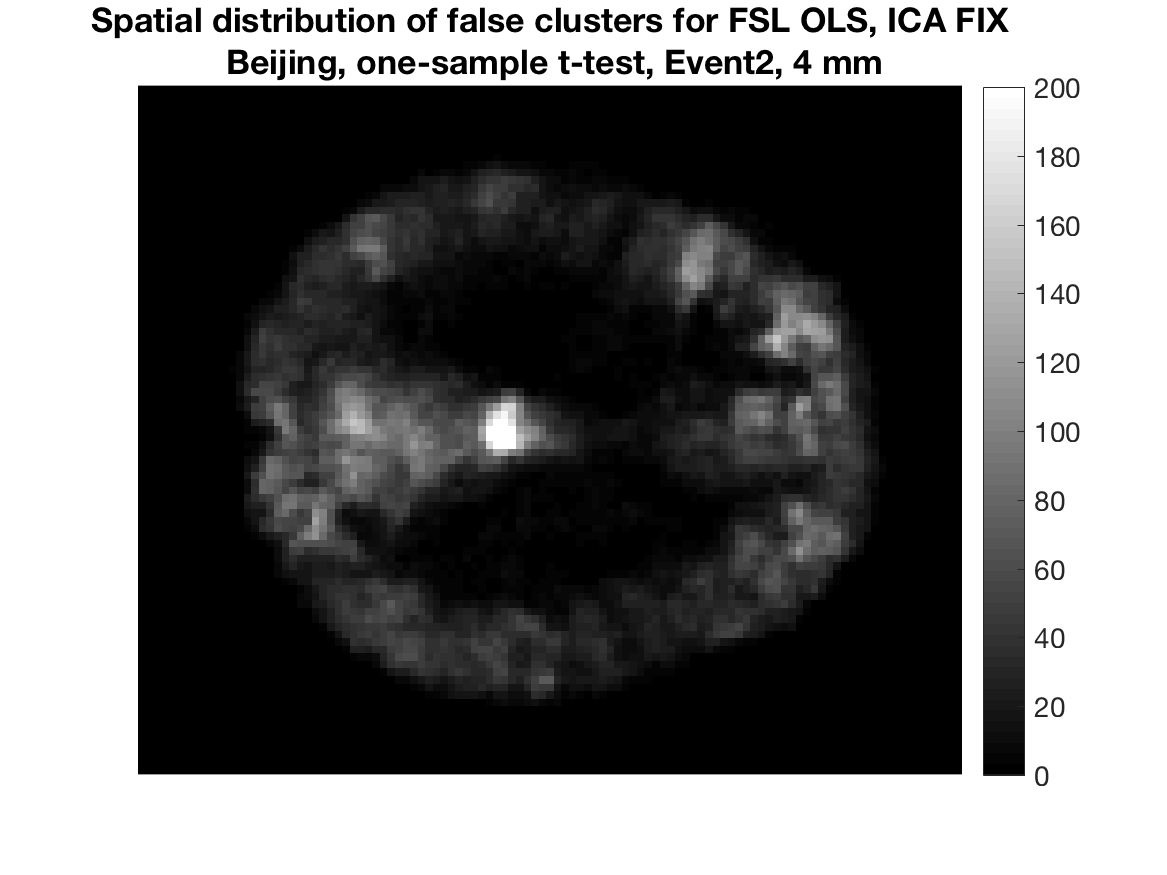}
}
\subfigure[]{
\includegraphics[scale=0.45]{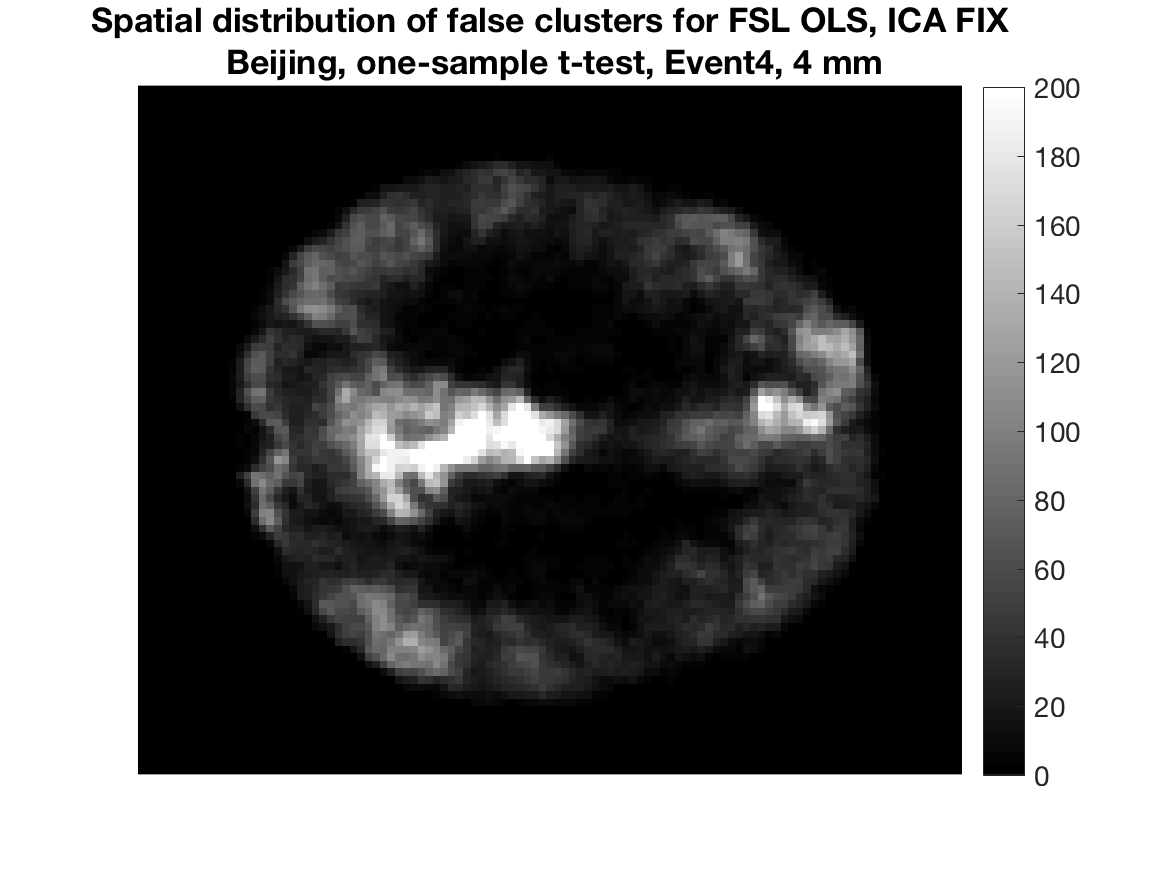}
}
\subfigure[]{
\includegraphics[scale=0.45]{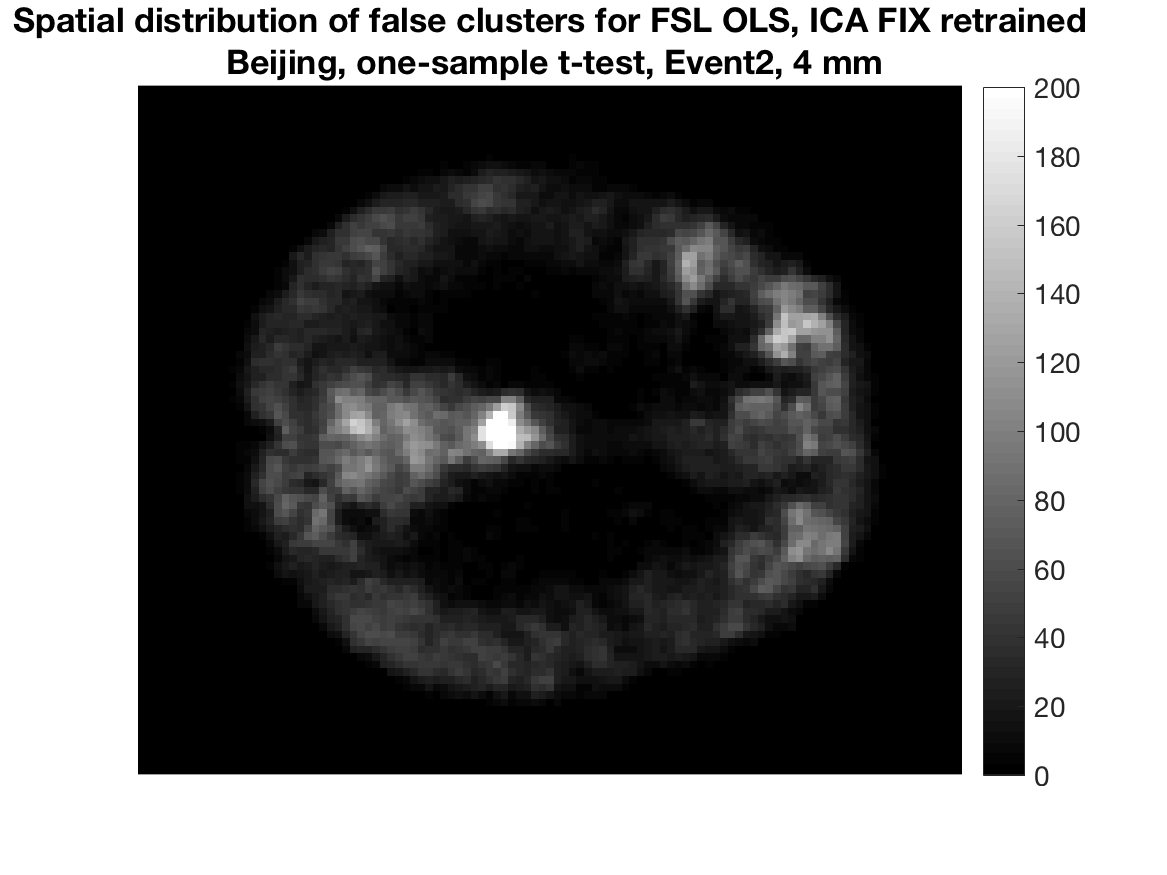}
}
\subfigure[]{
\includegraphics[scale=0.45]{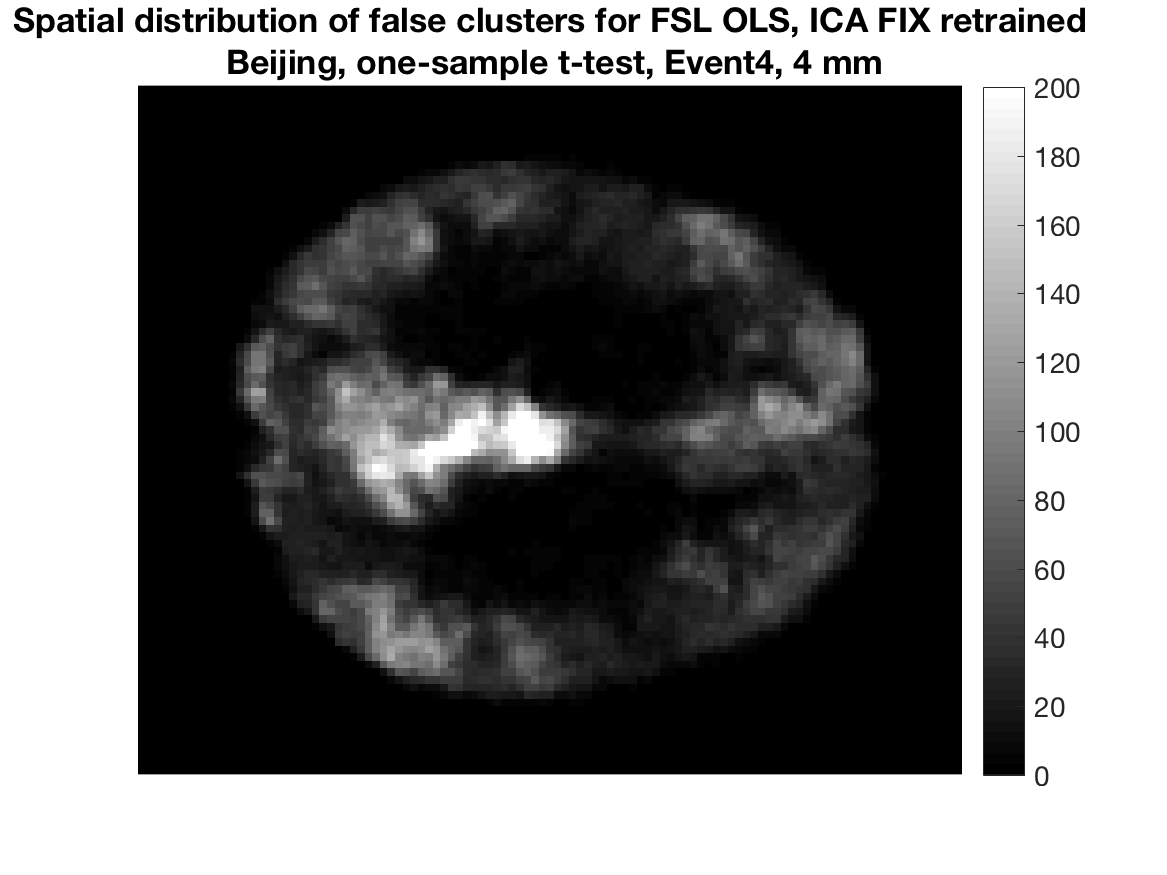}
}
\caption{\emph{The maps show voxel-wise incidence of false clusters for the Beijing data, for 2 of the 6 different first level designs (a,b) no ICA FIX (c,d) ICA FIX pre-trained (e,f) ICA FIX re-trained for Beijing. Left: results for design E2, Right: results for design E4. Image intensity is the number of times, out of 10,000 random analyses, a significant cluster occurred at a given voxel (CDT p = 0.01) for FSL OLS. Each analysis is a one-sample t-test using 20 subjects.}}
\label{fig:falseclusters_Beijing_icafix}
\end{figure*}
\begin{figure*}
\subfigure[]{
\includegraphics[scale=0.45]{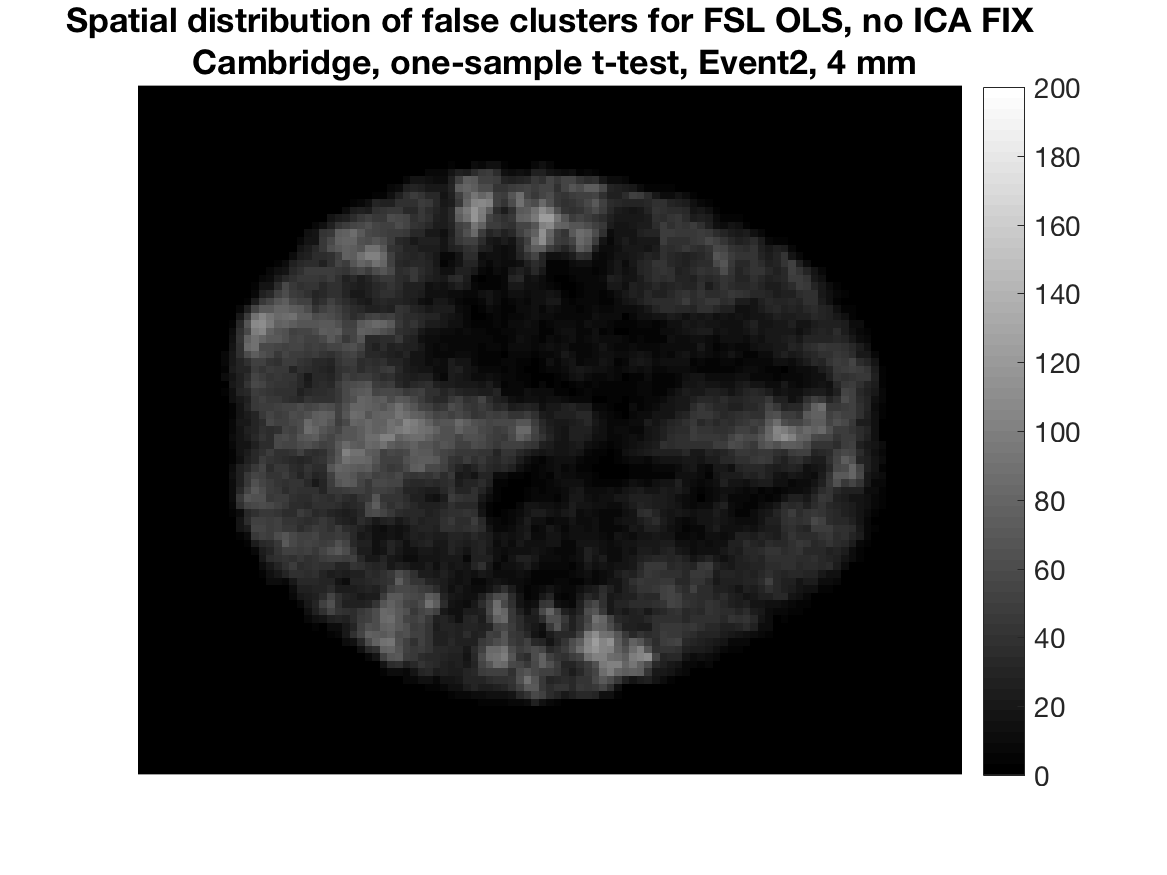}
}
\subfigure[]{
\includegraphics[scale=0.45]{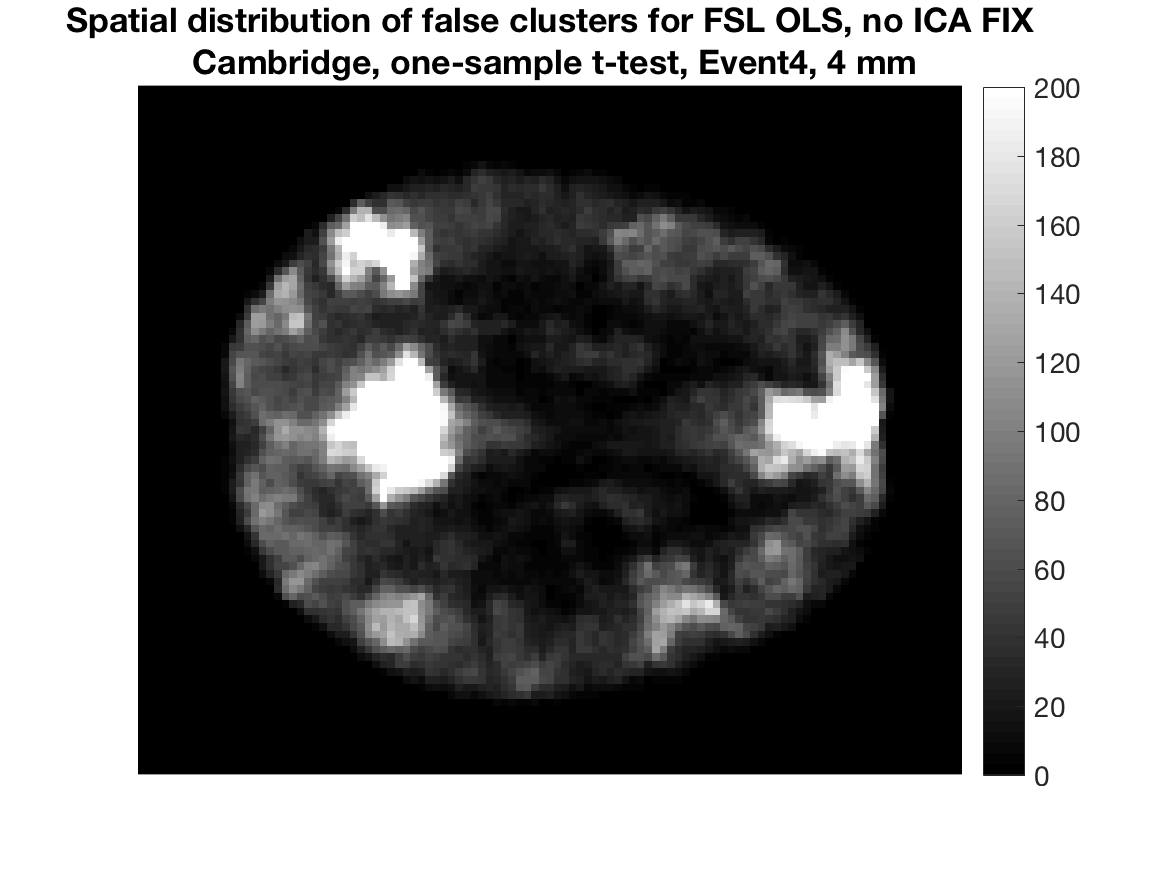}
}
\subfigure[]{
\includegraphics[scale=0.45]{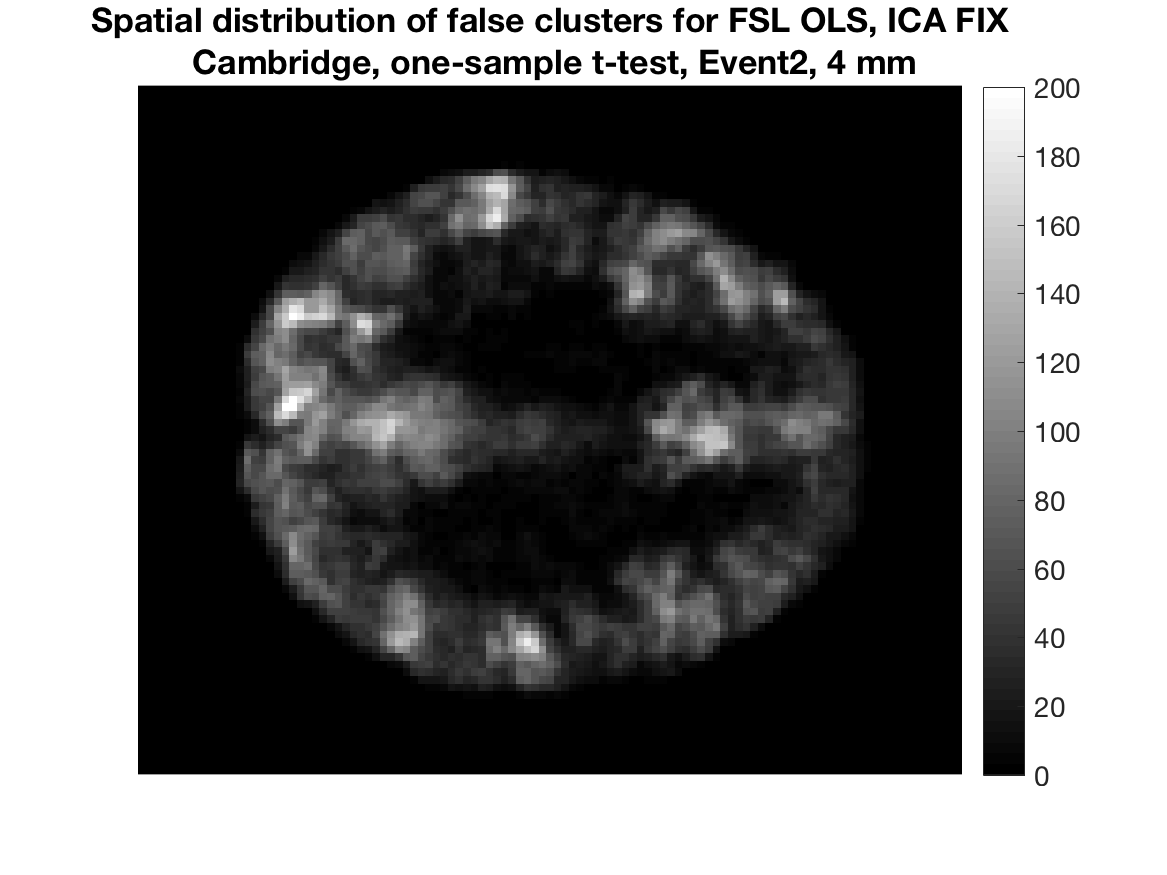}
}
\subfigure[]{
\includegraphics[scale=0.45]{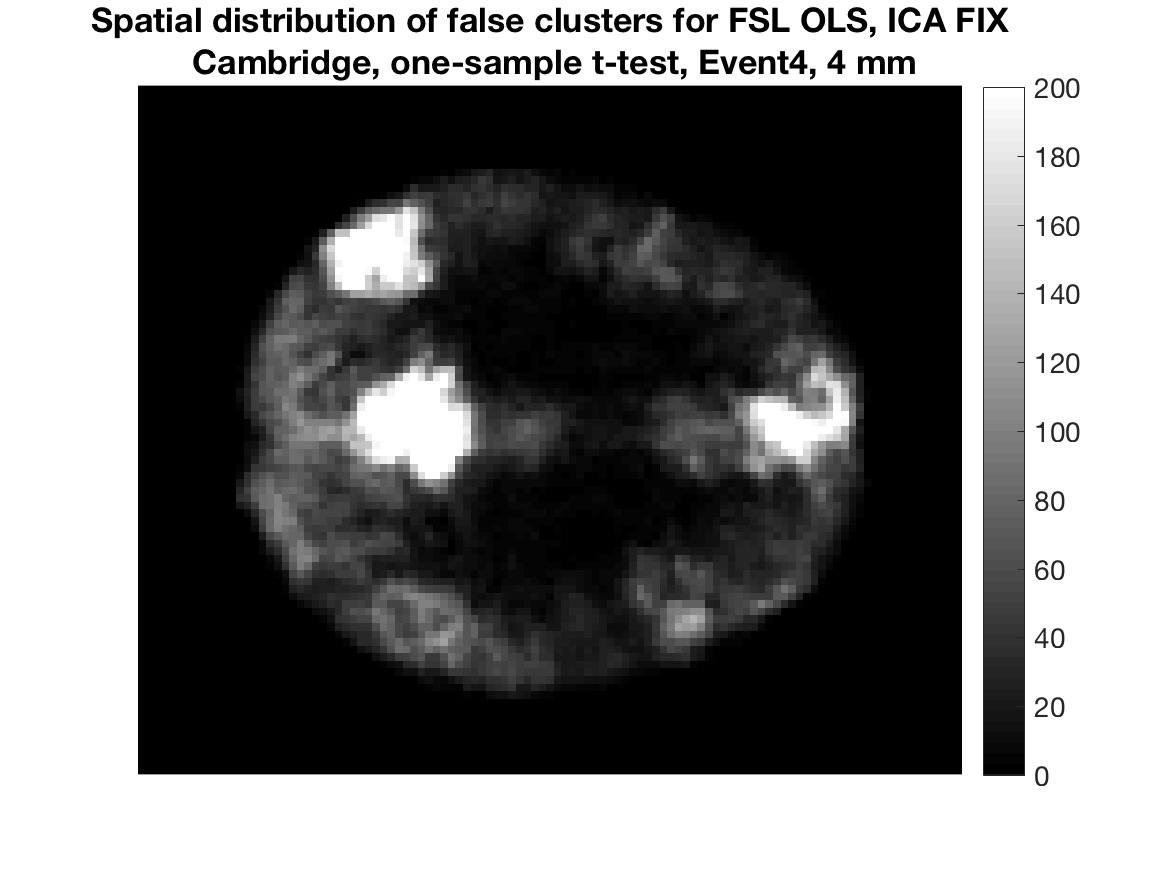}
}
\subfigure[]{
\includegraphics[scale=0.45]{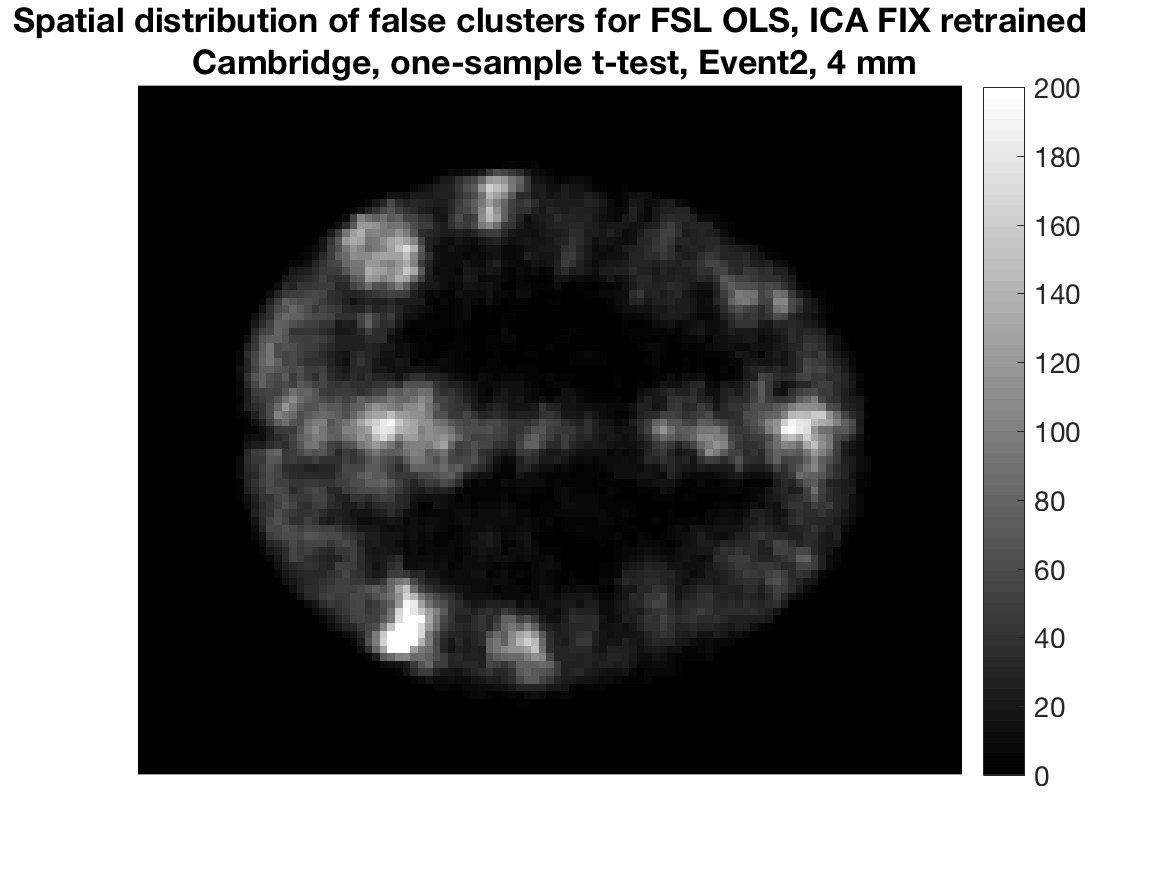}
}
\subfigure[]{
\includegraphics[scale=0.45]{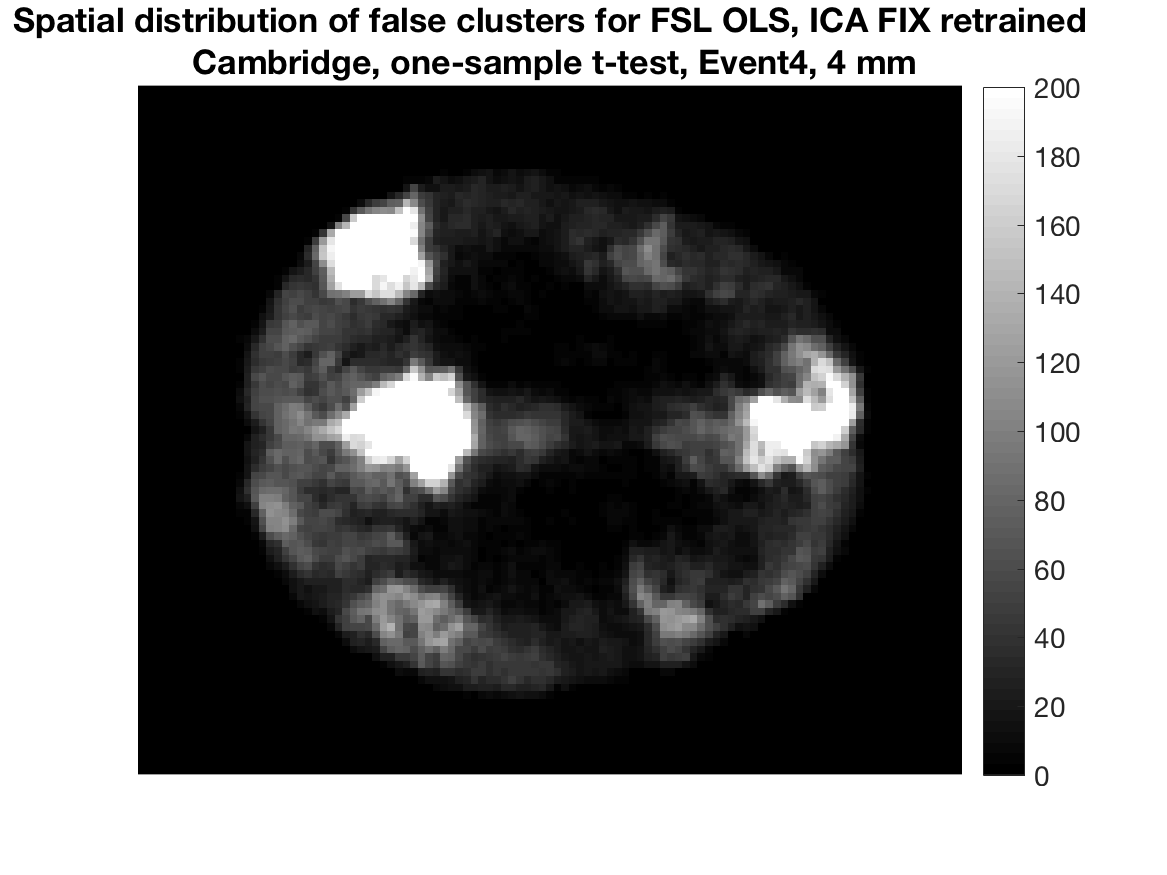}
}
\caption{\emph{The maps show voxel-wise incidence of false clusters for the Cambridge data, for 2 of the 6 different first level designs (a,b) no ICA FIX (c,d) ICA FIX pre-trained (e,f) ICA FIX re-trained for Cambridge. Left: results for design E2, Right: results for design E4. Image intensity is the number of times, out of 10,000 random analyses, a significant cluster occurred at a given voxel (CDT p = 0.01) for FSL OLS. Each analysis is a one-sample t-test using 20 subjects.}}
\label{fig:falseclusters_Cambridge_icafix}
\end{figure*}
\begin{figure*}
\subfigure[]{
\includegraphics[scale=0.45]{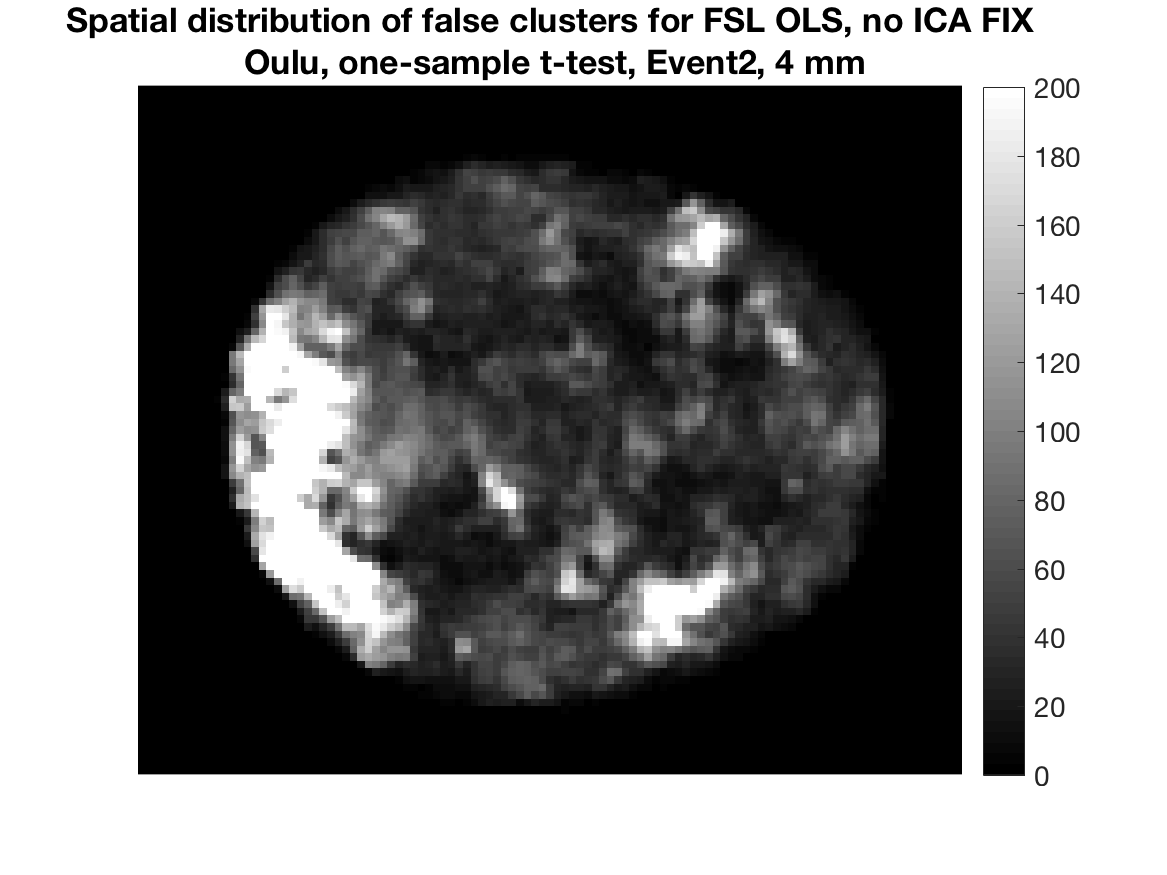}
}
\subfigure[]{
\includegraphics[scale=0.45]{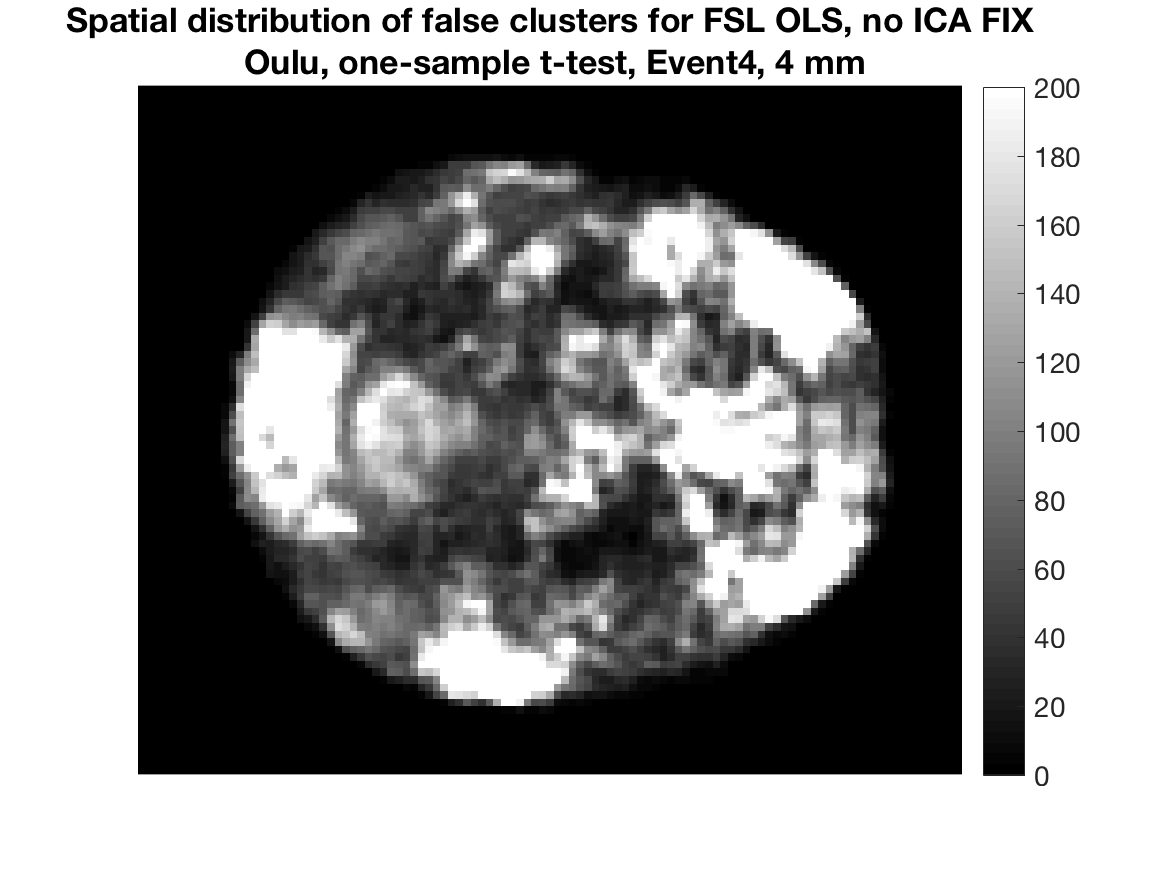}
}
\subfigure[]{
\includegraphics[scale=0.45]{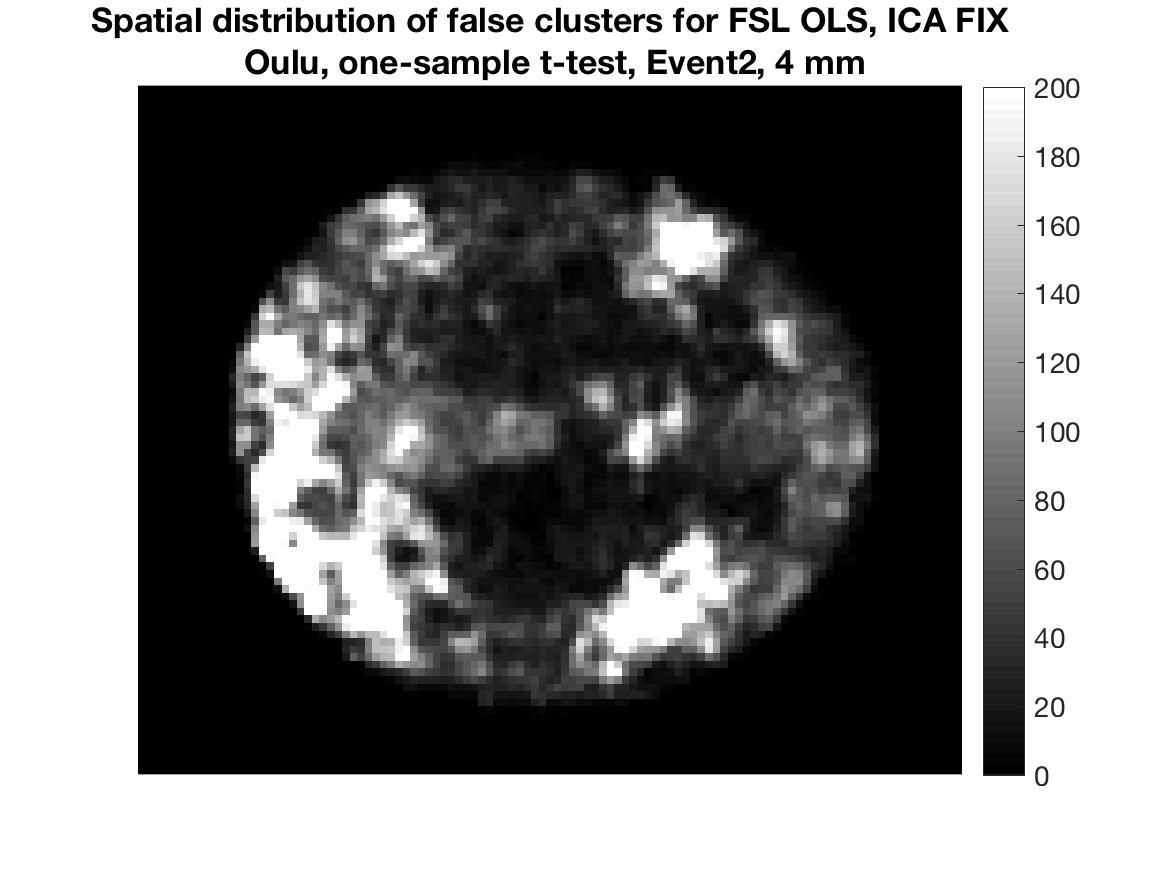}
}
\subfigure[]{
\includegraphics[scale=0.45]{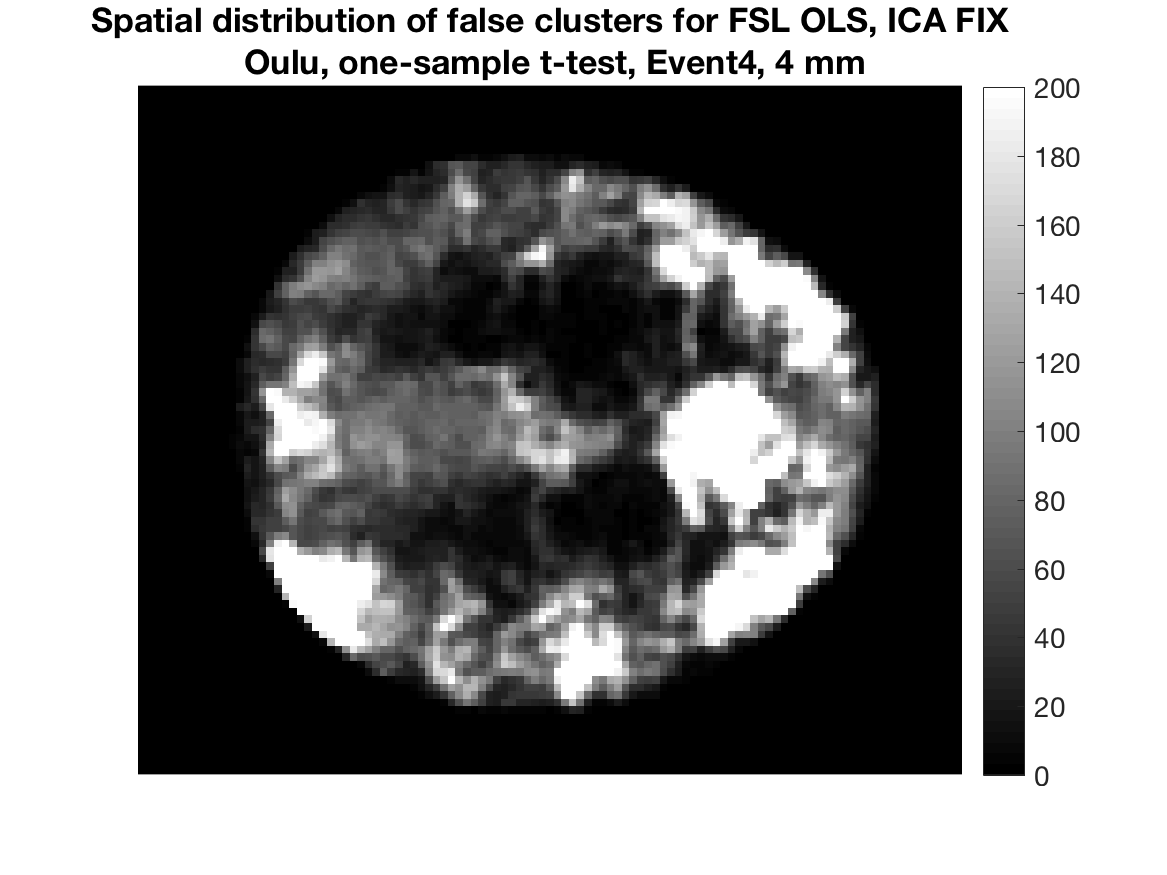}
}
\subfigure[]{
\includegraphics[scale=0.45]{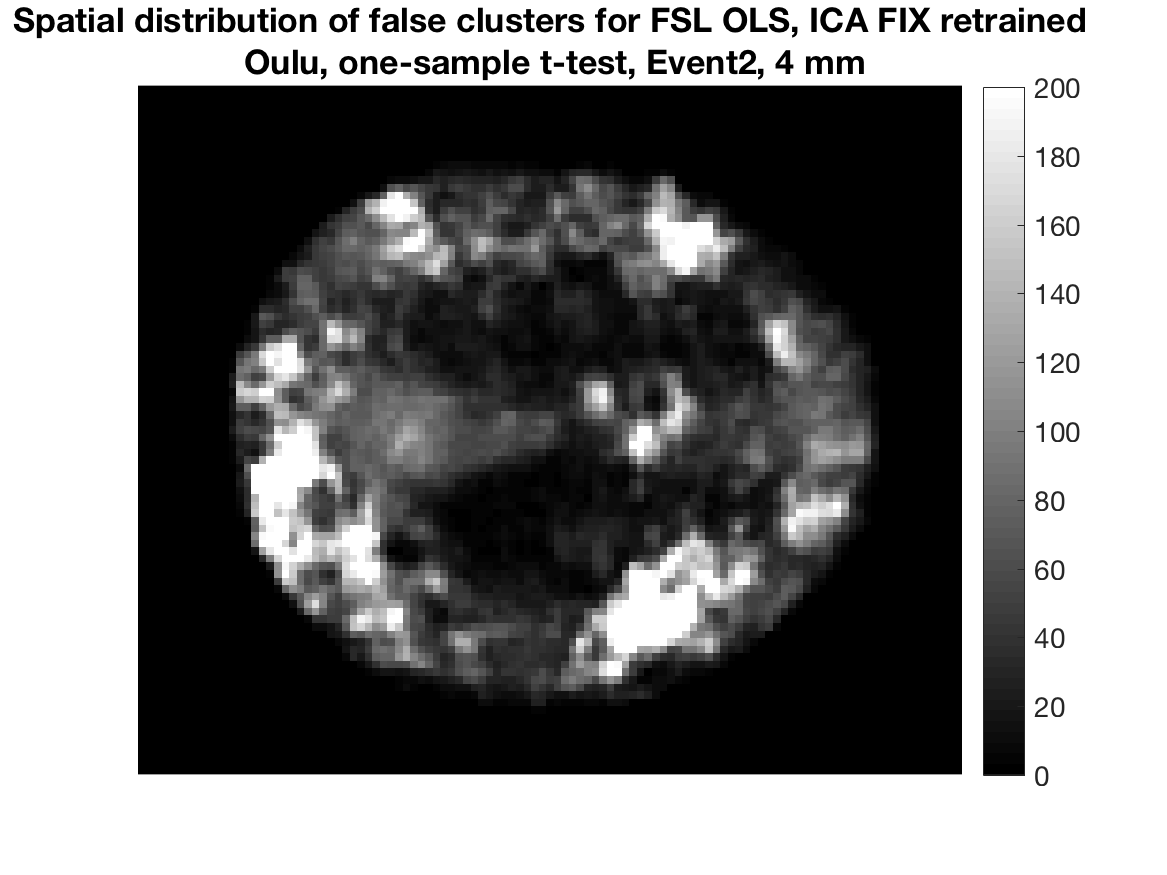}
}
\subfigure[]{
\includegraphics[scale=0.45]{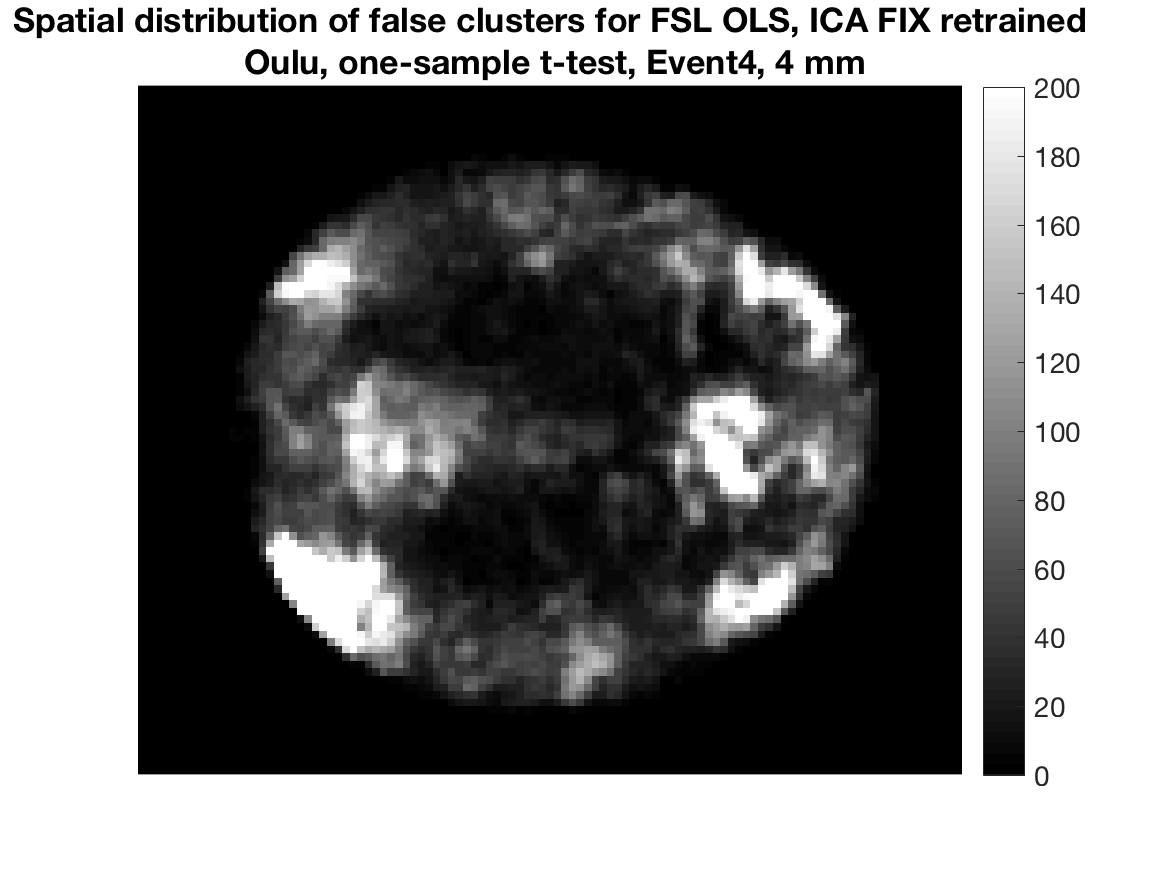}
}
\caption{\emph{The maps show voxel-wise incidence of false clusters for the Oulu data, for 2 of the 6 different first level designs (a,b) no ICA FIX (c,d) ICA FIX pre-trained (e,f) ICA FIX re-trained for Oulu. Left: results for design E2, Right: results for design E4. Image intensity is the number of times, out of 10,000 random analyses, a significant cluster occurred at a given voxel (CDT p = 0.01) for FSL OLS. Each analysis is a one-sample t-test using 20 subjects. The re-trained ICA FIX classifier is clearly better at suppressing artifacts compared to the pre-trained classifier, especially for design E4.}}
\label{fig:falseclusters_Oulu_icafix}
\end{figure*}
\begin{figure*}
\subfigure[]{
\includegraphics[scale=0.25]{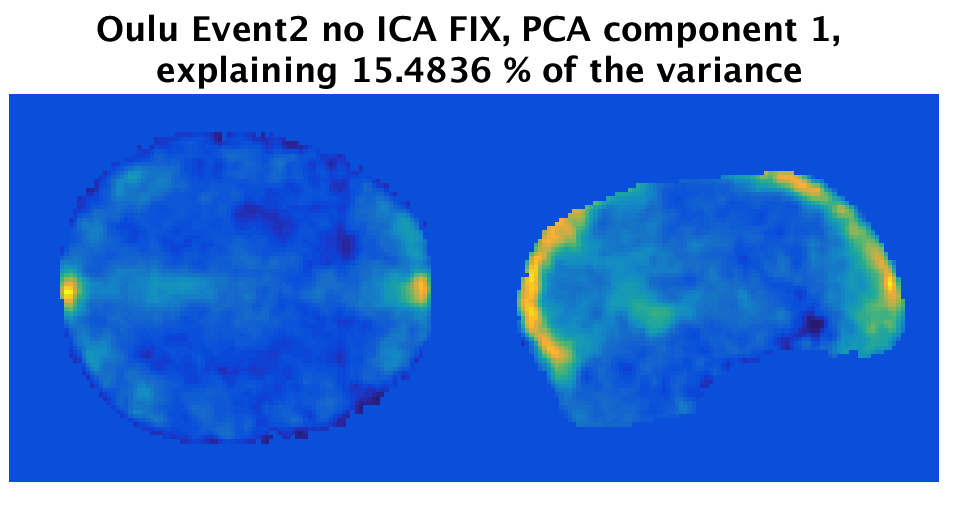}
}
\subfigure[]{
\includegraphics[scale=0.25]{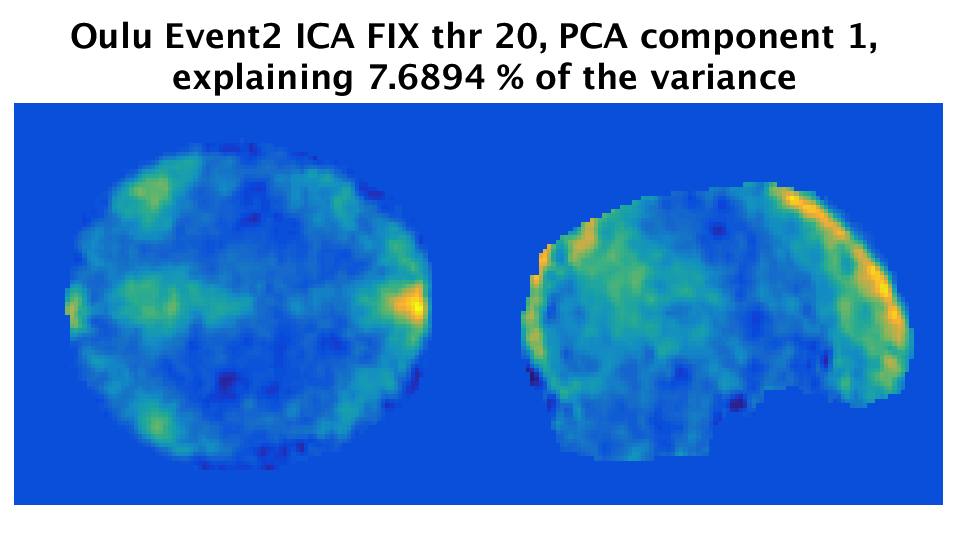}
}
\subfigure[]{
\includegraphics[scale=0.25]{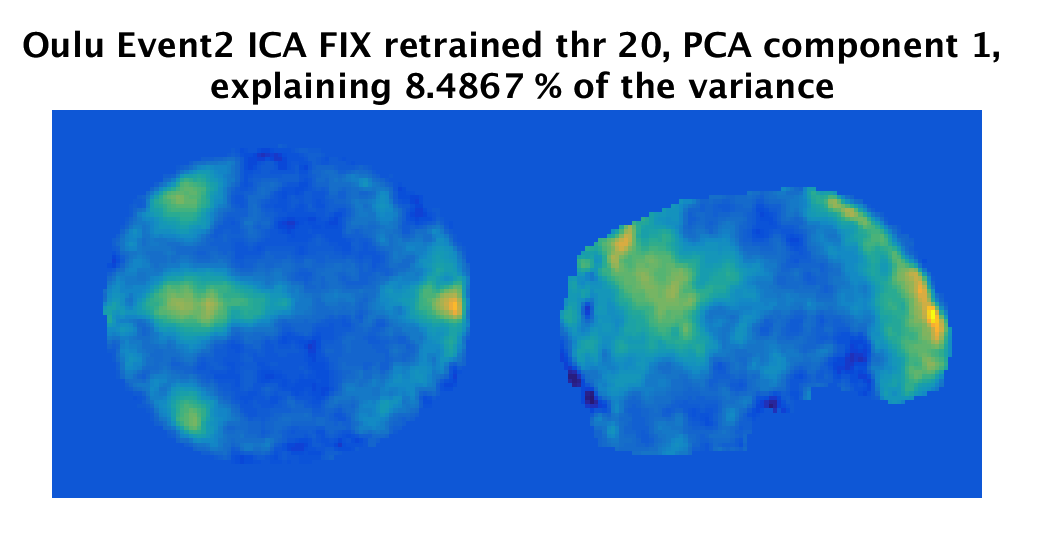}
}
\caption{\emph{The maps show an axial and a sagittal view of the first eigen component after running PCA on the 103 activity maps for Oulu E2, a) without ICA FIX b) with ICA FIX, using the pre-trained classifier, c) with ICA FIX, after retraining the ICA FIX classifier specifically for Oulu data. Using ICA FIX clearly suppresses the posterior part of the vein artefact in the superior sagittal sinus, but a portion of the artefact is still present. The retrained ICA FIX classifier is clearly better at suppressing the artefact. }}
\label{fig:PCA}
\end{figure*}
\begin{figure*}
\center
\subfigure[]{
\includegraphics[scale=0.75]{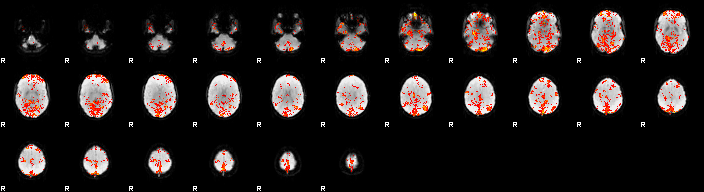}
}
\subfigure[]{
\includegraphics[scale=0.75]{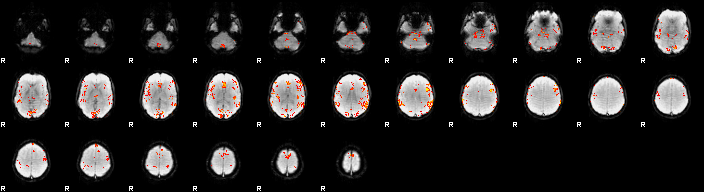}
}
\subfigure[]{
\includegraphics[scale=0.75]{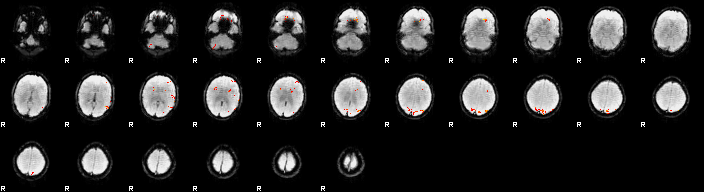}
}
\subfigure[]{
\includegraphics[scale=0.75]{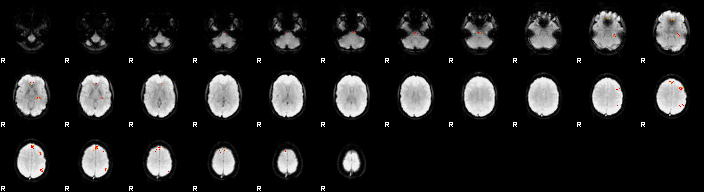}
}
\subfigure[]{
\includegraphics[scale=0.75]{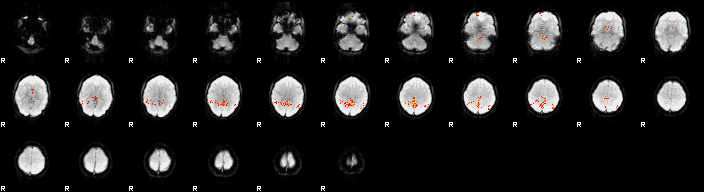}
}

\caption{\emph{Activity maps (thresholded at CDT p = 0.01 and cluster FWE corrected at p = 0.05, FSL default) for 5 Oulu subjects analyzed with 4 mm of smoothing and first level design E4. Despite testing for a difference between two random regressors, which are for design E4 also randomized over subjects, significant voxels are in several cases detected close to the superior sagittal sinus (indicating a vein artefact). Since many subjects have an activation difference in the same spatial location, this caused inflated false positive rates for the one-sample t-test. The two-sample t-test is not affected by these artefacts, since they cancel out when testing for a group difference.}}
\label{fig:Oulu_examples}
\end{figure*}

Figures~\ref{fig:falseclusters_Beijing_icafix} -  ~\ref{fig:falseclusters_Oulu_icafix} show cluster prevalence maps for group analyses without first running ICA FIX, with pre-trained ICA FIX and with re-trained ICA FIX, for first level designs E2 and E4. Using ICA FIX leads to false cluster maps that are more uniform across the brain, with fewer false clusters in white matter, and using ICA FIX made the biggest difference for the Oulu data. While Beijing and Cambridge sites have a concentration of clusters in posterior cingulate, frontal and parietal areas, Oulu has more clusters and a more diffuse pattern. Further inspection of these maps suggested a venous artefact, and running a PCA on the Oulu activity maps for design E2 finds substantial variation in the sagittal sinus picked up by the task regressor (see Figure~\ref{fig:PCA}). The posterior part of the artefact is suppressed by {\color{black} the pre-trained} ICA FIX {\color{black} classifier}, {\color{black} and the re-trained ICA FIX classifier is even better at suppressing the artefact}. Also see Figure~\ref{fig:Oulu_examples} for activation maps from 5 Oulu subjects, analyzed with design E4. In several cases, significant activity differences between two random task regressors are detected close to the superior sagittal sinus, indicating a vein artefact.

\clearpage

\section{Discussion}

We have presented results that support our original findings of inflated false positives with parametric cluster size inference. Specifically, new random null group task fMRI analyses, based on first level models with two fix regressors and models with two intersubject-randomized regressors, produced essentially the same results as the previous first level designs we considered. This argues against the charge that idiosyncratic attributes of our first level designs gave rise to our observed inflated false positives rates for cluster inference. Instead, we maintain that the best explanations for this behaviour are the long-tail spatial autocorrelation data (also present in MR phantom data~\citep{kriegeskorte}) and spatially-varying smoothness. {\color{black} Recently,~\citet{greve} showed that group analyses of cortical thickness, surface area and volume (using only the structural MRI data in the fcon1000 dataset~\citep{biswal2}) also lead to inflated false positive rates in some cases, indicating these issues affect structural analyses on the cortical surface as well, and thus is not specific to fMRI paradigms.}

It should be noted that AFNI provides another function for cluster thresholding, ETAC (equitable thresholding and clustering)~\citep{etac}, which performs better than the long-tail ACF function~\citep{coxbiorxiv} used here, but ETAC was not available when we started the new group analyses. AFNI also provides non-parametric group inference in the 3dttest++ function.

\subsection{Influence of artefacts on one-sample t-tests}

Another objective of this work was to understand and remediate the less-than-perfect false positive rate control for one-sample permutation tests. We tried various alternative modelling strategies, including data transformations and robust regression, but none yielded consistent control of FWE. It appears that (physiological) artifacts are a major problem for the Oulu data, although the MRIQC tool~\citep{esteban,gorgolewski} did not reveal any major quality differences between Beijing, Cambridge and Oulu. The contribution of physiological noise in fMRI depends on the spatial resolution (see e.g.~\citet{Bodurka}); larger voxels lead to a lower temporal signal to noise ratio. The Oulu data have a spatial resolution of 4 x 4 x 4.4 mm$^3$, compared to 3.13 x 3.13 x 3.6 mm$^3$ for Beijing and 3 x 3 x 3 mm$^3$ for Cambridge. Oulu voxels are thereby 2 times larger compared to Beijing voxels, and 2.6 times larger compared to Cambridge voxels, and this will make the Oulu data more prone to physiological noise. {\color{black} As mentioned in the Introduction, one can argue that a pure simulation~\citep{simulation} would avoid the problem of physiological noise, or that the Oulu data should be set aside, but we here opted to show results after denoising with ICA FIX, as many fMRI datasets have been collected without recordings of breathing and pulse.}

Some of our random regressors are strongly correlated with the fMRI data in specific brain regions (especially the superior sagittal sinus, the transverse sinus and the sigmoid sinus), which lead to inflated false positive rates. Other artifacts, such as CSF artifacts and susceptibility weighted artifacts, are also present in the data (compared to examples given by~\citet{griffanti}). For a two-sample t-test, artifacts in the same spatial location for all subjects cancel out, as one tests for a difference between two groups, but this is not the case for a one-sample t-test. Combining ICA FIX with a two-sided test led to nominal FWE rates for Beijing and Cambridge, but not for Oulu. As can be seen in Figures~\ref{fig:falseclusters_Beijing_icafix} -  ~\ref{fig:falseclusters_Oulu_icafix}, using ICA FIX clearly leads to false cluster maps which are more uniform accross the brain, with a lower number of false clusters in white matter. {\color{black} Re-training the ICA FIX classifier finally lead to nominal results for the Oulu data. A possible explanation is that the pre-trained classifier for standard fMRI data in ICA FIX is trained on fMRI data with a spatial resolution of 3.5 x 3.5 x 3.5 mm$^3$ (i.e. 1.6 times smaller voxels than Oulu).  {\color{black}Figure~\ref{fig:falseclusters_Oulu_icafix} shows that the re-trained classifier leads to more uniform false cluster maps, compared to the pre-trained classifier, for design E4 for Oulu.} As can be seen in Figure~\ref{fig:PCA}, the re-trained classifier is better at suppressing the artefact in the sagittal sinus, compared to the pre-trained classifier. We here trained the classifier for each dataset (Beijing, Cambridge, Oulu) using labeled ICA components from 10 subjects, as recommended by the ICA FIX user guide, and labeling components from more subjects can lead to even better results.} 


Using ICA FIX for resting state fMRI data is rather easy (but it currently requires a specific version of the R software), as pretrained weights are available for different kinds of fMRI data. However, using ICA FIX for task fMRI data will require more work, as it is necessary to first manually classify ICA components~\citep{griffanti} to provide training data for the classifier~\citep{icafix1}. An open database of manually classified fMRI ICA components, similar to NeuroVault~\citep{neurovault}, could potentially be used for fMRI researchers to automatically denoise their task fMRI data. A natural extension of MRIQC~\citep{esteban,gorgolewski} would then be to also measure the presence of artifacts in each fMRI dataset, by doing ICA and then comparing each component to the manually classified components in the open database. We also recommend researchers to collect physiological data, such that signal related to breathing and pulse can be modeled~\citep{retroicor,lund,respiratory,chang,bollmann}. This is especially important for 7T fMRI data, for which the physiological noise is often stronger compared to the thermal noise~\citep{7t1,7t2}. {\color{black} Alternatives to collecting physiological data, or using ICA FIX, include ICA AROMA~\citep{aroma} and FMRIPrep~\citep{fmriprep}. FMRIPrep can automatically generate nuisance regressors (e.g. from CSF and white matter) to be included in the statistical analysis. }

\clearpage
\subsection{Effect of multiband data on cluster inference}

We note that multiband MR sequences~\citep{multiband} are becoming increasingly common to improve temporal and/or spatial resolution, for example as provided by the Human Connectome Project~\citep{essen} and the enhanced NKI-Rockland sample~\citep{rockland}. Multiband data have a potentially complex spatial autocorrelation (see, e.g.~\citet{risk}), and an important topic for future work is establishing how this impacts parametric cluster inference. The non-parametric permutation test~\citep{winkler} does not make any assumption regarding the shape of the spatial autocorrelation function, and is therefore expected to perform well for any MR sequence. 

\subsection{Interpretation of affected studies}

In Appendix A we provide a rough bibliographic analysis to provide an estimate of how many articles used this particular CDT p = 0.01 setting. For a review conducted in January 2018, we estimated that out of 23,000 fMRI publications about 2,500, over 10\%, of all studies have used this most problematic setting with parametric inference. While this calculation suggests how the literature as a whole can be interpreted, a more practical question is how one individual affected study can be interpreted. When examining a study that uses CDT p = 0.01, or one that uses no correction at all, it is useful to consider three possible states of nature:

\begin{itemize}
\item[State 1:] Effect is truly present, and with revised methods, significance is retained
\item[State 2:] Effect is truly present, but with revised methods, significance is lost
\item[State 3:] Effect is truly null, absent; the study's detection is a false positive
\end{itemize}

In each of these, the statement about `truth' reflects presence or absence of the effect in the population from which the subjects were drawn. When considering heterogeneity of different populations used for research, we could also add a fourth state:

\begin{itemize}
\item[State 4:] Effect is truly null in population sampled, and this study's detection is a false positive; but later studies find and replicate the effect in other populations.
\end{itemize}

These could be summarised as ``State 1: Robust true positive,'' ``State 2: Fragile true positive,'' ``State 3: False positive'' and ``State 4: Idiosyncratic false positive.'' 

Unfortunately we can never know the true state of an effect, and, because of a lack of data archiving and sharing, we will mostly never know whether significance is retained or lost with reanalysis.  All we can do is make qualitative judgments on individual works. To this end we can suggest that findings with no form of corrected significance receive the greatest skepticism; likewise, CDT p = 0.01 cluster size inference cluster p-values that {\em just} barely fall below 5\% FWE significance should be judged with great skepticism. In fact, given small perturbations arising from a range of methodological choices, {\em all} research findings on the edge of a significance threshold deserves such skepticism. On the other hand, findings based on large clusters with P-values far below 0.05 could possibly survive a re-analysis with improved methods.

\section{Conclusions}

To summarize, our new results confirm that inflated familywise error rates for parametric cluster inference are also present when testing for a difference between two tasks, and when randomizing the task over subjects. Furthermore, the inflated familywise error rates for the non-parametric one-sample t-tests are due to random correlations with artifacts in the fMRI data, which for Beijing and Cambridge we found could be suppressed using {\color{black} the pre-trained} ICA FIX {\color{black} classifier for standard fMRI data}. The Oulu data were collected with a lower spatial resolution, and are therefore more prone to physiological noise. {\color{black} By re-training the ICA FIX classifier specifically for the Oulu data, nominal results were finally obtained for Oulu as well. Data cleaning is clearly important for task fMRI, and not only for resting state fMRI.}

\section*{Acknowledgements}

The authors have no conflict of interest to declare. The authors would like to thank Jeanette Mumford for fruitful discussions. This study was supported by Swedish research council grants 2013-5229 and 2017-04889. Funding was also provided by the Center for Industrial Information Technology (CENIIT) at Link\"{o}ping University, and the Knut and Alice Wallenberg foundation project "Seeing organ function". Thomas E. Nichols was supported by the Wellcome Trust (100309/Z/12/Z) and the NIH (R01 EB015611). The Nvidia Corporation, who donated the Nvidia Titan X Pascal graphics card used to run all permutation tests, is also acknowledged. This study would not be possible without the recent data-sharing initiatives in the neuroimaging field. We therefore thank the Neuroimaging Informatics Tools and Resources Clearinghouse and all of the researchers who have contributed with resting-state data to the 1,000 Functional Connectomes Project.

\clearpage

\appendix

\section{Results for CDT p = 0.01}

\begin{figure*}
\subfigure[]{
\includegraphics[scale=0.425]{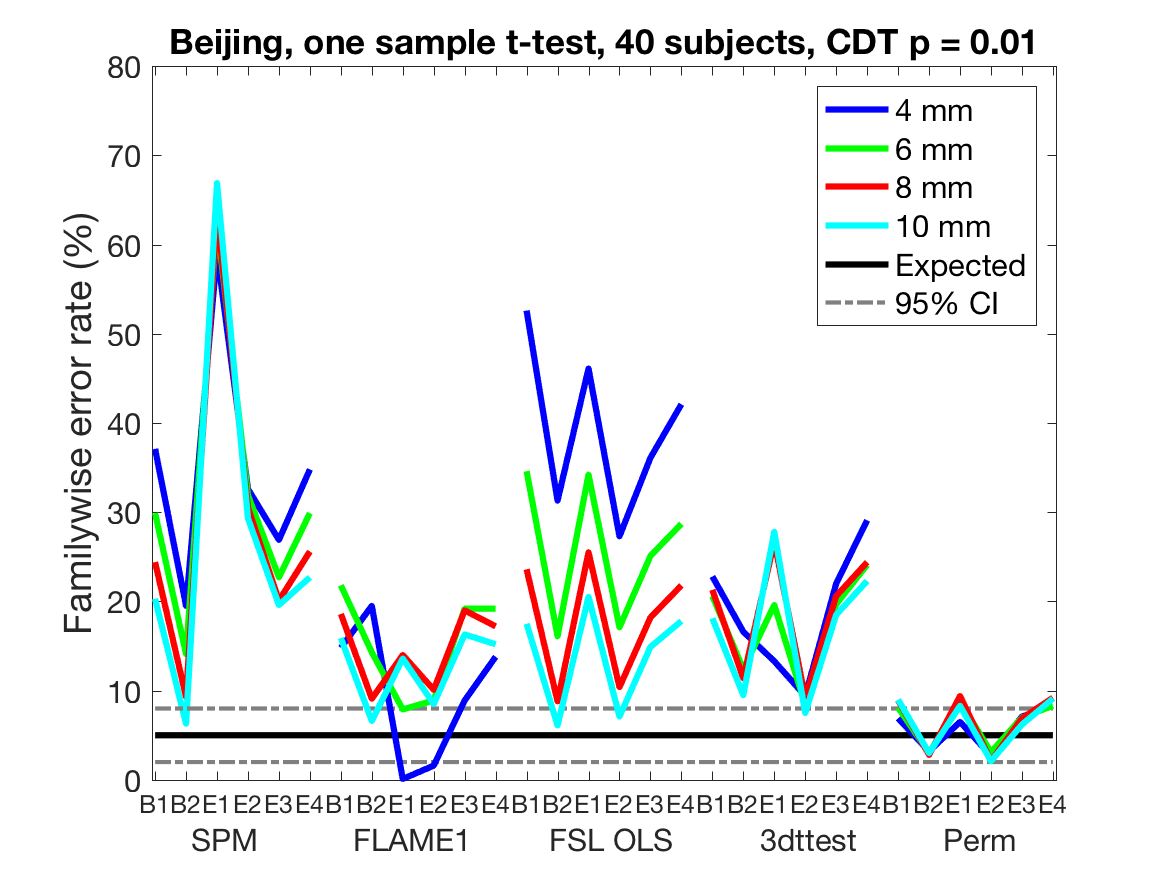}
}
\subfigure[]{
\includegraphics[scale=0.425]{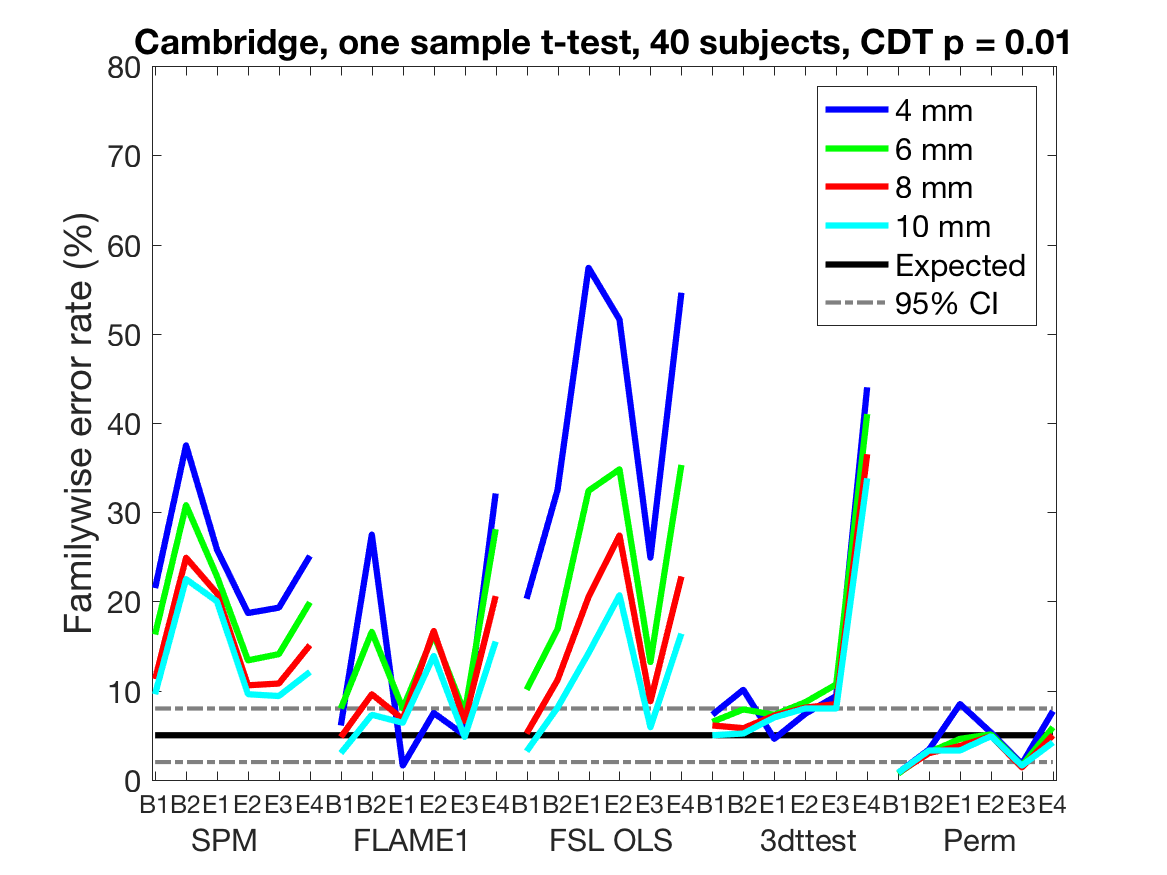}
}
\subfigure[]{
\includegraphics[scale=0.425]{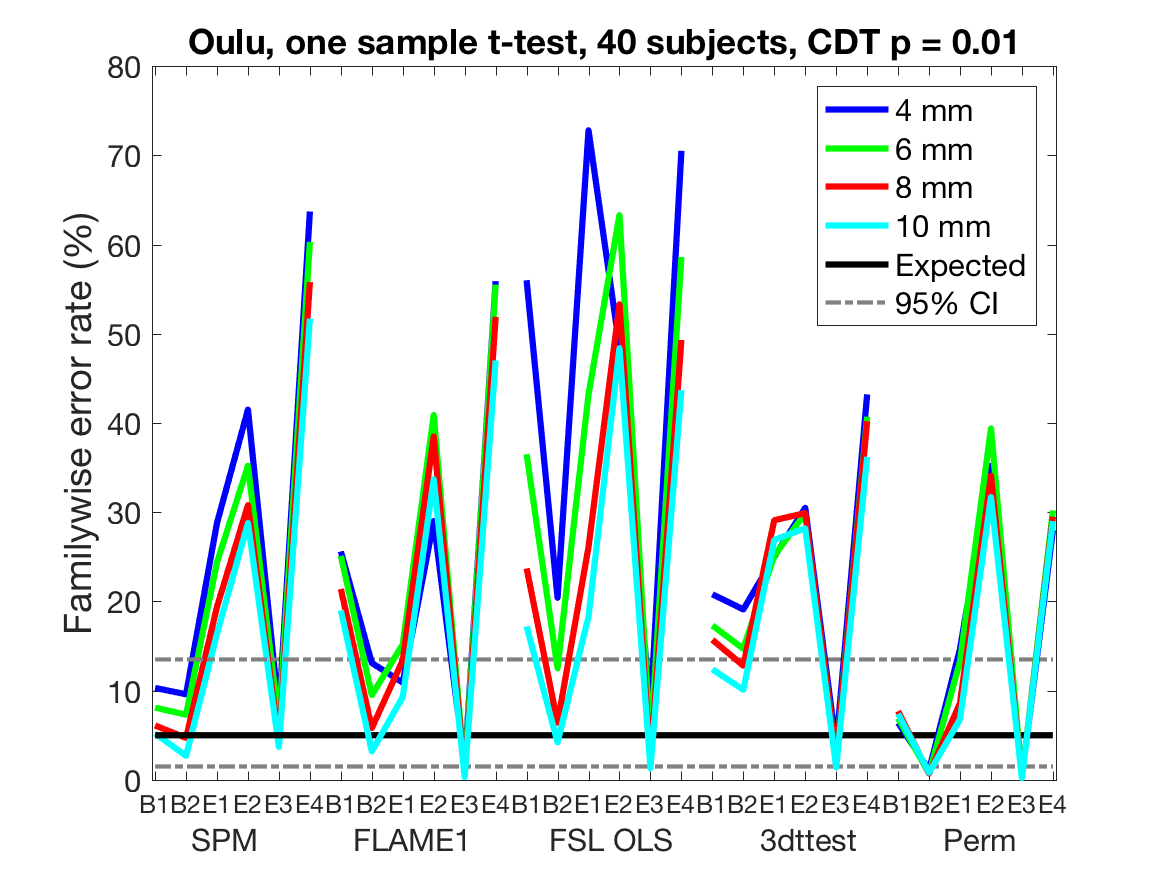}
}
\caption{\emph{Results for one sample t-test and cluster-wise inference using a cluster defining threshold (CDT) of p = 0.01, showing estimated familywise error rates for 4 - 10 mm of smoothing and six different activity paradigms (old paradigms B1, B2, E1, E2 and new paradigms E3, E4), for SPM, FSL, AFNI and a permutation test. These results are for a group size of 40. Each statistic map was first thresholded using a CDT of p = 0.01, uncorrected for multiple comparisons, and the surviving clusters were then compared to a FWE-corrected cluster extent threshold, $p_{FWE} = 0.05$. The estimated familywise error rates are simply the number of analyses with any significant group activations divided by the number of analyses (1,000). \textbf{(a)} results for Beijing data \textbf{(b)} results for Cambridge data \textbf{(c)} results for Oulu data.}}
\label{fig:fwe_cluster_onesample_40_subjects_cdt1}
\end{figure*}

\begin{figure*}
\subfigure[]{
\includegraphics[scale=0.425]{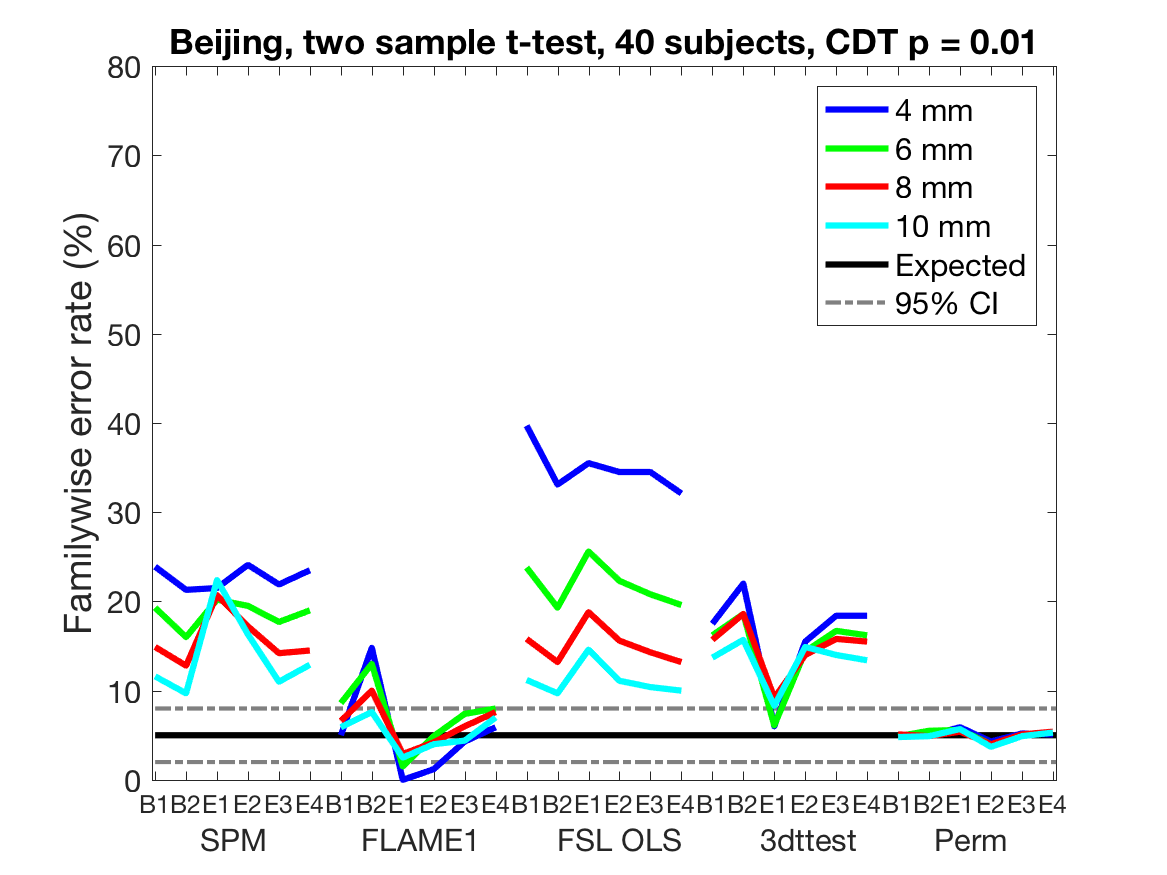}
}
\subfigure[]{
\includegraphics[scale=0.425]{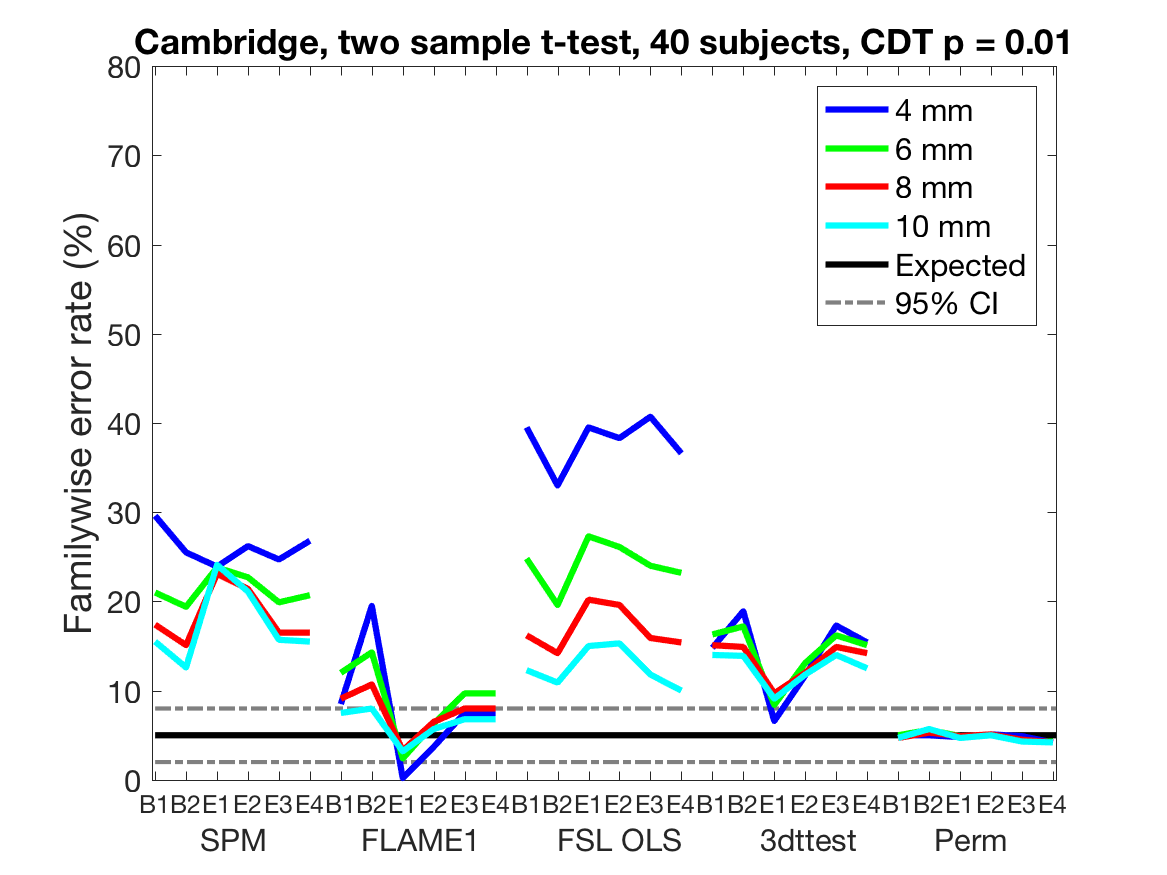}
}
\subfigure[]{
\includegraphics[scale=0.425]{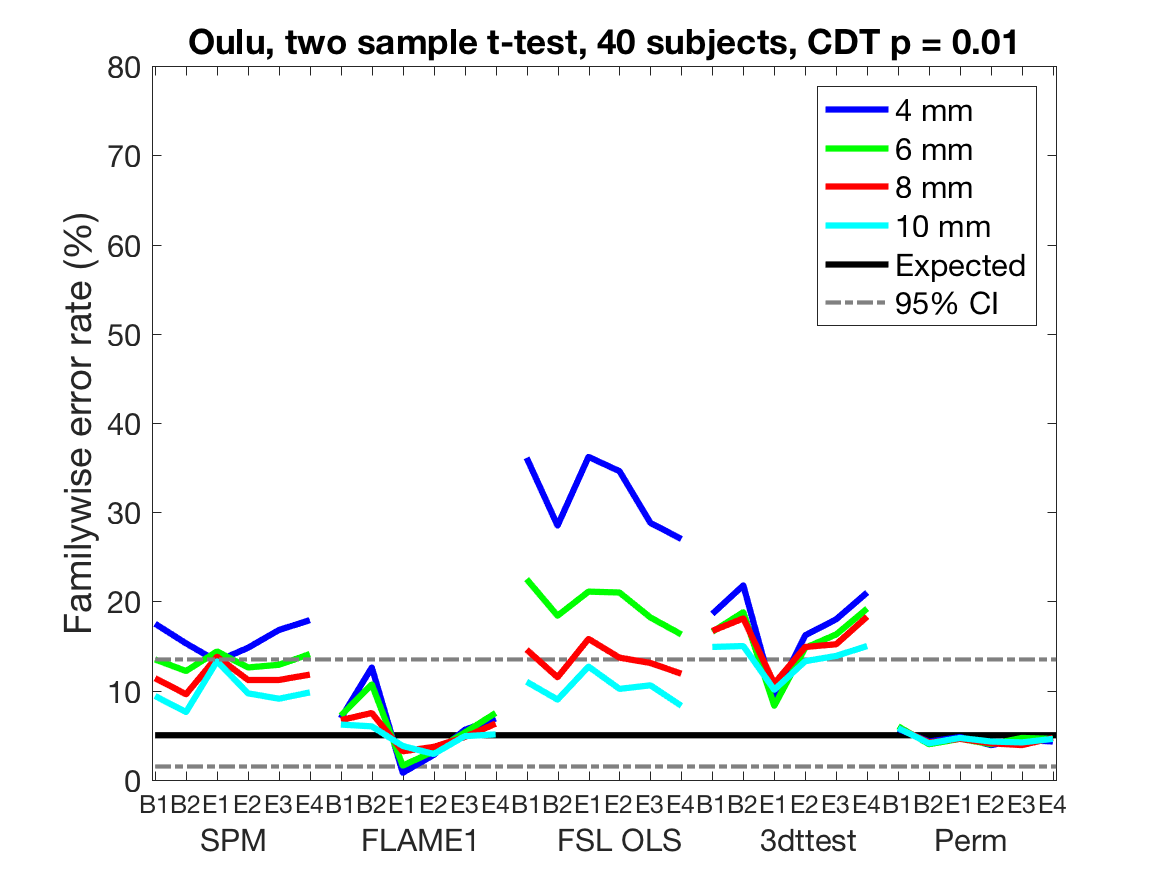}
}
\caption{\emph{Results for two sample t-test and cluster-wise inference using a cluster defining threshold (CDT) of p = 0.01, showing estimated familywise error rates for 4 - 10 mm of smoothing and six different activity paradigms (old paradigms B1, B2, E1, E2 and new paradigms E3, E4), for SPM, FSL, AFNI and a permutation test. These results are for a group size of 20, giving a total of 40 subjects. Each statistic map was first thresholded using a CDT of p = 0.01, uncorrected for multiple comparisons, and the surviving clusters were then compared to a FWE-corrected cluster extent threshold, $p_{FWE} = 0.05$. The estimated familywise error rates are simply the number of analyses with any significant group activations divided by the number of analyses (1,000). \textbf{(a)} results for Beijing data \textbf{(b)} results for Cambridge data \textbf{(c)} results for Oulu data.}}
\label{fig:fwe_cluster_twosample_40_subjects_cdt1}
\end{figure*}

\section{Bibliometrics of Cluster Inference}

\subsection{Number of affected studies}

Here we conduct a biblibographic analysis to obtain an estimate of how much of the literature depends on our most troubling result, the severe inflation of FWE for a cluster-defining threshold (CDT) of p = 0.01.

We use the results of a systematic review of the fMRI literature conducted by~\citet{carp} and~\citet{woo}, which provides essential statistics on prevalence of cluster inference techniques.~\citet{carp} defined a search for fMRI publications that today finds about $N(\mathrm{fMRI})=23,000$ publications\footnote{22,629 hits for Pubmed search text ``(((((((fmri[title/abstract] OR functional MRI[title/abstract]) OR functional Magnetic Resonance Imaging[title/abstract]) AND brain[title/abstract]))) AND humans[MeSH Terms])) NOT systematic [sb]'', conducted 30th January, 2018.}.  Drawing on a sample of 300 publications\footnote{Carp (2012) further constrained his search to publications with full text available in the open-access database PubMed Central (PMC).} published 2007 -- 2012, Carp found $P(\mathrm{HasData})=241/300=80\%$ contained original data, and among these\\$P(\mathrm{Corrected}|\mathrm{HasData})=59\%$ used some form of correction for multiple comparisons\footnote{``Although a majority of studies (59\%) reported the use of some variety of correction for multiple comparisons, a substantial minority did not.''}.~\citet{woo}, considering a sample of 815 papers\footnote{Woo et al. used ``fmri" and ``threshold" as keywords on original fMRI research papers published between in {\em Cerebral Cortex}, {\em Nature}, {\em Nature Neuroscience}, {\em NeuroImage}, {\em Neuron}, {\em PNAS}, and {\em Science}, yielding over 1500 papers; then following exclusion criterion were applied ``(1) non-human studies, (2) lesion studies, (3) studies inwhich a threshold or correction method could not be clarified, (4) voxel-based morphometry studies, (5) studies primarily about methodology, and (6) machine-learning based studies"} published 2010 -- 2011; of these, they found $P(\mathrm{ClusterInference}|\mathrm{HasData})=607/814=75\%$; noting that 6\% (Fig. 1, Woo et al.) of the 814 include studies with no correction, we also compute \\$P(\mathrm{ClusterInference}|\mathrm{HasData},\mathrm{Corrected})$ $ \\=607/(814-0.06\times814)=79\%$.  Finally, with data from Figure 2(B) of~\citet{woo} (shown in Table~\ref{tab:woo}, kindly supplied by the authors) from the 480 studies that used cluster inference with correction (and had sufficient detail) we can compute  $$P(\mathrm{CDT(}P\geq0.01\mathrm{)}|\mathrm{ClusterInference},\mathrm{HasData},\mathrm{Corrected})=(35+80)/480=24\%.$$

Thus we can finally estimate the number of published fMRI studies using corrected cluster inference as
$$
N(\mathrm{HasData},\mathrm{Corrected},\mathrm{ClusterInference}) =
23,000 \times 0.59 \times 0.79 = 10,720,
$$
and, among those, $10,720\times0.24=2,573$ used a CDT of P = 0.01 or higher, or about 10\% of publications reporting original fMRI results. 

There are many caveats to this calculation, starting with different sampling criterion used in the two studies, and the ever-changing patterns of practice in neuroimaging. However, a recent survey of task fMRI papers published in early 2017 found that 270/388 = 69.6\% used cluster inference, and 72/270 = 26.7\% used a CDT of p = 0.01 or higher, suggesting that the numbers above remain representative~\citep{yeung}.

\clearpage
\newpage

\renewcommand{\thetable}{A.\arabic{table}}

\begin{table}
\begin{center}
\caption{Upon request, the authors of Woo et al. (2014) provided a detailed cross-tabulation of the frequencies of different cluster defining thresholds (the data presented in Figure 2(B) of their paper).  Among the 607 studies that used cluster thresholding, they found 480 studies for which sufficient detail could be obtained to record the software and the particular cluster defining threshold used.}

\begin{tabular}{c|ccccc|c}
       CDT & AFNI &BrainVoyager& FSL &   SPM  & Others & TOTAL\\ \hline
$>$.01	& ~9	& ~5	& ~9	& ~~8	& ~4	& ~35\\
.01	    & ~9	& ~4	& 44	& ~20	& ~3	& ~80\\
.005	& 24	& ~6	& ~1	& ~48	& ~3    & ~82	\\
.001	& 13	& 20	& 11	& 206	& ~5	& 255\\
$<$.001	& ~2	& ~5	& ~3	& ~16	& ~2	& ~28\\ \hline
TOTAL   & 57    & 40    & 68    & 298   & 17    & 480
\end{tabular}
\end{center}
\label{tab:woo}
\end{table}

\clearpage
\newpage

\bibliography{references_full}

\begin{thebibliography}{61}
\expandafter\ifx\csname natexlab\endcsname\relax\def\natexlab#1{#1}\fi
\providecommand{\bibinfo}[2]{#2}
\ifx\xfnm\relax \def\xfnm[#1]{\unskip,\space#1}\fi
\bibitem[{Ashburner(2012)}]{spm}
\bibinfo{author}{Ashburner, J.} (\bibinfo{year}{2012}).
\newblock \bibinfo{title}{Spm: A history}.
\newblock {\it \bibinfo{journal}{NeuroImage}\/},  {\it \bibinfo{volume}{62}\/},
  \bibinfo{pages}{791 -- 800}.
\bibitem[{Beckmann \& Smith(2004)}]{melodic}
\bibinfo{author}{Beckmann, C.}, \& \bibinfo{author}{Smith, S.}
  (\bibinfo{year}{2004}).
\newblock \bibinfo{title}{{Probabilistic independent component analysis for
  functional magnetic resonance imaging}}.
\newblock {\it \bibinfo{journal}{{IEEE Transactions on Medical Imaging}}\/},
  {\it \bibinfo{volume}{23}\/}, \bibinfo{pages}{137 -- 152}.
\bibitem[{Birn et~al.(2006)Birn, Diamond, Smith \& Bandettini}]{respiratory}
\bibinfo{author}{Birn, R.}, \bibinfo{author}{Diamond, J.},
  \bibinfo{author}{Smith, M.}, \& \bibinfo{author}{Bandettini, P.}
  (\bibinfo{year}{2006}).
\newblock \bibinfo{title}{Separating respiratory-variation-related fluctuations
  from neuronal-activity-related fluctuations in {fMRI}}.
\newblock {\it \bibinfo{journal}{NeuroImage}\/},  {\it \bibinfo{volume}{31}\/},
  \bibinfo{pages}{1536 -- 1548}.
\bibitem[{Biswal et~al.(2010)Biswal, Mennes, ... \& Milham}]{biswal2}
\bibinfo{author}{Biswal, B.}, \bibinfo{author}{Mennes, M.},
  \bibinfo{author}{..., X.~Z.}, \& \bibinfo{author}{Milham, M.}
  (\bibinfo{year}{2010}).
\newblock \bibinfo{title}{Toward discovery science of human brain function}.
\newblock {\it \bibinfo{journal}{PNAS}\/},  {\it \bibinfo{volume}{107}\/},
  \bibinfo{pages}{4734--4739}.
\bibitem[{Bodurka et~al.(2007)Bodurka, Ye, Petridou, Murphy \&
  Bandettini}]{Bodurka}
\bibinfo{author}{Bodurka, J.}, \bibinfo{author}{Ye, F.},
  \bibinfo{author}{Petridou, N.}, \bibinfo{author}{Murphy, K.}, \&
  \bibinfo{author}{Bandettini, P.} (\bibinfo{year}{2007}).
\newblock \bibinfo{title}{{Mapping the MRI voxel volume in which thermal noise
  matches physiological noise - Implications for fMRI}}.
\newblock {\it \bibinfo{journal}{NeuroImage}\/},  {\it \bibinfo{volume}{34}\/},
  \bibinfo{pages}{542 -- 549}.
\bibitem[{Bollmann et~al.(2018)Bollmann, Puckett, Cunnington \&
  Barth}]{bollmann}
\bibinfo{author}{Bollmann, S.}, \bibinfo{author}{Puckett, A.},
  \bibinfo{author}{Cunnington, R.}, \& \bibinfo{author}{Barth, M.}
  (\bibinfo{year}{2018}).
\newblock \bibinfo{title}{{Serial correlations in single-subject fMRI with
  sub-second TR}}.
\newblock {\it \bibinfo{journal}{NeuroImage}\/},  {\it
  \bibinfo{volume}{166}\/}, \bibinfo{pages}{152 -- 166}.
\bibitem[{Carp(2012)}]{carp}
\bibinfo{author}{Carp, J.} (\bibinfo{year}{2012}).
\newblock \bibinfo{title}{The secret lives of experiments: Methods reporting in
  the {fMRI} literature}.
\newblock {\it \bibinfo{journal}{NeuroImage}\/},  {\it \bibinfo{volume}{63}\/},
  \bibinfo{pages}{289--300}.
\bibitem[{Chang \& Glover(2009)}]{chang}
\bibinfo{author}{Chang, C.}, \& \bibinfo{author}{Glover, G.}
  (\bibinfo{year}{2009}).
\newblock \bibinfo{title}{Effects of model-based physiological noise correction
  on default mode network anti-correlations and correlations}.
\newblock {\it \bibinfo{journal}{NeuroImage}\/},  {\it \bibinfo{volume}{47}\/},
  \bibinfo{pages}{1448 -- 1459}.
\bibitem[{Chen et~al.(2018)Chen, Cox, Glen, Rajendra, Reynolds \&
  Taylor}]{twosided}
\bibinfo{author}{Chen, G.}, \bibinfo{author}{Cox, R.~W.},
  \bibinfo{author}{Glen, D.~R.}, \bibinfo{author}{Rajendra, J.~K.},
  \bibinfo{author}{Reynolds, R.~C.}, \& \bibinfo{author}{Taylor, P.~A.}
  (\bibinfo{year}{2018}).
\newblock \bibinfo{title}{{A tail of two sides: Artificially doubled false
  positive rates in neuroimaging due to the sidedness choice with t-tests}}.
\newblock {\it \bibinfo{journal}{bioRxiv,10.1101/328567}\/}, .
\bibitem[{Cox et~al.(2017{\natexlab{a}})Cox, Chen, Glen, Reynolds \&
  Taylor}]{coxPNAS}
\bibinfo{author}{Cox, R.}, \bibinfo{author}{Chen, G.}, \bibinfo{author}{Glen,
  D.}, \bibinfo{author}{Reynolds, R.}, \& \bibinfo{author}{Taylor, P.}
  (\bibinfo{year}{2017}{\natexlab{a}}).
\newblock \bibinfo{title}{{FMRI Clustering and False Positive Rates}}.
\newblock {\it \bibinfo{journal}{PNAS}\/},  {\it \bibinfo{volume}{114}\/},
  \bibinfo{pages}{E3370--E3371}.
\bibitem[{Cox et~al.(2017{\natexlab{b}})Cox, Chen, Glen, Reynolds \&
  Taylor}]{coxbiorxiv}
\bibinfo{author}{Cox, R.}, \bibinfo{author}{Chen, G.}, \bibinfo{author}{Glen,
  D.}, \bibinfo{author}{Reynolds, R.}, \& \bibinfo{author}{Taylor, P.}
  (\bibinfo{year}{2017}{\natexlab{b}}).
\newblock \bibinfo{title}{{FMRI Clustering in AFNI: False-Positive Rates
  Redux}}.
\newblock {\it \bibinfo{journal}{Brain Connectivity}\/},  {\it
  \bibinfo{volume}{7}\/}, \bibinfo{pages}{152--171}.
\bibitem[{Cox(1996)}]{afni}
\bibinfo{author}{Cox, R.~W.} (\bibinfo{year}{1996}).
\newblock \bibinfo{title}{{AFNI}: Software for analysis and visualization of
  functional magnetic resonance neuroimages}.
\newblock {\it \bibinfo{journal}{Computers and Biomedical Research}\/},  {\it
  \bibinfo{volume}{29}\/}, \bibinfo{pages}{162--173}.
\bibitem[{Cox(2018)}]{etac}
\bibinfo{author}{Cox, R.~W.} (\bibinfo{year}{2018}).
\newblock \bibinfo{title}{Equitable thresholding and clustering}.
\newblock {\it \bibinfo{journal}{bioRxiv, 10.1101/295931}\/}, .
\bibitem[{Eklund et~al.(2014)Eklund, Dufort, Villani \& LaConte}]{broccoli}
\bibinfo{author}{Eklund, A.}, \bibinfo{author}{Dufort, P.},
  \bibinfo{author}{Villani, M.}, \& \bibinfo{author}{LaConte, S.}
  (\bibinfo{year}{2014}).
\newblock \bibinfo{title}{{BROCCOLI: Software for fast fMRI analysis on
  many-core CPUs and GPUs}}.
\newblock {\it \bibinfo{journal}{Frontiers in Neuroinformatics}\/},  {\it
  \bibinfo{volume}{8:24}\/}.
\bibitem[{Eklund et~al.(2016{\natexlab{a}})Eklund, Nichols \&
  Knutsson}]{eklundPNAS}
\bibinfo{author}{Eklund, A.}, \bibinfo{author}{Nichols, T.}, \&
  \bibinfo{author}{Knutsson, H.} (\bibinfo{year}{2016}{\natexlab{a}}).
\newblock \bibinfo{title}{Cluster failure: why {fMRI} inferences for spatial
  extent have inflated false positive rates}.
\newblock {\it \bibinfo{journal}{PNAS}\/},  {\it \bibinfo{volume}{113}\/},
  \bibinfo{pages}{7900--7905}.
\bibitem[{Eklund et~al.(2016{\natexlab{b}})Eklund, Nichols \&
  Knutsson}]{eklundPNAScorrection}
\bibinfo{author}{Eklund, A.}, \bibinfo{author}{Nichols, T.}, \&
  \bibinfo{author}{Knutsson, H.} (\bibinfo{year}{2016}{\natexlab{b}}).
\newblock \bibinfo{title}{Correction for {Eklund} et al., {Cluster} failure:
  why {fMRI} inferences for spatial extent have inflated false positive rates}.
\newblock {\it \bibinfo{journal}{PNAS}\/},  {\it \bibinfo{volume}{113}\/},
  \bibinfo{pages}{E4929}.
\bibitem[{Eklund et~al.(2017)Eklund, Nichols \& Knutsson}]{eklundPNAS2}
\bibinfo{author}{Eklund, A.}, \bibinfo{author}{Nichols, T.}, \&
  \bibinfo{author}{Knutsson, H.} (\bibinfo{year}{2017}).
\newblock \bibinfo{title}{{Reply to Brown and Behrmann, Cox et al. and Kessler
  et al.: Data and code sharing is the way forward for fMRI}}.
\newblock {\it \bibinfo{journal}{PNAS}\/},  {\it \bibinfo{volume}{114}\/},
  \bibinfo{pages}{E3374--E3375}.
\bibitem[{Essen et~al.(2013)Essen, Smith, Barch, Behrens, Yacoub \&
  Ugurbil}]{essen}
\bibinfo{author}{Essen, D.~V.}, \bibinfo{author}{Smith, S.},
  \bibinfo{author}{Barch, D.}, \bibinfo{author}{Behrens, T.},
  \bibinfo{author}{Yacoub, E.}, \& \bibinfo{author}{Ugurbil, K.}
  (\bibinfo{year}{2013}).
\newblock \bibinfo{title}{{The WU-Minn Human Connectome Project: An overview}}.
\newblock {\it \bibinfo{journal}{NeuroImage}\/},  {\it \bibinfo{volume}{80}\/},
  \bibinfo{pages}{62--79}.
\bibitem[{Esteban et~al.(2017)Esteban, Birman, Schaer, Koyejo, Poldrack \&
  Gorgolewski}]{esteban}
\bibinfo{author}{Esteban, O.}, \bibinfo{author}{Birman, D.},
  \bibinfo{author}{Schaer, M.}, \bibinfo{author}{Koyejo, O.},
  \bibinfo{author}{Poldrack, R.}, \& \bibinfo{author}{Gorgolewski, K.}
  (\bibinfo{year}{2017}).
\newblock \bibinfo{title}{{MRIQC: Advancing the automatic prediction of image
  quality in MRI from unseen sites}}.
\newblock {\it \bibinfo{journal}{PLOS ONE}\/},  {\it \bibinfo{volume}{12}\/},
  \bibinfo{pages}{1--21}.
\bibitem[{Esteban et~al.(2018)Esteban, Markiewicz, Blair, Moodie, Isik,
  Erramuzpe~Aliaga, Kent, Goncalves, DuPre, Snyder, Oya, Ghosh, Wright, Durnez,
  Poldrack \& Gorgolewski}]{fmriprep}
\bibinfo{author}{Esteban, O.}, \bibinfo{author}{Markiewicz, C.},
  \bibinfo{author}{Blair, R.~W.}, \bibinfo{author}{Moodie, C.},
  \bibinfo{author}{Isik, A.~I.}, \bibinfo{author}{Erramuzpe~Aliaga, A.},
  \bibinfo{author}{Kent, J.}, \bibinfo{author}{Goncalves, M.},
  \bibinfo{author}{DuPre, E.}, \bibinfo{author}{Snyder, M.},
  \bibinfo{author}{Oya, H.}, \bibinfo{author}{Ghosh, S.},
  \bibinfo{author}{Wright, J.}, \bibinfo{author}{Durnez, J.},
  \bibinfo{author}{Poldrack, R.}, \& \bibinfo{author}{Gorgolewski, K.~J.}
  (\bibinfo{year}{2018}).
\newblock \bibinfo{title}{{FMRIPrep: a robust preprocessing pipeline for
  functional MRI}}.
\newblock {\it \bibinfo{journal}{bioRxiv, 10.1101/306951}\/}, .
\bibitem[{Flandin \& Friston(2017)}]{friston}
\bibinfo{author}{Flandin, G.}, \& \bibinfo{author}{Friston, K.}
  (\bibinfo{year}{2017}).
\newblock \bibinfo{title}{Analysis of family-wise error rates in statistical
  parametric mapping using random field theory}.
\newblock {\it \bibinfo{journal}{Human Brain Mapping}\/}, .
\bibitem[{Friman et~al.(2003)Friman, Borga, Lundberg \& Knutsson}]{friman}
\bibinfo{author}{Friman, O.}, \bibinfo{author}{Borga, M.},
  \bibinfo{author}{Lundberg, P.}, \& \bibinfo{author}{Knutsson, H.}
  (\bibinfo{year}{2003}).
\newblock \bibinfo{title}{Adaptive analysis of {fMRI} data}.
\newblock {\it \bibinfo{journal}{NeuroImage}\/},  {\it \bibinfo{volume}{19}\/},
  \bibinfo{pages}{837 -- 845}.
\bibitem[{Genovese et~al.(2002)Genovese, Lazar \& Nichols}]{fdr}
\bibinfo{author}{Genovese, C.}, \bibinfo{author}{Lazar, N.}, \&
  \bibinfo{author}{Nichols, T.} (\bibinfo{year}{2002}).
\newblock \bibinfo{title}{{Thresholding of Statistical Maps in Functional
  Neuroimaging Using the False Discovery Rate}}.
\newblock {\it \bibinfo{journal}{NeuroImage}\/},  {\it \bibinfo{volume}{15}\/},
  \bibinfo{pages}{870--878}.
\bibitem[{Glover et~al.(2000)Glover, Li \& Ress}]{retroicor}
\bibinfo{author}{Glover, G.}, \bibinfo{author}{Li, T.}, \&
  \bibinfo{author}{Ress, D.} (\bibinfo{year}{2000}).
\newblock \bibinfo{title}{{Image-based method for retrospective correction of
  physiological motion effects in fMRI: RETROICOR}}.
\newblock {\it \bibinfo{journal}{Magnetic Resonance in Medicine}\/},  {\it
  \bibinfo{volume}{44}\/}, \bibinfo{pages}{162--167}.
\bibitem[{Gopinath et~al.(2018{\natexlab{a}})Gopinath, Krishnamurthy, Lacey \&
  Sathian}]{Kaundinya2}
\bibinfo{author}{Gopinath, K.}, \bibinfo{author}{Krishnamurthy, V.},
  \bibinfo{author}{Lacey, S.}, \& \bibinfo{author}{Sathian, K.}
  (\bibinfo{year}{2018}{\natexlab{a}}).
\newblock \bibinfo{title}{{Accounting for Non-Gaussian Sources of Spatial
  Correlation in Parametric Functional Magnetic Resonance Imaging Paradigms II:
  A Method to Obtain First-Level Analysis Residuals with Uniform and Gaussian
  Spatial Autocorrelation Function and Independent and Identically Distributed
  Time-Series}}.
\newblock {\it \bibinfo{journal}{Brain Connectivity}\/},  {\it
  \bibinfo{volume}{8}\/}, \bibinfo{pages}{10--21}.
\bibitem[{Gopinath et~al.(2018{\natexlab{b}})Gopinath, Krishnamurthy \&
  Sathian}]{Kaundinya1}
\bibinfo{author}{Gopinath, K.}, \bibinfo{author}{Krishnamurthy, V.}, \&
  \bibinfo{author}{Sathian, K.} (\bibinfo{year}{2018}{\natexlab{b}}).
\newblock \bibinfo{title}{{Accounting for Non-Gaussian Sources of Spatial
  Correlation in Parametric Functional Magnetic Resonance Imaging Paradigms I:
  Revisiting Cluster-Based Inferences}}.
\newblock {\it \bibinfo{journal}{Brain Connectivity}\/},  {\it
  \bibinfo{volume}{8}\/}, \bibinfo{pages}{1--9}.
\bibitem[{Gorgolewski et~al.(2017)Gorgolewski, Alfaro-Almagro, Auer, Bellec,
  CapotaÉ, Chakravarty, Churchill, Cohen, Craddock, Devenyi, Eklund, Esteban,
  Flandin, Ghosh, Guntupalli, Jenkinson, Keshavan, Kiar, Liem, Raamana,
  Raffelt, Steele, Quirion, Smith, Strother, Varoquaux, Wang, Yarkoni \&
  Poldrack}]{gorgolewski}
\bibinfo{author}{Gorgolewski, K.}, \bibinfo{author}{Alfaro-Almagro, F.},
  \bibinfo{author}{Auer, T.}, \bibinfo{author}{Bellec, P.},
  \bibinfo{author}{CapotaÉ, M.}, \bibinfo{author}{Chakravarty, M.},
  \bibinfo{author}{Churchill, N.}, \bibinfo{author}{Cohen, A.},
  \bibinfo{author}{Craddock, C.}, \bibinfo{author}{Devenyi, G.},
  \bibinfo{author}{Eklund, A.}, \bibinfo{author}{Esteban, O.},
  \bibinfo{author}{Flandin, G.}, \bibinfo{author}{Ghosh, S.},
  \bibinfo{author}{Guntupalli, S.}, \bibinfo{author}{Jenkinson, M.},
  \bibinfo{author}{Keshavan, A.}, \bibinfo{author}{Kiar, G.},
  \bibinfo{author}{Liem, F.}, \bibinfo{author}{Raamana, P.},
  \bibinfo{author}{Raffelt, D.}, \bibinfo{author}{Steele, C.},
  \bibinfo{author}{Quirion, P.}, \bibinfo{author}{Smith, R.},
  \bibinfo{author}{Strother, S.}, \bibinfo{author}{Varoquaux, G.},
  \bibinfo{author}{Wang, Y.}, \bibinfo{author}{Yarkoni, T.}, \&
  \bibinfo{author}{Poldrack, R.} (\bibinfo{year}{2017}).
\newblock \bibinfo{title}{{BIDS Apps: I}mproving ease of use, accessibility,
  and reproducibility of neuroimaging data analysis methods}.
\newblock {\it \bibinfo{journal}{PLOS Computational Biology}\/},  {\it
  \bibinfo{volume}{13}\/}, \bibinfo{pages}{1--16}.
\bibitem[{Gorgolewski et~al.(2015)Gorgolewski, Varoquaux, Rivera, Schwarz,
  Ghosh, Maumet, Sochat, Nichols, Poldrack, Poline, Yarkoni \&
  Margulies}]{neurovault}
\bibinfo{author}{Gorgolewski, K.}, \bibinfo{author}{Varoquaux, G.},
  \bibinfo{author}{Rivera, G.}, \bibinfo{author}{Schwarz, Y.},
  \bibinfo{author}{Ghosh, S.}, \bibinfo{author}{Maumet, C.},
  \bibinfo{author}{Sochat, V.}, \bibinfo{author}{Nichols, T.},
  \bibinfo{author}{Poldrack, R.}, \bibinfo{author}{Poline, J.},
  \bibinfo{author}{Yarkoni, T.}, \& \bibinfo{author}{Margulies, D.}
  (\bibinfo{year}{2015}).
\newblock \bibinfo{title}{{NeuroVault.org}: A repository for sharing
  unthresholded statistical maps, parcellations, and atlases of the human
  brain}.
\newblock {\it \bibinfo{journal}{NeuroImage}\/},  {\it
  \bibinfo{volume}{124}\/}, \bibinfo{pages}{1242--1244}.
\bibitem[{Greve \& Fischl(2018)}]{greve}
\bibinfo{author}{Greve, D.}, \& \bibinfo{author}{Fischl, B.}
  (\bibinfo{year}{2018}).
\newblock \bibinfo{title}{False positive rates in surface-based anatomical
  analysis}.
\newblock {\it \bibinfo{journal}{NeuroImage}\/},  {\it
  \bibinfo{volume}{171}\/}, \bibinfo{pages}{6--14}.
\bibitem[{Greve et~al.(2013)Greve, Brown, Mueller, Glover \& Liu}]{grevenoise}
\bibinfo{author}{Greve, D.~N.}, \bibinfo{author}{Brown, G.~G.},
  \bibinfo{author}{Mueller, B.~A.}, \bibinfo{author}{Glover, G.}, \&
  \bibinfo{author}{Liu, T.~T.} (\bibinfo{year}{2013}).
\newblock \bibinfo{title}{A survey of the sources of noise in fmri}.
\newblock {\it \bibinfo{journal}{Psychometrika}\/},  {\it
  \bibinfo{volume}{78}\/}, \bibinfo{pages}{396--416}.
\bibitem[{Griffanti et~al.(2017)Griffanti, Douaud, Bijsterbosch, Evangelisti,
  Alfaro-Almagro, Glasser, Duff, Fitzgibbon, Westphal, Carone, Beckmann \&
  Smith}]{griffanti}
\bibinfo{author}{Griffanti, L.}, \bibinfo{author}{Douaud, G.},
  \bibinfo{author}{Bijsterbosch, J.}, \bibinfo{author}{Evangelisti, S.},
  \bibinfo{author}{Alfaro-Almagro, F.}, \bibinfo{author}{Glasser, M.},
  \bibinfo{author}{Duff, E.}, \bibinfo{author}{Fitzgibbon, S.},
  \bibinfo{author}{Westphal, R.}, \bibinfo{author}{Carone, D.},
  \bibinfo{author}{Beckmann, C.}, \& \bibinfo{author}{Smith, S.}
  (\bibinfo{year}{2017}).
\newblock \bibinfo{title}{Hand classification of {fMRI ICA} noise components}.
\newblock {\it \bibinfo{journal}{NeuroImage}\/},  {\it
  \bibinfo{volume}{154}\/}, \bibinfo{pages}{188 -- 205}.
\bibitem[{Griffanti et~al.(2014)Griffanti, Salimi-Khorshidi, Beckmann,
  Auerbach, Douaud, Sexton, Zsoldos, Ebmeier, Filippini, Mackay, Moeller, Xu,
  Yacoub, Baselli, Ugurbil, Miller \& Smith}]{icafix2}
\bibinfo{author}{Griffanti, L.}, \bibinfo{author}{Salimi-Khorshidi, G.},
  \bibinfo{author}{Beckmann, C.}, \bibinfo{author}{Auerbach, E.},
  \bibinfo{author}{Douaud, G.}, \bibinfo{author}{Sexton, C.},
  \bibinfo{author}{Zsoldos, E.}, \bibinfo{author}{Ebmeier, K.},
  \bibinfo{author}{Filippini, N.}, \bibinfo{author}{Mackay, C.},
  \bibinfo{author}{Moeller, S.}, \bibinfo{author}{Xu, J.},
  \bibinfo{author}{Yacoub, E.}, \bibinfo{author}{Baselli, G.},
  \bibinfo{author}{Ugurbil, K.}, \bibinfo{author}{Miller, K.}, \&
  \bibinfo{author}{Smith, S.} (\bibinfo{year}{2014}).
\newblock \bibinfo{title}{{ICA-based} artefact removal and accelerated {fMRI}
  acquisition for improved resting state network imaging}.
\newblock {\it \bibinfo{journal}{NeuroImage}\/},  {\it \bibinfo{volume}{95}\/},
  \bibinfo{pages}{232 -- 247}.
\bibitem[{Hutton et~al.(2011)Hutton, Josephs, Stadler, Featherstone, Reid,
  Speck, Bernarding \& Weiskopf}]{7t2}
\bibinfo{author}{Hutton, C.}, \bibinfo{author}{Josephs, O.},
  \bibinfo{author}{Stadler, J.}, \bibinfo{author}{Featherstone, E.},
  \bibinfo{author}{Reid, A.}, \bibinfo{author}{Speck, O.},
  \bibinfo{author}{Bernarding, J.}, \& \bibinfo{author}{Weiskopf, N.}
  (\bibinfo{year}{2011}).
\newblock \bibinfo{title}{The impact of physiological noise correction on {fMRI
  at 7T}}.
\newblock {\it \bibinfo{journal}{NeuroImage}\/},  {\it \bibinfo{volume}{57}\/},
  \bibinfo{pages}{101 -- 112}.
\bibitem[{Jenkinson et~al.(2012)Jenkinson, Beckmann, Behrens, Woolrich \&
  Smith}]{fsl}
\bibinfo{author}{Jenkinson, M.}, \bibinfo{author}{Beckmann, C.},
  \bibinfo{author}{Behrens, T.}, \bibinfo{author}{Woolrich, M.}, \&
  \bibinfo{author}{Smith, S.} (\bibinfo{year}{2012}).
\newblock \bibinfo{title}{{FSL}}.
\newblock {\it \bibinfo{journal}{NeuroImage}\/},  {\it \bibinfo{volume}{62}\/},
  \bibinfo{pages}{782--790}.
\bibitem[{Kessler et~al.(2017)Kessler, Angstadt \& Sripada}]{kessler}
\bibinfo{author}{Kessler, D.}, \bibinfo{author}{Angstadt, M.}, \&
  \bibinfo{author}{Sripada, C.} (\bibinfo{year}{2017}).
\newblock \bibinfo{title}{{Reevaluating "cluster failure"Äù in fMRI using
  nonparametric control of the false discovery rate}}.
\newblock {\it \bibinfo{journal}{PNAS}\/},  {\it \bibinfo{volume}{114}\/},
  \bibinfo{pages}{E3372--E3373}.
\bibitem[{Kriegeskorte et~al.(2008)Kriegeskorte, Bodurka \&
  Bandettini}]{kriegeskorte}
\bibinfo{author}{Kriegeskorte, N.}, \bibinfo{author}{Bodurka, J.}, \&
  \bibinfo{author}{Bandettini, P.} (\bibinfo{year}{2008}).
\newblock \bibinfo{title}{Artifactual time-course correlations in echo-planar
  {fMRI} with implications for studies of brain function}.
\newblock {\it \bibinfo{journal}{International Journal of Imaging Systems and
  Technology}\/},  {\it \bibinfo{volume}{18}\/}, \bibinfo{pages}{345--349}.
\bibitem[{Lund et~al.(2006)Lund, Madsen, Sidaros, Luo \& Nichols}]{lund}
\bibinfo{author}{Lund, T.}, \bibinfo{author}{Madsen, K.},
  \bibinfo{author}{Sidaros, K.}, \bibinfo{author}{Luo, W.}, \&
  \bibinfo{author}{Nichols, T.} (\bibinfo{year}{2006}).
\newblock \bibinfo{title}{{Non-white noise in fMRI: Does modelling have an
  impact?}}
\newblock {\it \bibinfo{journal}{NeuroImage}\/},  {\it \bibinfo{volume}{29}\/},
  \bibinfo{pages}{54 -- 66}.
\bibitem[{Moeller et~al.(2010)Moeller, Yacoub, Olman, Auerbach, Strupp, Harel
  \& Ugüurbil}]{multiband}
\bibinfo{author}{Moeller, S.}, \bibinfo{author}{Yacoub, E.},
  \bibinfo{author}{Olman, C.}, \bibinfo{author}{Auerbach, E.},
  \bibinfo{author}{Strupp, J.}, \bibinfo{author}{Harel, N.}, \&
  \bibinfo{author}{Ugüurbil, K.} (\bibinfo{year}{2010}).
\newblock \bibinfo{title}{{Multiband multislice GE-EPI at 7 tesla, with 16-fold
  acceleration using partial parallel imaging with application to high spatial
  and temporal whole-brain fMRI}}.
\newblock {\it \bibinfo{journal}{Magnetic Resonance in Medicine}\/},  {\it
  \bibinfo{volume}{63}\/}, \bibinfo{pages}{1144--1153}.
\bibitem[{Mueller et~al.(2017)Mueller, Lepsien, M\"{o}ller \&
  Lohmann}]{mueller}
\bibinfo{author}{Mueller, K.}, \bibinfo{author}{Lepsien, J.},
  \bibinfo{author}{M\"{o}ller, H.}, \& \bibinfo{author}{Lohmann, G.}
  (\bibinfo{year}{2017}).
\newblock \bibinfo{title}{Commentary: {Cluster failure: Why fMRI} inferences
  for spatial extent have inflated false-positive rates}.
\newblock {\it \bibinfo{journal}{Frontiers in Human Neuroscience}\/},  {\it
  \bibinfo{volume}{11}\/}.
\bibitem[{Mumford(2017)}]{mumford}
\bibinfo{author}{Mumford, J.} (\bibinfo{year}{2017}).
\newblock \bibinfo{title}{{A comprehensive review of group level model
  performance in the presence of heteroscedasticity: Can a single model control
  Type I errors in the presence of outliers?}}
\newblock {\it \bibinfo{journal}{NeuroImage}\/},  {\it
  \bibinfo{volume}{147}\/}, \bibinfo{pages}{658 -- 668}.
\bibitem[{Murphy et~al.(2009)Murphy, Birn, Handwerker, Jones \&
  Bandettini}]{gsr1}
\bibinfo{author}{Murphy, K.}, \bibinfo{author}{Birn, R.},
  \bibinfo{author}{Handwerker, D.}, \bibinfo{author}{Jones, T.}, \&
  \bibinfo{author}{Bandettini, P.} (\bibinfo{year}{2009}).
\newblock \bibinfo{title}{{The impact of global signal regression on resting
  state correlations: Are anti-correlated networks introduced?}}
\newblock {\it \bibinfo{journal}{NeuroImage}\/},  {\it \bibinfo{volume}{44}\/},
  \bibinfo{pages}{893 -- 905}.
\bibitem[{Murphy \& Fox(2017)}]{gsr2}
\bibinfo{author}{Murphy, K.}, \& \bibinfo{author}{Fox, M.}
  (\bibinfo{year}{2017}).
\newblock \bibinfo{title}{Towards a consensus regarding global signal
  regression for resting state functional connectivity {MRI}}.
\newblock {\it \bibinfo{journal}{NeuroImage}\/},  {\it
  \bibinfo{volume}{154}\/}, \bibinfo{pages}{169 -- 173}.
\bibitem[{Nichols et~al.(2017)Nichols, Eklund \& Knutsson}]{nichols}
\bibinfo{author}{Nichols, T.}, \bibinfo{author}{Eklund, A.}, \&
  \bibinfo{author}{Knutsson, H.} (\bibinfo{year}{2017}).
\newblock \bibinfo{title}{A defense of using resting state {fMRI} as null data
  for estimating false positive rates}.
\newblock {\it \bibinfo{journal}{Cognitive Neuroscience}\/},  {\it
  \bibinfo{volume}{8}\/}, \bibinfo{pages}{144--149}.
\bibitem[{Nooner et~al.(2012)Nooner, Colcombe, Tobe, Mennes, Benedict, Moreno,
  Panek, Brown, Zavitz, Li, Sikka, Gutman, Bangaru, Schlachter, Kamiel, Anwar,
  Hinz, Kaplan, Rachlin, Adelsberg, Cheung, Khanuja, Yan, Craddock, Calhoun,
  Courtney, King, Wood, Cox, Kelly, DiMartino, Petkova, Reiss, Duan, Thompsen,
  Biswal, Coffey, Hoptman, Javitt, Pomara, Sidtis, Koplewicz, Castellanos,
  Leventhal \& Milham}]{rockland}
\bibinfo{author}{Nooner, K.}, \bibinfo{author}{Colcombe, S.},
  \bibinfo{author}{Tobe, R.}, \bibinfo{author}{Mennes, M.},
  \bibinfo{author}{Benedict, M.}, \bibinfo{author}{Moreno, A.},
  \bibinfo{author}{Panek, L.}, \bibinfo{author}{Brown, S.},
  \bibinfo{author}{Zavitz, S.}, \bibinfo{author}{Li, Q.},
  \bibinfo{author}{Sikka, S.}, \bibinfo{author}{Gutman, D.},
  \bibinfo{author}{Bangaru, S.}, \bibinfo{author}{Schlachter, R.},
  \bibinfo{author}{Kamiel, S.}, \bibinfo{author}{Anwar, A.},
  \bibinfo{author}{Hinz, C.}, \bibinfo{author}{Kaplan, M.},
  \bibinfo{author}{Rachlin, A.}, \bibinfo{author}{Adelsberg, S.},
  \bibinfo{author}{Cheung, B.}, \bibinfo{author}{Khanuja, R.},
  \bibinfo{author}{Yan, C.}, \bibinfo{author}{Craddock, C.},
  \bibinfo{author}{Calhoun, V.}, \bibinfo{author}{Courtney, W.},
  \bibinfo{author}{King, M.}, \bibinfo{author}{Wood, D.}, \bibinfo{author}{Cox,
  C.}, \bibinfo{author}{Kelly, C.}, \bibinfo{author}{DiMartino, A.},
  \bibinfo{author}{Petkova, E.}, \bibinfo{author}{Reiss, P.},
  \bibinfo{author}{Duan, N.}, \bibinfo{author}{Thompsen, D.},
  \bibinfo{author}{Biswal, B.}, \bibinfo{author}{Coffey, B.},
  \bibinfo{author}{Hoptman, M.}, \bibinfo{author}{Javitt, D.},
  \bibinfo{author}{Pomara, N.}, \bibinfo{author}{Sidtis, J.},
  \bibinfo{author}{Koplewicz, H.}, \bibinfo{author}{Castellanos, F.},
  \bibinfo{author}{Leventhal, B.}, \& \bibinfo{author}{Milham, M.}
  (\bibinfo{year}{2012}).
\newblock \bibinfo{title}{{The NKI-Rockland Sample: A Model for Accelerating
  the Pace of Discovery Science in Psychiatry}}.
\newblock {\it \bibinfo{journal}{Frontiers in Neuroscience}\/},  {\it
  \bibinfo{volume}{6}\/}, \bibinfo{pages}{152}.
\bibitem[{Poldrack et~al.(2013)Poldrack, Barch, Mitchell, Wager, Wagner,
  Devlin, Cumba, Koyejo \& Milham}]{openfmri}
\bibinfo{author}{Poldrack, R.}, \bibinfo{author}{Barch, D.},
  \bibinfo{author}{Mitchell, J.}, \bibinfo{author}{Wager, T.},
  \bibinfo{author}{Wagner, A.}, \bibinfo{author}{Devlin, J.},
  \bibinfo{author}{Cumba, C.}, \bibinfo{author}{Koyejo, O.}, \&
  \bibinfo{author}{Milham, M.} (\bibinfo{year}{2013}).
\newblock \bibinfo{title}{Toward open sharing of task-based {fMRI} data: the
  {OpenfMRI} project}.
\newblock {\it \bibinfo{journal}{Frontiers in Neuroinformatics}\/},  {\it
  \bibinfo{volume}{7}\/}.
\bibitem[{Pruim et~al.(2015)Pruim, Mennes, van Rooij, Llera, Buitelaar \&
  Beckmann}]{aroma}
\bibinfo{author}{Pruim, R.~H.}, \bibinfo{author}{Mennes, M.},
  \bibinfo{author}{van Rooij, D.}, \bibinfo{author}{Llera, A.},
  \bibinfo{author}{Buitelaar, J.~K.}, \& \bibinfo{author}{Beckmann, C.~F.}
  (\bibinfo{year}{2015}).
\newblock \bibinfo{title}{{ICA-AROMA: A robust ICA-based strategy for removing
  motion artifacts from fMRI data}}.
\newblock {\it \bibinfo{journal}{NeuroImage}\/},  {\it
  \bibinfo{volume}{112}\/}, \bibinfo{pages}{267--277}.
\bibitem[{Risk et~al.(2018)Risk, Kociuba \& Rowe}]{risk}
\bibinfo{author}{Risk, B.}, \bibinfo{author}{Kociuba, M.}, \&
  \bibinfo{author}{Rowe, D.} (\bibinfo{year}{2018}).
\newblock \bibinfo{title}{Impacts of simultaneous multislice acquisition on
  sensitivity and specificity in {fMRI}}.
\newblock {\it \bibinfo{journal}{NeuroImage}\/},  {\it
  \bibinfo{volume}{172}\/}, \bibinfo{pages}{538--553}.
\bibitem[{Salimi-Khorshidi et~al.(2014)Salimi-Khorshidi, Douaud, Beckmann,
  Glasser, Griffanti \& Smith}]{icafix1}
\bibinfo{author}{Salimi-Khorshidi, G.}, \bibinfo{author}{Douaud, G.},
  \bibinfo{author}{Beckmann, C.}, \bibinfo{author}{Glasser, M.},
  \bibinfo{author}{Griffanti, L.}, \& \bibinfo{author}{Smith, S.}
  (\bibinfo{year}{2014}).
\newblock \bibinfo{title}{Automatic denoising of functional {MRI} data:
  Combining independent component analysis and hierarchical fusion of
  classifiers}.
\newblock {\it \bibinfo{journal}{NeuroImage}\/},  {\it \bibinfo{volume}{90}\/},
  \bibinfo{pages}{449 -- 468}.
\bibitem[{Slotnick(2016)}]{slotnick}
\bibinfo{author}{Slotnick, S.} (\bibinfo{year}{2016}).
\newblock \bibinfo{title}{{Resting-state fMRI data reflects default network
  activity rather than null data: A defense of commonly employed methods to
  correct for multiple comparisons}}.
\newblock {\it \bibinfo{journal}{Cognitive Neuroscience}\/},  {\it
  \bibinfo{volume}{8}\/}, \bibinfo{pages}{141--143}.
\bibitem[{Slotnick(2017)}]{slotnick2}
\bibinfo{author}{Slotnick, S.} (\bibinfo{year}{2017}).
\newblock \bibinfo{title}{{Cluster success: fMRI inferences for spatial extent
  have acceptable false-positive rates}}.
\newblock {\it \bibinfo{journal}{Cognitive Neuroscience}\/},  {\it
  \bibinfo{volume}{8}\/}, \bibinfo{pages}{150--155}.
\bibitem[{Smith \& Nichols(2009)}]{tfce}
\bibinfo{author}{Smith, S.}, \& \bibinfo{author}{Nichols, T.}
  (\bibinfo{year}{2009}).
\newblock \bibinfo{title}{Threshold-free cluster enhancement: Addressing
  problems of smoothing, threshold dependence and localisation in cluster
  inference}.
\newblock {\it \bibinfo{journal}{NeuroImage}\/},  {\it \bibinfo{volume}{44}\/},
  \bibinfo{pages}{83--98}.
\bibitem[{Stelzer et~al.(2013)Stelzer, Chen \& Turner}]{stelzer}
\bibinfo{author}{Stelzer, J.}, \bibinfo{author}{Chen, Y.}, \&
  \bibinfo{author}{Turner, R.} (\bibinfo{year}{2013}).
\newblock \bibinfo{title}{{Statistical inference and multiple testing
  correction in classification-based multi-voxel pattern analysis (MVPA):
  Random permutations and cluster size control}}.
\newblock {\it \bibinfo{journal}{NeuroImage}\/},  {\it \bibinfo{volume}{65}\/},
  \bibinfo{pages}{69 -- 82}.
\bibitem[{Triantafyllou et~al.(2005)Triantafyllou, Hoge, Krueger, Wiggins,
  Potthast, Wiggins \& Wald}]{7t1}
\bibinfo{author}{Triantafyllou, C.}, \bibinfo{author}{Hoge, R.},
  \bibinfo{author}{Krueger, G.}, \bibinfo{author}{Wiggins, C.},
  \bibinfo{author}{Potthast, A.}, \bibinfo{author}{Wiggins, G.}, \&
  \bibinfo{author}{Wald, L.} (\bibinfo{year}{2005}).
\newblock \bibinfo{title}{{Comparison of physiological noise at 1.5 T, 3 T and
  7 T and optimization of fMRI acquisition parameters}}.
\newblock {\it \bibinfo{journal}{NeuroImage}\/},  {\it \bibinfo{volume}{26}\/},
  \bibinfo{pages}{243 -- 250}.
\bibitem[{Wager et~al.(2005)Wager, Keller, Lacey \& Jonides}]{wager}
\bibinfo{author}{Wager, T.}, \bibinfo{author}{Keller, M.},
  \bibinfo{author}{Lacey, S.}, \& \bibinfo{author}{Jonides, J.}
  (\bibinfo{year}{2005}).
\newblock \bibinfo{title}{Increased sensitivity in neuroimaging analyses using
  robust regression}.
\newblock {\it \bibinfo{journal}{NeuroImage}\/},  {\it \bibinfo{volume}{15}\/},
  \bibinfo{pages}{99--113}.
\bibitem[{Wald \& Polimeni(2017)}]{wald}
\bibinfo{author}{Wald, L.}, \& \bibinfo{author}{Polimeni, J.}
  (\bibinfo{year}{2017}).
\newblock \bibinfo{title}{{Impacting the effect of fMRI noise through hardware
  and acquisition choices - implications for controlling false positive
  rates}}.
\newblock {\it \bibinfo{journal}{NeuroImage}\/},  {\it
  \bibinfo{volume}{154}\/}, \bibinfo{pages}{15--22}.
\bibitem[{Welvaert \& Rosseel(2014)}]{simulation}
\bibinfo{author}{Welvaert, M.}, \& \bibinfo{author}{Rosseel, Y.}
  (\bibinfo{year}{2014}).
\newblock \bibinfo{title}{A review of fmri simulation studies}.
\newblock {\it \bibinfo{journal}{{PLOS ONE}}\/},  {\it \bibinfo{volume}{9}\/},
  \bibinfo{pages}{1--10}.
\bibitem[{Winkler et~al.(2014)Winkler, Ridgway, Webster, Smith \&
  Nichols}]{winkler}
\bibinfo{author}{Winkler, A.}, \bibinfo{author}{Ridgway, G.},
  \bibinfo{author}{Webster, M.}, \bibinfo{author}{Smith, S.}, \&
  \bibinfo{author}{Nichols, T.} (\bibinfo{year}{2014}).
\newblock \bibinfo{title}{Permutation inference for the general linear model}.
\newblock {\it \bibinfo{journal}{NeuroImage}\/},  {\it \bibinfo{volume}{92}\/},
  \bibinfo{pages}{381--397}.
\bibitem[{Woo et~al.(2014)Woo, Krishnan \& Wager}]{woo}
\bibinfo{author}{Woo, C.}, \bibinfo{author}{Krishnan, A.}, \&
  \bibinfo{author}{Wager, T.} (\bibinfo{year}{2014}).
\newblock \bibinfo{title}{Cluster-extent based thresholding in {fMRI} analyses:
  {Pitfalls} and recommendations}.
\newblock {\it \bibinfo{journal}{NeuroImage}\/},  {\it \bibinfo{volume}{91}\/},
  \bibinfo{pages}{412 -- 419}.
\bibitem[{Woolrich(2008)}]{woolrich}
\bibinfo{author}{Woolrich, M.} (\bibinfo{year}{2008}).
\newblock \bibinfo{title}{Robust group analysis using outlier inference}.
\newblock {\it \bibinfo{journal}{NeuroImage}\/},  {\it \bibinfo{volume}{41}\/},
  \bibinfo{pages}{286--301}.
\bibitem[{Yeo \& Johnson(2000)}]{yeo}
\bibinfo{author}{Yeo, I.}, \& \bibinfo{author}{Johnson, R.}
  (\bibinfo{year}{2000}).
\newblock \bibinfo{title}{{A New Family of Power Transformations to Improve
  Normality or Symmetry}}.
\newblock {\it \bibinfo{journal}{Biometrika}\/},  {\it \bibinfo{volume}{87}\/},
  \bibinfo{pages}{954 -- 959}.
\bibitem[{Yeung(2018)}]{yeung}
\bibinfo{author}{Yeung, A.} (\bibinfo{year}{2018}).
\newblock \bibinfo{title}{An updated survey on statistical thresholding and
  sample size of {fMRI} studies}.
\newblock {\it \bibinfo{journal}{Frontiers in Human Neuroscience}\/},  {\it
  \bibinfo{volume}{12:16}\/}.

\end{thebibliography}

\end{document}